%% file: OWNS-S_final.tex
\documentclass[a4paper,fleqn]{cas-sc}

\usepackage[authoryear]{natbib}

\def\tsc#1{\csdef{#1}{\textsc{\lowercase{#1}}\xspace}}
\tsc{WGM}
\tsc{QE}
\tsc{EP}
\tsc{PMS}
\tsc{BEC}
\tsc{DE}

\usepackage{graphicx}
\usepackage{epstopdf,epsfig}
\usepackage{newtxtext}

\usepackage{amsmath}
\usepackage{dsfont}

\usepackage{wrapfig}
\usepackage{xcolor}
\usepackage{tikz}
\usepackage{subcaption}

\usepackage{mathtools}

\usepackage{gensymb}

\usepackage{pgfplots}
\usepackage{algpseudocode}
\usepackage{algorithm}

\usetikzlibrary{arrows.meta}
\usepackage{cancel}
\usepackage{soul}

\usepackage{tabularx,array}

\renewcommand{\arraystretch}{1.05}
\pgfplotsset{compat=1.18}

\ExplSyntaxOn
\cs_gset:Npn \__first_footerline:
  {
    \group_begin:
    \small\sffamily
    \__short_authors:
    \group_end:
  }
\ExplSyntaxOff

\makeatletter
\def\fps@figure{htbp}
\def\fps@table{htbp}
\makeatother

\begin{document}
\let\WriteBookmarks\relax
\shorttitle{OWNS-Summation}
\shortauthors{E.J. Badcock \& S. Mughal}

\title [mode = title]{Stable and Efficient One-Way Modelling of Convective Disturbances in Laminar Boundary Layers: OWNS-Summation} 




\author[]{Elliot J. Badcock}[orcid=0009-0008-4977-2286]
\cormark[1]
\ead{ejb321@ic.ac.uk; ej.badcock@outlook.com}
\ead[url]{https://profiles.imperial.ac.uk/e.badcock21}
\credit{Conceptualisation, Methodology, Software, Investigation, Writing - Original Draft}

\affiliation[]{organization={Department of Mathematics, Imperial College London},
                addressline={South Kensington Campus}, 
                city={London},
                postcode={SW7 2AZ}, 
                country={United Kingdom}}

\author[]{Shahid Mughal}[orcid=0000-0002-1675-489X]
\ead{s.mughal@imperial.ac.uk}
\ead[URL]{https://profiles.imperial.ac.uk/s.mughal}
\credit{Supervision, Project administration, Writing - Review \& Editing}

\cortext[cor1]{Corresponding author}

\begin{abstract}
One-way spatial marching methods separate and remove upstream- from
downstream-propagating disturbances by applying a rational approximation of a
spectral projector. Among existing one-way Navier--Stokes (OWNS)
formulations the recursive variant, OWNS-R, is the most economical, evaluating
that approximation as a product of $N$ resolvent factors. The product amplifies rounding error multiplicatively, so the
approximation order is bounded above by a flow-dependent threshold that cannot be determined in advance.
 
We reformulate the same approximation as an additive partial-fraction sum
(OWNS-Summation, OWNS-S). In exact arithmetic, with identical weights and poles, the two evaluations are equivalent; in floating point they are not. Each of the $N$ resolvent solves
acts on the same input state and contributes one term of a weighted sum, so
errors are never amplified multiplicatively: the approximation order $N$ becomes a
pure convergence parameter, and the solves may be executed in parallel.
Supplied with the same auxiliary poles as OWNS-R, the summation remains
accurate in every configuration tested, identifying the recursive
evaluation rather than the poles as the dominant source of the instability. The OWNS-S procedure presented in this paper overcomes the numerical
instability in existing OWNS-R approaches. 
To further strengthen the method, a paired greedy parameter-selection
procedure is presented whose candidates are drawn from analytic estimates
of the upstream and downstream spectral regions, so that no
eigen-decomposition is required during the numerical march.
OWNS-S is validated on incompressible,
hypersonic and transonic boundary layers. In the transonic case the
disturbance spectrum reorganises from a subsonic to a supersonic topology
during the numerical march, and the one-way computation is carried continuously through
the transition at a streamwise resolution {{that}} the parabolised stability equations
cannot {attain}.
\end{abstract}


\maketitle

\section{Introduction}\label{sec:introduction}

Accurate prediction of laminar--turbulent transition is central to the design of
transonic and hypersonic vehicles, swept wings, and atmospheric re-entry
systems.  Classical linear stability theory (LST) provides a low-cost
description of disturbance growth by treating the baseflow as locally parallel
and analysing individual normal eigenmodes at successive streamwise stations.
This approximation is effective for determining the local stability of
individual modes, but it lacks disturbance history, neglects non-parallel
effects, and does not provide the initial- or boundary-value structure needed to
track disturbance evolution through a developing flow.  These limitations are
removed with the linearised harmonic Navier--Stokes (LHNS) equations, which
retain the streamwise development of the baseflow and govern the spatial
evolution of disturbances as a time-harmonic, forced boundary-value problem.
The LHNS equations are elliptic in the streamwise direction: they support
downstream-propagating vortical, entropic and acoustic waves as well as
upstream-propagating acoustic waves.  In general they must therefore be treated
with global or elliptic
solvers~\citep{theofilisGlobalLinearInstability2011,appelBoundaryLayerInstabilities2021},
which are substantially more expensive than local analysis.

For many developing boundary layers, however, the instabilities of interest are
convective, and their physically relevant evolution is predominantly
downstream.  Flows in which the disturbance dynamics are absolutely unstable,
as may occur in the presence of laminar separation bubbles (LSBs), fall outside
this class and are not considered here.
This makes one-way spatial marching methods attractive: they advance the
disturbance field station by station, replacing the global solve with a sequence
of local ones, thereby significantly reducing both computational cost and memory
requirements.  The central difficulty is that the upstream-propagating modes
supported by the LHNS equations can destabilise a na\"ive downstream march.
Practical one-way formulations must therefore remove or suppress the
upstream-travelling components while retaining the downstream dynamics of
interest.

The parabolised stability equations (PSE) introduced by
\citet{herbertBertolotti1987} achieve a stabilised downstream march through a
wave-carrier decomposition that separates disturbances into slowly varying shape
functions and rapidly oscillating phases.  
Although successful across subsonic and supersonic
regimes~\citep{malikNumericalMethodsHypersonic1990,mughalactive}, the PSE are
not truly parabolic, since the underlying operator retains the elliptic
character of the LHNS.  Stability of the march therefore requires a
sufficiently large streamwise step, which damps the residual elliptic
component~\citep{liMathematicalNatureParabolized1995}. The resulting minimum
step size acts as a floor: the streamwise resolution cannot be refined below it,
so a step-size refinement study cannot be carried out at all.  Baseflows can vary strongly and rapidly, for example under rapid
pressure-gradient change or a change in surface geometry (see \S\ref{sec:swift}). Tracking the
disturbance accurately can then require spatial scales finer than that floor
permits, so the restriction is not merely a matter of efficiency.

The One-Way Navier--Stokes (OWNS) framework takes a different approach.
Rather than suppressing upstream modes through wave-carrier factoring and
step-size control, OWNS methods construct a \emph{projection operator} that
removes upstream-propagating components of the LHNS operator, retaining only
the downstream-propagating subspace.
Building on non-reflecting
boundary-condition
theory~\citep{engquistAbsorbingBoundaryConditions1977,thompsonTimeDependentBoundary1987a},
\citet{towneOnewaySpatialIntegration2015} introduced the first practical
one-way formulation, the outflow method (OWNS-O), subsequently extended to
viscous boundary-layer stability~\citep{rigasOneWayNavierStokes2017}.  The
projection formulation (OWNS-P), which splits the state explicitly into
upstream- and downstream-propagating components and admits a forcing term, was
developed by \citet{towneEfficientGlobalResolvent2022} for resolvent analysis.
Both operate in characteristic variables.

The OWNS-Recursion (OWNS-R)
method~\citep{zhuRecursiveOnewayNavierStokes2023} realises the rational
approximation through a \emph{recursive product} of resolvent factors,
constructed directly from the pencil without imposing the
characteristic-variable transformation employed by OWNS-O and OWNS-P.  This
formulation is computationally efficient and memory economical, with the
spectral filtering behaviour controlled by $N$ tunable ``recursion-parameter''
pairs. A wave-carrier-factored extension of OWNS-R relaxes the streamwise step
restriction further~\citep{badcocktcfd2026}.  The spectral content of interest
is shifted towards a prescribed carrier wavenumber, so that the residual shape
function varies slowly and a larger step resolves the same disturbance detail.

The accuracy of OWNS-R, however, deteriorates once the approximation order
grows beyond a certain size.
\citet{zhuRecursiveOnewayNavierStokes2023} identify the deterioration and
attribute it to rounding error in the polynomial solve that returns the
auxiliary poles, recommending that those poles be computed in extended
precision.  \citet{sleemanGreedyRecursionParameter2025a} (first posted 2025) adopt that remedy and report that the
deterioration survives it: with the poles computed in quadruple precision the projection can still deteriorate when the order $N$ is sufficiently large.
The structural role of the recursive product is not analysed in detail there, and the
practical response is to keep the order small.
The order~$N$ is an approximation parameter rather than
a physical one: it sets the fidelity with which the rational filter reproduces
the exact spectral projector, and the LHNS operator itself is unchanged by it.
Its consequences are nonetheless felt in the computed physics, and in two
distinct ways. 
Upstream content that the filter fails to suppress grows under the downstream
march and destabilises it, unless the discretisation damps it first.  A
strongly retained mode terminates the computation, and a weak one may leave a
localised signature, often in the pressure.  Failure of this kind is
therefore visible when it occurs. Downstream modes that the filter fails to retain faithfully are
attenuated or distorted instead, and the march then returns disturbance
amplitudes that are wrong while remaining smooth and plausible. 

Large~$N$ is required whenever the parameter placement is not sufficient to separate the spectrum. Some spectral topologies can be difficult to separate, for example: in subsonic boundary-layers near pinch-point coalescence, where
the upstream and downstream subsets approach arbitrarily closely; where
discrete upstream-propagating acoustic modes exist and migrate during the numerical march, so that
a placement exact at one station is inexact at the next; and through rapid
spectral reorganisation, where no single placement serves the whole domain.
These situations arise in the incompressible, hypersonic and transonic boundary
layers considered here, respectively, and it is precisely there that the
deterioration becomes problematic.

This paper introduces the \emph{OWNS-Summation} (OWNS-S) framework, which
evaluates the same rational approximation of the one-way projector as an
\emph{additive} partial-fraction expansion rather than a recursive product.
The distinction is elementary.  OWNS-R applies $N$ resolvent factors in
sequence, each acting on the output of the last; OWNS-S applies $N$ resolvents to the same input state and forms a weighted sum of the results. 
In exact
arithmetic the two return the same operator.  In floating point they do not,
because a product accumulates local errors multiplicatively and a sum
accumulates them additively.  The amplification mechanism that limits OWNS-R is
therefore absent from the summation, and large approximation orders become
usable.

That difference also identifies the mechanism.  Both evaluations are supplied
with the same auxiliary poles from the same polynomial solve, so any
inaccuracy in those poles is common to the two: the poles are held fixed
while the evaluation is varied.  Under that control the summation produces
accurate projections in every configuration considered here, at orders well
past those at which the product has failed.  What the product does and the sum
does not is propagate the error committed at one solve through the solves that
follow it, and that propagation is what \S\ref{sec:forward_error} sets out and measures.
A second OWNS-S variant (hereafter OWNS-S~(LS)) separates the pole locations
from the weights derived from them. A least-squares recalibration obtains the weights by enforcing the
separation conditions at the computed poles rather than by evaluating
residues, so any degradation of those conditions by rounding error in
the polynomial solve is repaired rather than reproduced. 

That an additive rational filter accumulates error more benignly than a
multiplicative one is not itself new, and partial-fraction evaluation of
contour-integral projectors is standard in contour-based
eigen-solvers~\citep{polizziDensitymatrixbasedAlgorithmSolving2009,
nakatsukasaComputingFundamentalMatrix2016}.  
The contribution here is not that observation but its bearing on one-way
marching.  Previously the OWNS-R filter has been evaluated as a product, the
large-$N$ breakdown has been documented and mitigated by holding the order
small, and the product structure has not been confirmed as its leading source.

The key consequence is that the approximation order in OWNS-S is a pure
convergence parameter.  It may be raised until the computed solution ceases to
change, so that a converged result can be certified in the ordinary way.  In
OWNS-R it cannot, because the order at which the product begins to amplify is
not known in advance and raising the order to test convergence is the operation
that destroys it.  Where PSE imposes a floor on the streamwise step and OWNS-R
imposes an unknown ceiling on the approximation order $N$, OWNS-S imposes neither.

The independence of the solves has a second consequence, which is
computational.  Since no solve depends on another, the $N$ solves at a station
may be distributed across $n_\mathrm{t}$ parallel threads, and the per-station
critical path falls from $1+N$ solves to $1+\lceil N/n_\mathrm{t}\rceil$. 
This is hugely beneficial for an OWNS practitioner and makes the study of significantly larger domains practicable. 

Alongside the summation reformulation we introduce a greedy
parameter-selection strategy, extending~\citet{sleemanGreedyRecursionParameter2025a}, which places
the parameters by minimising a scalar measure of the projection error over a set
of candidates.  The candidates of~\citet{sleemanGreedyRecursionParameter2025a} are computed
eigenvalues: a full eigen-decomposition is required at the inlet and whenever
the count of upstream and downstream characteristics changes, and the selected
eigenvalues are tracked between stations by local sparse eigen-solves. 
We instead form the candidate pool by oversampling a modified heuristic
parameter placement given in previous works \citep{badcocktcfd2026, ejb2025}.
That placement is constructed from analytic estimates of the upstream and
downstream regions of the spectrum, so neither an eigen-decomposition nor
eigenvalue tracking is required during the numerical march.  The pool retains one
spectral input: an approximate location for the upstream discrete acoustic
modes, taken from a computed spectrum at a single station
(\S\ref{sec:heuristic}).
To confirm that nothing is lost by this substitution,
one configuration is additionally run with the pool seeded from eigenvalues
computed by the QZ algorithm, the candidate set being the only quantity changed. 

The paper is organised as follows.  \S\ref{sec:mathframework} presents the
mathematical framework, including spectral structure and projection theory.
\S\ref{sec:owns_s} develops the OWNS-S projection operator and weight
calculation, and introduces the least-squares weight recalibration.
\S\ref{sec:owns_r} contrasts it with OWNS-R and analyses the finite-precision
error accumulation that distinguishes the two.
\S\ref{sec:parameter_selection} discusses retention and removal parameter
placement and the greedy algorithm.  \S\ref{sec:results} validates OWNS-S
across incompressible, hypersonic, and transonic disturbance configurations.
\S\ref{sec:conclusion} summarises the main findings.

\section{Mathematical Formulation}\label{sec:mathframework}

\subsection{Governing Equations and Pencil Formulation}\label{sec:pencil}

We decompose the total state into a steady, spanwise-invariant baseflow
$Q(x,y)$ and a small harmonic perturbation with spanwise
wavenumber~$\beta$ and frequency~$\omega$,
\begin{equation}\label{eqn:dist_decomp}
    q^{\mathrm{tot}}(x,y,z,t)
    = Q(x,y) + q(x,y)\,\exp\!\left({i\beta z - i\omega t}\right).
\end{equation}
Substitution into the Navier--Stokes equations and linearisation
with respect to~$q$ yields the linearised harmonic Navier--Stokes
(LHNS) equations,
\begin{equation}\label{eqn:LHNS}
    \mathrm{A}\!\left(\frac{\partial}{\partial y};\,\beta\right)
    \frac{\partial q}{\partial x}
    = \mathrm{B}\!\left(\frac{\partial}{\partial y};\,\beta,\,\omega\right) q
    + \mathrm{C}\,\frac{\partial^2 q}{\partial x^2},
\end{equation}
where $\mathrm{A}$, $\mathrm{B}$ and $\mathrm{C}$ are $5\times5$
linear operators in the wall-normal coordinate.  The baseflow~$Q$ is
obtained from a boundary-layer solver~\citep{mughalactive}.

Discretisation in the wall-normal direction with $n_y\in\mathbb{N}$
points yields the semi-discrete system
\begin{equation}\label{eqn:two-way2}
    \mathbf{A}\frac{\mathrm{d}\phi}{\mathrm{d} x}
    = \mathbf{B}\,\phi
    + \mathbf{C}\frac{\mathrm{d}^2\phi}{\mathrm{d} x^2},
\end{equation}
where $\mathbf{A}$, $\mathbf{B}$ and $\mathbf{C}$ are
discretisations of the corresponding continuous operators.

Consistent with boundary-layer OWNS formulations in prior
work~\citep{zhuRecursiveOnewayNavierStokes2023,
sleemanNonlinearStabilityWallbounded2023,
badcocktcfd2026}, the projection operator is
constructed from the first-order pencil
\begin{equation}
    \mathbf{\Pi} = (\mathbf{B},\mathbf{A})
\end{equation}
only.  This requires care, as $\mathbf{C}$ is not discarded from the
problem.  Eq.~\eqref{eqn:two-way2} is marched in full (albeit projected) with $\mathbf{C}$ included.  The
justification is that the contribution of $\mathbf{C}$ carries the
reciprocal-Reynolds-number factor, and its neglect has not been observed to
affect the upstream--downstream separation, here or in previously published
OWNS work.

The approximation is nonetheless real.  Since $\mathbf{C}$ does not commute
with the projection operator, the operator built from the first-order pencil is not an
exact spectral splitting of the second-order system it is applied to.
Constructing one would require a projector for the full quadratic pencil whose additional parasitic
branches complicate the upstream--downstream partition substantially.

The spatial modes of $\mathbf{\Pi}$ satisfy
\begin{equation}\label{eq:gevp}
    \mathbf{B}\,v_k = i\alpha_k\,\mathbf{A}\,v_k,
\end{equation}
where $\alpha_k\in\mathbb{C}$ are the spatial eigenvalues and $v_k$
the corresponding right eigenvectors.  We write
    $\sigma(\mathbf{\Pi})$
for the finite spectrum of the pencil, the set of all such finite eigenvalues $\alpha_k$.  The
singularity of~$\mathbf{A}$, induced by the boundary conditions, places some
of them at infinity; these are excluded from $\sigma(\mathbf{\Pi})$ and correspond to algebraic constraints rather than to
propagating disturbances.

\subsection{Spectral Structure}\label{sec:spec_structure}

Equation~\eqref{eqn:two-way2} is a two-way evolution equation whose
spectrum accommodates disturbances propagating in both directions.
Through Briggs'
criterion~\citep{briggsElectronStreamInteractionPlasmas1964a}, the
finite spectrum of~$\mathbf{\Pi}$ is partitioned as
\begin{equation}\label{eqn:twoway_spec}
    \sigma\!\left(\mathbf{\Pi}\right) =
    \begin{cases}
        \text{downstream-propagating eigenvalues,} \\
        \text{upstream-propagating eigenvalues,} \\
        \text{pinch-point eigenvalues (neither).}
    \end{cases}
\end{equation}
The third category arises only at isolated coalescence points of the
continuous spectrum.  In subsonic boundary layers, the downstream and
upstream acoustic branches meet at such a pinch-point where the two
spectral subsets are arbitrarily close in the continuous problem.
After discretisation the eigenvalues near the pinch are distinct and
are classified individually by Briggs' criterion as upstream or
downstream, so the third category is empty in practice.  However, the
proximity of the downstream and upstream eigenvalues near the pinch
makes accurate spectral separation challenging.

\begin{figure}[htbp!]
\centering
\begin{subfigure}{5.5cm}
\centering
\input{figs2/tikz/spec1}
\caption{Subsonic disturbance}
\label{fig:spec_cartoon1a}
\end{subfigure}
\hspace{1.5cm}
\begin{subfigure}{5.5cm}
\centering
\input{figs2/tikz/spec3}
\caption{Supersonic disturbance}
\label{fig:spec_cartoon1b}
\end{subfigure}
\caption{(a) Subsonic disturbance  at
($M=0.5$, $\beta=0$) showing an acoustic pinch-point and
both upstream and downstream continua; the Tollmien--Schlichting
discrete mode is denoted by T-S.  (b) Supersonic disturbance ($M=4.5$, $\beta=0$) with
purely downstream-propagating acoustic branches.}
\label{fig:spec_cartoon1}
\end{figure}
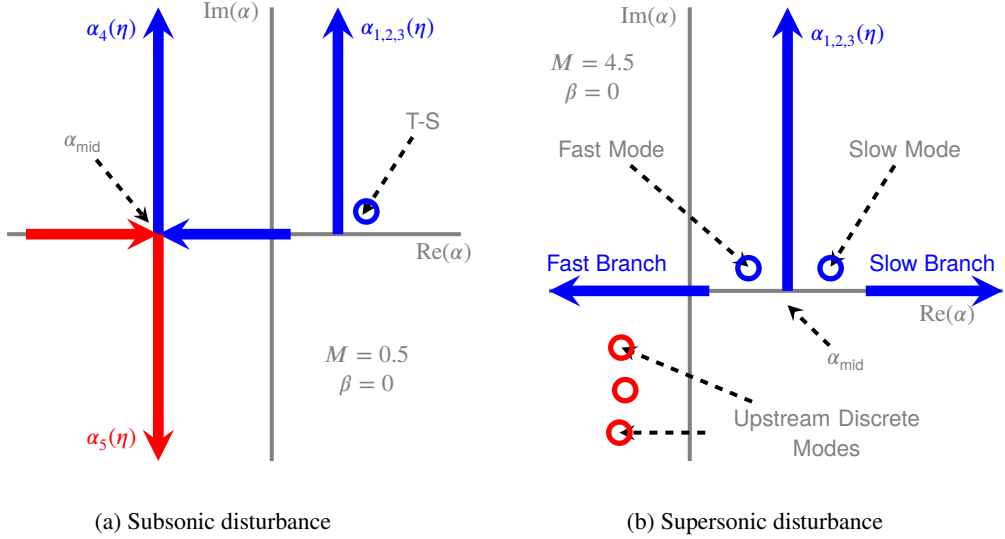

A frozen-coefficient analysis of the LHNS equations yields five
continuous spectral branches: two vorticity and one entropy branch, which
are always downstream-propagating, and two acoustic branches, which
may propagate either upstream or downstream depending on the flow
conditions~\citep{schmidStabilityTransitionShear2001}; explicit
formulas are given in Appendix~\ref{sec:continuous_spectra}.  These
branches are unbounded in the complex $\alpha$-plane.  After
discretisation their finite extent depends on wall-normal resolution and both the continuous branches and the discrete eigenvalues may
evolve during the streamwise march.  Representative $\alpha$-plane
spectra for subsonic and supersonic disturbances are shown in
Fig.~\ref{fig:spec_cartoon1}.

We define a spectral reference point $\alpha_{\mathrm{mid}}$ whose
construction depends on the disturbance regime
(Eq.~\eqref{eqn:mid_app}, Appendix~\ref{app:Cayley}).  For subsonic
disturbances it is the midpoint of the two acoustic branch endpoints,
coinciding with the acoustic pinch-point when one exists.  For
supersonic disturbances, where the acoustic branches do not coalesce,
it is taken at the common endpoint of the vorticity and entropy
branches, displaced slightly below the real axis so that the
continuous branches fall on the correct side of the separating
contour under the map below.

The unbounded extent of the continuous branches makes the geometric
separation between the two spectral subsets difficult to assess in
the $\alpha$-plane.  
A Cayley (M\"obius) map $\xi(\alpha)$, centred so
that $\alpha_{\mathrm{mid}}$ is mapped to $\xi_{\mathrm{mid}} = i$,
compactifies the upstream spectrum onto a bounded region of the
$\xi$-plane: upstream eigenvalues are generally mapped inside the unit
circle and downstream eigenvalues outside, so that the unit circle
serves as an estimate for a separating contour. The map is eigenvector-preserving and only the eigenvalues themselves change, so the
projectors defined below are identical in both representations; its
construction and properties are detailed in
Appendix~\ref{app:Cayley}.  Fig.~\ref{fig:spec_cartoon2} shows the
Cayley-transformed spectra for the same subsonic and supersonic
configurations as Fig.~\ref{fig:spec_cartoon1}: the unbounded continuous acoustic spectra are now fully observable. 

\begin{figure}[!htbp]
\centering
\begin{subfigure}{5.5cm}
\centering
\input{figs2/tikz/spec2}
\caption{Subsonic disturbance}
\label{fig:spec_cartoon2a}
\end{subfigure}
\hspace{1.5cm}
\begin{subfigure}{5.5cm}
\centering
\input{figs2/tikz/spec4}
\caption{Supersonic disturbance}
\label{fig:spec_cartoon2b}
\end{subfigure}
\caption{Cayley-transformed spectra for (a) a subsonic disturbance at
$M=0.5$, $\beta=0$ and (b) a supersonic disturbance at $M=4.5$,
$\beta=0$.  Upstream modes lie inside the unit circle; downstream
modes lie outside.}
\label{fig:spec_cartoon2}
\end{figure}
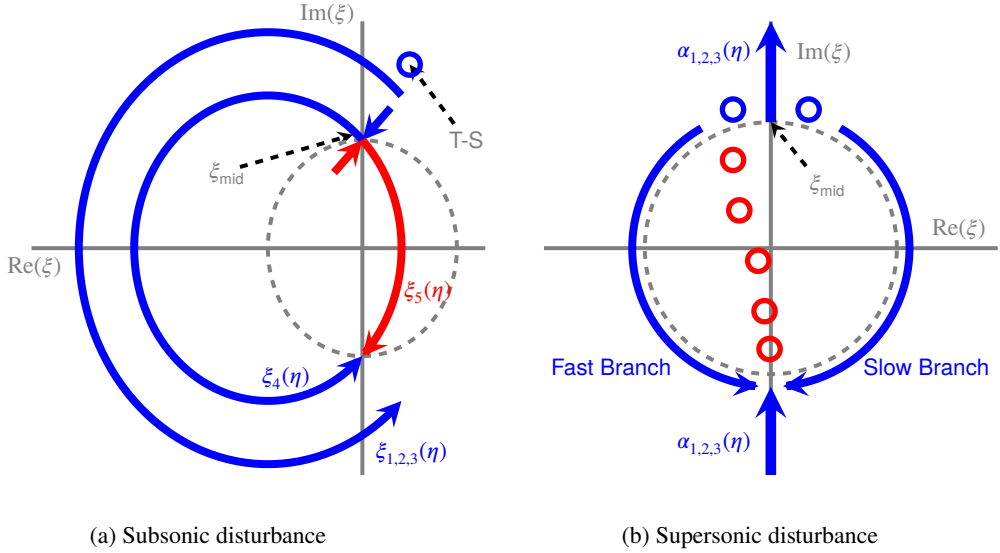

\subsection{One-Way Projection}\label{sec:projection}
We denote by~$\Gamma^+$ a simple closed contour in the complex
$\alpha$-plane that encloses only the downstream eigenvalues, and
define the downstream open spectral region $\Omega^+$ as the interior
of~$\Gamma^+$ and the upstream closed spectral region as
$\Omega^- = \mathbb{C}\setminus{\Omega^+}$.  When a pinch-point exists and
lies on~$\Gamma^+$ it is formally excluded from~$\Omega^+$ but is
retained in~$\Omega^-$.  The contour is not unique: the projector
defined below depends only on which eigenvalues $\Gamma^+$ encloses, so
any simple closed curve separating the two subsets defines the same
operator.  This freedom is what the constructions of
\S\ref{sec:implicit} and \S\ref{sec:parameter_selection} exploit, and
simplicity of the curve is what makes the partition into $\Omega^+$
and~$\Omega^-$ meaningful.

The exact one-way projector onto the downstream subspace is defined
by the contour integral
\begin{equation}\label{eqn:contour}
    \mathbf{P}^+
    = \frac{1}{2\pi}\oint_{\Gamma^+}
      \!\left(\mathbf{B} - i\alpha\,\mathbf{A}\right)^{-1}
      \mathbf{A}\,\mathrm{d}\alpha,
\end{equation}
which up to a sign convention (dependent on the contour orientation)
is the standard spectral projector
of~\citet{katoPerturbationTheoryLinear1966} applied to the matrix
pencil~$\mathbf{\Pi}$; related constructions underlie
contour-integral eigensolvers such as
FEAST~\citep{polizziDensitymatrixbasedAlgorithmSolving2009}.  The
$\mathbf{A}$ factor in the integrand arises from the Lancaster
derivative of the
pencil~\citep{lancasterEigenvaluesMatricesDependent1964}, which
relates the resolvent of~$\mathbf{\Pi}$ to the standard resolvent and
absorbs the factor of~$i$ from the prefactor.

The projector~$\mathbf{P}^+$ is idempotent,
$(\mathbf{P}^+)^2 = \mathbf{P}^+$, and the complementary projector
$\mathbf{P}^- = \mathbf{I} - \mathbf{P}^+$ isolates the upstream
subspace.  We define the projected pencils
\begin{equation}\label{eqn:project_pencil}
    \mathbf{\Pi}^\pm = (\mathbf{B}\mathbf{P}^\pm,\,\mathbf{A}),
\end{equation}
which retain the eigenvectors of~$\mathbf{\Pi}$ while modifying its
eigenvalues: modes propagating in the selected direction are
preserved, and all others are mapped to zero.  Thus
\begin{equation}\label{eqn:one_way_spec}
    \sigma\!\left(\mathbf{\Pi}^+\right) =
    \begin{cases}
        \text{downstream-propagating eigenvalues,} \\
        \text{zero eigenvalues.}
    \end{cases}
\end{equation}
These projected pencils provide the formal mathematical target that
OWNS approximations seek to reproduce.

As discussed in \S\ref{sec:pencil}, the pencil also possesses infinite eigenvalues associated with algebraic constraints. These arise from the singularity of $\mathbf A$ and are not part of the finite spectrum $\sigma\left(\mathbf \Pi\right)$. Consequently, the projector defined by Eq.~\eqref{eqn:contour} acts only on the finite eigenvalues, leaving the infinite eigenvalues unchanged.

\subsection{Rational Approximation of the Projector}\label{sec:quadrature}

The exact projector $\mathbf{P}^+$ defined by
Eq.~\eqref{eqn:contour} selects the downstream invariant subspace
of~$\mathbf{\Pi}$.  We define $\mathcal{P}^\pm$ as the characteristic
function satisfying
\begin{equation}\label{eqn:char_func}
    \mathcal{P}^+(\alpha) =
    \begin{cases}
        1, & \alpha \in \Omega^+,\\
        0, & \alpha \notin \Omega^+,
    \end{cases}
    \qquad
    \mathcal{P}^- = 1 - \mathcal{P}^+.
\end{equation}
If the pencil admits a biorthogonal basis with left eigenfunctions
$u_k$ satisfying $u_k^* \mathbf{A} v_j = \delta_{jk}$, the matrix
projector is recovered via the spectral sum
\begin{equation}\label{eqn:spectral_sum}
    \mathbf{P}^+
    = \sum_{\alpha_k \in \sigma(\mathbf{\Pi})}
      \mathcal{P}^+(\alpha_k)\, v_k\, u_k^{*} \,\mathbf{A},
\end{equation}
where ${\cdot}^*$ denotes the conjugate transpose.

The challenge is to replace this discontinuous characteristic
function by a smooth rational approximation that can be evaluated
without explicit eigen-decomposition.  This is the same object as the
rational approximation of the matrix sign function used to separate
spectra in numerical linear algebra, where near-optimal approximants
follow from Zolotarev's solution of the corresponding minimax
problem~\citep{nakatsukasaComputingFundamentalMatrix2016}; the FEAST
eigensolver~\citep{polizziDensitymatrixbasedAlgorithmSolving2009}
exploits the same structure.

\subsubsection{Rational filter}\label{sec:rational_filter}

We seek a rational function $\mathcal{P}^+_N$ that approximates the
characteristic function~\eqref{eqn:char_func} on the spectrum of
$\mathbf{\Pi}$.  One such function with $N$ prescribed simple poles
$\beta_k^* \in \mathbb{C}$ takes the
partial-fraction form
\begin{equation}\label{eqn:quadrature_scalar}
    \mathcal{P}^+_N(\alpha)
    = w_0 + \sum_{k=1}^{N}\frac{w_k}{i\alpha - i\beta_k^*},
\end{equation}
with weights $w_k \in \mathbb{C}$ $(k\in\{0,\dots,N\})$.  Lifting this scalar expression to
the matrix pencil $(\mathbf{B},\mathbf{A})$ gives the approximate
projection operator
\begin{equation}\label{eqn:quadrature_matrix}
    \mathbf{P}^+_N
    = w_0\,\mathbf{I}
    + \sum_{k=1}^N w_k\bigl(\mathbf{B} - i\beta_k^*\,\mathbf{A}\bigr)^{-1}\mathbf{A}.
\end{equation}
The connection between the two is direct: if $(\alpha_k, v_k)$ is an
eigenpair of $\mathbf{\Pi}$, then
$\mathbf{P}^+_N v_k = \mathcal{P}^+_N(\alpha_k)\,v_k$.  That is, the
matrix operator applies the scalar filter eigenvalue-by-eigenvalue.
The scalar function $\mathcal{P}^+_N$ therefore serves as a
numerically inexpensive surrogate for testing candidate pole--weight
configurations before committing to the full matrix computation.  The
correspondence is exact only mode-by-mode, however: the pencil is
non-normal, so the norm of the matrix operator is not controlled by
the scalar filter values alone, and the resolvent factors
in~\eqref{eqn:quadrature_matrix} may be large even where the scalar
filter is well behaved.  The scalar surrogate guides pole and weight
selection; the finite-precision behaviour of the matrix evaluation is
a separate question, taken up in \S\ref{sec:forward_error}.

Each term in Eq.~\eqref{eqn:quadrature_matrix} requires a resolvent
evaluation at an auxiliary pole $\beta_k^*$.  With appropriate pole
and weight selection, $\mathcal{P}^+_N(\alpha) \approx 1$ for
downstream eigenvalues and $\mathcal{P}^+_N(\alpha) \approx 0$ for
upstream eigenvalues, so that $\mathbf{P}^+_N$ reproduces the action
of the exact projector on all dynamically relevant modes.

The poles and weights in
Eqs.~\eqref{eqn:quadrature_scalar}--\eqref{eqn:quadrature_matrix} are
free parameters of the rational ansatz.  Although the
contour-integral definition~\eqref{eqn:contour} motivates why a
rational function of the resolvent yields a projector, the poles
$\beta_k^*$ need not correspond to quadrature nodes on any particular
contour.  How the resulting filter is \emph{evaluated} also matters
in finite precision, and is analysed in \S\ref{sec:forward_error}.

\subsubsection{Implicit pole placement}\label{sec:implicit}

The Cayley unit circle identifies where a separation could occur, but it
does not by itself fix the poles of the rational filter, nor establish that it
is the best contour to use.  The direct route is to take
Eq.~\eqref{eqn:contour} at face value and evaluate it by quadrature on an
explicit contour, the poles then being the quadrature nodes.  We implemented
this for an incompressible boundary layer, and using the unit circle with a small indentation around the pinch-point as a contour found it to require prohibitively large~$N$ there.  It is worth setting
out why, since the reasons motivate the construction that follows.

Quadrature requires two choices to be made separately, and neither is easy.
The first is the contour itself.  The second is the distribution of nodes
along it: on a circle, which angles to use and how strongly to cluster them
where the two spectra approach.  Neither choice is informed by the other, and
a contour that is well placed does not tell one where its nodes should sit.
The third difficulty is more fundamental.  Both choices are made against the
\emph{regions} $\Omega^\pm$ rather than against the spectrum itself, so a
contour that separates the two regions cleanly may still leave nodes at which
the resolvent is large, and the accuracy of the filter on the eigenvalues that
matter is not controlled by the geometry of the curve.  For a non-normal
pencil this is not a minor caveat.

Where the upstream and downstream spectra are widely separated, none of this
bites: the contour has room, the node distribution is uncorrelated, and
quadrature is a reasonable and simple approach.  The incompressible and transonic configurations of
interest in this paper are the opposite case.  The construction below therefore
abandons the explicit contour and combines the two choices into one, by
specifying approximate spectral locations directly and letting the poles
follow from them algebraically. 

The eigenvalues of~$\mathbf{\Pi}$ are not available without computation, and
that computation is not cheap.  Instead, approximate spectral
locations are obtained from analytical estimates of the continuous
branches~\citep{schmidStabilityTransitionShear2001}.  
Sampling these estimates yields two finite point sets, referred to as
\emph{recursion parameters}~\citep{towneOnewaySpatialIntegration2015}:
downstream points $\beta^+ = \{\beta_k^+\}_{k=1}^N \subset \Omega^+$ and
upstream points $\beta^-=\{\beta_k^-\}_{k=1}^N \subset \Omega^-$.  In
this work we call these sets \emph{retention}~$\beta^+$ and
\emph{removal}~$\beta^-$ parameters to elucidate their role.  What is
required of them is membership of the correct spectral region, not
proximity to any particular eigenvalue.  Placing a parameter close to an
eigenvalue is often desirable, and near the pinch-point or discrete mode(s) deliberately so,
but this is a choice made for accuracy rather than a condition for
admissibility.  The problem is then to infer, from these two sets,
auxiliary poles $\beta^*=\{\beta_k^*\}_{k=1}^N \subset \mathbb{C}$ that
lie on a curve separating~$\Omega^+$ from~$\Omega^-$.

\citet{zhuRecursiveOnewayNavierStokes2023} provide such a
construction through the polynomial identity
\begin{equation}\label{eqn:poly_general}
    c\prod_{k=1}^{N}(\alpha-\beta_k^+)
    + \prod_{k=1}^{N}(\alpha-\beta_k^-)
    = (c+1)\prod_{k=1}^{N}(\alpha-\beta_k^*),
\end{equation}
originally derived from the recursive product structure of OWNS-R.  The
identity admits a purely geometric reading: the polynomial has $N$
roots, exhibited on the right as the~$\beta_k^*$, and these interleave between
the two input sets and trace a separating curve in the complex plane.  The
parameter~$c>0$ controls how close that curve sits to one set or the other:
$c<1$ favours retention and $c>1$ favours removal, as discussed in
Appendix~\ref{sec:C}.  The construction makes a single choice, of where the
spectral subsets lie, and obtains the poles from it, in place of the two
independent choices that quadrature requires.

The quality of the resulting contour depends on the relationship between
the two input sets.  When the retention and removal parameters are well
distributed relative to the spectral geometry, the resulting
$\beta_k^*$~poles should remain sufficiently distant from the physical
eigenvalues and the associated resolvents well conditioned.  The curve should
also remain simple, with no closed loops or significant defects, which is
attainable because the two spectral subsets are themselves separable by a
simple curve, in the subsonic case at every point but the pinch.  Both properties depend on how the two sets are
related to each other rather than on where either sits individually:
choosing them independently, even with each correctly placed in its own
region, leaves the induced contour unconstrained, whereas generating the
removal parameters from the retention parameters by a fixed pairing (or vice versa)
keeps it balanced about the $\Omega^\pm$ boundary.  See \S\ref{sec:heuristic} for a discussion on parameter pairings.
 
Near pinch-points, or in flows where the spectrum evolves strongly
during the streamwise march, the coupling in
Eq.~\eqref{eqn:poly_general} can place some poles undesirably close to
physical eigenvalues, producing ill-conditioning.  The difficulty is
compounded when the spectrum reorganises during the numerical march, since
$\Omega^\pm$ are defined by a contour we choose: a placement admissible
at one station need not remain so once the parameters have migrated
relative to that boundary.  A further complication is that since the pencil is
non-normal, maximising the distance between a pole and its nearest
eigenvalue is not sufficient; the relevant quantity, not pursued in this
work, is instead the resolvent norm at the pole (a costly metric to compute).

When using the implicit pole placement, the $\beta^*$ poles trace the contour $\Gamma^+$ in the spectrum. As such, we highly recommend using a Cayley or similarly transformed spectra to see the full behaviour of the contour and its dependence on $\beta^\pm$.

\section{OWNS-Summation}\label{sec:owns_s}
 
Given the auxiliary poles $\{\beta_k^*\}_{k=1}^N$ generated by the
implicit pole-placement framework of \S\ref{sec:implicit} or otherwise, OWNS-Summation
(OWNS-S) evaluates the rational filter~\eqref{eqn:quadrature_matrix}
directly in its partial-fraction form. Its action on a state
vector~$\phi$ is
\begin{subequations}\label{eqn:owns_s_apply}
\begin{align}
    (\mathbf B - i\beta_k^*\mathbf A)\,\psi_k &= \mathbf A\,\phi,
    \qquad k=1,\ldots,N, \label{eqn:owns_s_solve}\\
    \mathbf{P}_{N}^+ \phi &= w_0\,\phi + \sum_{k=1}^N w_k\,\psi_k,
    \label{eqn:owns_s_sum}
\end{align}
\end{subequations}
and $\mathbf{P}_{N}^+ \phi$ replaces the state at each station of the
spatial march (\S\ref{sec:numerics}). Each solve in~\eqref{eqn:owns_s_solve} depends only on
the original state~$\phi$, the pencil $(\mathbf B,\mathbf A)$, and the
corresponding auxiliary pole~$\beta_k^*$. The intermediate
solutions~$\psi_k$ are mutually independent; no pole contribution is
used as input to another.

This structure has two practical consequences. 
First, the sum carries no ordering. Neither the factorisation of the shifted
operators $(\mathbf B - i\beta_k^*\mathbf A)$ nor the back-substitutions
that follow them depends on the index~$k$, so the terms
of~\eqref{eqn:owns_s_sum} may be formed in any order and no ordering of the
poles need be selected. Thus, the $N$ matrix inversions can be
accomplished in parallel across poles. Second,
numerical noise incurred in the solve at pole~$\beta_k^*$ enters the
projected state only through the single term $w_k\psi_k$
of~\eqref{eqn:owns_s_sum}. The accuracy of OWNS-S is therefore
controlled by the conditioning of the individual resolvent solves and by
the numerical behaviour of the weighted sum.

\subsection{Weight calculation}\label{sec:weights}

The weights $\{w_k\}_{k=0}^N$ in~\eqref{eqn:owns_s_sum} are determined by
requiring that the scalar filter~\eqref{eqn:quadrature_scalar}
approximate the desired modal separation, taking the value $1$ on
downstream eigenvalues and $0$ on upstream eigenvalues.  We enforce this
on a finite set of \emph{test points}
$\mathcal Z = \{\zeta_m\}_{m=1}^{M}$,
\begin{equation}\label{eqn:interp_conditions}
    \mathcal P^+_N(\zeta_m) =
    \begin{cases}
        1, & \zeta_m \in \Omega^+,\\
        0, & \zeta_m \notin \Omega^+,
    \end{cases}
    \qquad m = 1,\ldots,M .
\end{equation}

The natural default, used throughout this section, is to test at the
retention/removal parameters themselves,
$\mathcal Z = \{\beta_k^+\}_{k=1}^N \cup \{\beta_k^-\}_{k=1}^N$ with
$M = 2N$, giving $2N$ conditions for $N+1$ unknowns.  We write
$\mathcal Z^+ = \mathcal Z\cap\Omega^+$ and
$\mathcal Z^- = \mathcal Z\cap\Omega^-$ for the two subsets, so that the
conditions~\eqref{eqn:interp_conditions} read
$\mathcal P^+_N = 1$ on $\mathcal Z^+$ and
$\mathcal P^+_N = 0$ on $\mathcal Z^-$.

With exact
auxiliary poles the polynomial identity~\eqref{eqn:poly_general}
guarantees that the filter satisfies all $2N$ conditions exactly, so the
system is consistent despite being overdetermined.  Test sets enlarged
beyond the retention and removal parameters are taken up with the greedy
selection of \S\ref{sec:greedy}.  Two constructions are used.

\paragraph*{Residue weights.}
When $\mathcal Z$ consists of the retention/removal parameters, the weights are
available in closed form as the residues of the partial-fraction
expansion,
\begin{equation}\label{eqn:standard_weights}
    w_0 = \frac{1}{c+1}, \qquad
    w_k = \frac{i(\beta_k^* - \beta_k^-)}{c+1}
    \prod_{\substack{j=1 \\ j\neq k}}^N
    \frac{\beta_k^* - \beta_j^-}{\beta_k^* - \beta_j^*},
    \qquad k = 1,\ldots,N .
\end{equation}
When the poles are well separated this formula is accurate.
Its product-of-ratios structure degrades in two situations: when
auxiliary poles cluster, the denominators $\beta_k^*-\beta_j^*$ become
small and rounding error in each ratio is amplified through the product;
and when the retention and removal parameters are widely separated, the
residues acquire large modulus and alternating sign, exposing the
sum~\eqref{eqn:owns_s_sum} to cancellation.  Both are symptoms of
ill-conditioning in the underlying Cauchy interpolation problem.

At large $N$ a separate, purely representational difficulty arises.  The
numerator and denominator of Eq.~\eqref{eqn:standard_weights} are each
products of $N$ pole separations spanning the whole spectrum, and formed
directly they overflow or underflow the double-precision exponent range long
before relative accuracy is exhausted.  The residues are therefore evaluated
in logarithmic form: the logarithms of the factors are accumulated for
numerator and denominator separately and the difference exponentiated once,
which is exact for the product and remains in range at every order used
here.

A further consideration is the polynomial
identity~\eqref{eqn:poly_general} itself.
\citet{zhuRecursiveOnewayNavierStokes2023} report that at large $N$ the
root-finding used to obtain $\beta^*$ carries rounding error sufficient
to warrant extended precision, a remedy adopted by
\citet{sleemanGreedyRecursionParameter2025a}.  The residue formula is
not the casualty of this error: Eq.~\eqref{eqn:standard_weights} is
exact for whatever poles it is given, returning the coefficients for
which the partial-fraction form~\eqref{eqn:quadrature_scalar} equals
the ratio
$\prod_{k}(\alpha-\beta_k^-)\big/\bigl[(c+1)\prod_{k}(\alpha-\beta_k^*)\bigr]$
at the computed poles, so the pair $(\beta_k^*,w_k)$ remains mutually
consistent.  What the pole error degrades is the separation
conditions~\eqref{eqn:interp_conditions}.  Removal survives, the zeros
of the filter being the removal parameters themselves, but the
retention value $\mathcal P^+_N(\beta_k^+)=1$ holds only through the
identity and fails by its residual once the poles are perturbed.  The
residue weights therefore reproduce, faithfully, a filter whose
retention is degraded.  This is what the least-squares construction
addresses.

\paragraph*{Least-squares weights.}
The least-squares construction, denoted OWNS-S~(LS), takes the computed
poles as given and finds the weights that best enforce the separation
conditions~\eqref{eqn:interp_conditions} for those poles, so that no
assumption about the identity enters the weight calculation.
Substituting the summation form~\eqref{eqn:quadrature_scalar} into each
condition gives one linear equation in $(w_0,\ldots,w_N)$ per test point,
and collecting all $M$ conditions yields
\begin{equation}\label{eqn:ls_system}
    \mathbf D\, w = \mathbf b,
    \qquad
    \mathbf D \in \mathbb C^{M\times(N+1)},
    \qquad
    w=[w_0,w_1,\ldots,w_N]^T,
\end{equation}
with entries
\begin{equation}\label{eqn:ls_matrix}
    D_{mj} =
    \begin{cases}
        1, & j = 0,\\[4pt]
        \displaystyle\frac{1}{i\zeta_m - i\beta_j^*}, & j = 1,\ldots,N ,
    \end{cases}
\end{equation}
and $b_m = 1$ at downstream test points, $b_m = 0$ at upstream test
points.  The weights are taken as the minimiser of
$\lVert\mathbf D w-\mathbf b\rVert_2$, computed by singular-value
decomposition. 

The matrix $\mathbf D$ is Cauchy-like, and its conditioning reflects the
same sensitivities as the residue formula.  Where the retention and
removal parameters are very densely distributed, $\mathbf D$ becomes
severely ill conditioned in double precision and the least-squares
weights are unreliable; the residue weights are used instead in that
case.  Extended precision removes the difficulty but is too expensive to
apply at every station of a march.

In the configurations reported here the two constructions give closely
similar results. The pole error
\citep{zhuRecursiveOnewayNavierStokes2023,sleemanGreedyRecursionParameter2025a} is therefore not large
enough in these cases to degrade the separation conditions appreciably,
and the least-squares construction is optional rather than required. In \S\ref{sec:results}, OWNS-S (LS) improves the convergence of OWNS-S at low-to-moderate $N$.

Independently of this correction, taking the poles as given allows the
separation conditions to be enforced at any $\mathcal Z$, so the filter
may be fitted around chosen spectral regions rather than only at the
retention/removal parameters.  The greedy selection of \S\ref{sec:greedy}
exploits this.

\subsection*{OWNS-S Algorithm}

\begin{algorithm}[!t]
\caption{OWNS-S projection at one streamwise station}
\begin{algorithmic}[1]
\Require pencil $(\mathbf B,\mathbf A)$ at station $x$; retention/removal
         parameters $\{\beta_k^+,\beta_k^-\}_{k=1}^N$; balance parameter $c$; state $\phi$ found from implicitly solving the discretised equation 

\State solve the polynomial identity Eq.~\eqref{eqn:poly_general} for the auxiliary poles
       $\{\beta_k^*\}_{k=1}^N$
\State form the weights $\{w_k\}_{k=0}^N$: residues, or least squares on
       $\mathcal Z$
\State $r \gets \mathbf A\phi$

\For{$k=1,\ldots,N$} \Comment{independent across $k$}
    \State factorise $\mathbf B-i\beta_k^*\mathbf A$ and back-substitute on $r$
           to obtain $\psi_k$
\EndFor

\State $\mathbf P_N^+\phi \gets
w_0\phi+\displaystyle\sum_{k=1}^N w_k\psi_k$
\end{algorithmic}
\end{algorithm}

\section{Comparison with OWNS-Recursion and Error Accumulation}\label{sec:owns_r}

OWNS-Recursion (OWNS-R) evaluates the approximate projection operator
Eq.~\eqref{eqn:quadrature_matrix} through a sequence of recursive
resolvent products.  This is the recursive formulation introduced by
\citet{zhuRecursiveOnewayNavierStokes2023} and used in later
applications to boundary-layer
flows~\citep{IUTAM2024_ejbmug,badcocktcfd2026,sleemanGreedyRecursionParameter2025a}.
That literature operates at moderate approximation orders with carefully
tuned retention/removal parameters.  Once the order is raised, however, the
evaluation deteriorates, and it does so in a way that is a property of the
product structure rather than of any particular flow.  This section sets out the product representation, identifies the mechanism,
and measures it with the poles held fixed.

The product representation is obtained as
follows. Define
\begin{equation}\label{eqn:R_def}
    R(\alpha) =
    \prod_{k=1}^N
    \frac{\alpha-\beta_k^+}{\alpha-\beta_k^-}.
\end{equation}
Using the polynomial identity~\eqref{eqn:poly_general}, the scalar
filter can be written as
\begin{equation}\label{eqn:product}
    \mathcal P^+_N(\alpha)
    =
    \bigl(1+cR(\alpha)\bigr)^{-1}
    =
    \frac{1}{c+1}
    \prod_{k=1}^N
    \frac{\alpha-\beta_k^-}{\alpha-\beta_k^*},
\end{equation}
with matrix counterpart
\begin{equation}\label{eqn:R_matrix}
    \mathbf P^+_N
    =
    \frac{1}{c+1}
    \prod_{k=1}^N
    (\mathbf B-i\beta_k^*\mathbf A)^{-1}
    (\mathbf B-i\beta_k^-\mathbf A).
\end{equation}
Its action on a state vector $\phi$ is applied recursively:
\begin{subequations}\label{eqn:owns_r_apply}
\begin{align}
    \phi_0 &= \frac{1}{c+1}\,\phi, \\
    (\mathbf B-i\beta_k^*\mathbf A)\,\phi_k
    &=
    (\mathbf B-i\beta_k^-\mathbf A)\,\phi_{k-1},
    \qquad k=1,\ldots,N, \\
    \mathbf{P}_{N}^+ \phi &= \phi_N .
\end{align}
\end{subequations}
Each stage takes its input from the output of the previous stage, so
the recursion is inherently sequential. OWNS-R requires the same $N$
resolvent solves per projection as OWNS-S, but each stage additionally
requires the matrix--vector product
$(\mathbf B-i\beta_k^-\mathbf A)\phi_{k-1}$ to form its right-hand
side: $N$ such products per projection, against the single product
$\mathbf A\phi$ shared by all OWNS-S solves.

\subsection{Finite-precision error accumulation}\label{sec:forward_error}

With the residue weights~\eqref{eqn:standard_weights}, OWNS-S and
OWNS-R define the same operator in exact arithmetic; they differ in
the floating-point computation that evaluates it.

In OWNS-S the computed projection has the form
\begin{equation}\label{eqn:owns_s_error}
    \mathbf{P}_{N}^+ \phi
    =
    w_0\phi
    +
    \sum_{k=1}^N
    w_k\left(\psi_k+\delta\psi_k\right),
\end{equation}
where $\delta\psi_k$ is the error produced in the $k$-th resolvent
solve. Each error is multiplied by a scalar and added once. No error
is used to generate another, and no operator acts on any of them. The
accuracy of OWNS-S is therefore controlled by the conditioning of the
individual solves, which does not explicitly depend on $N$, and by the rounding
of the weighted sum.

In OWNS-R the computed stage operator is
\begin{equation}\label{eqn:stage_op}
    \mathbf M_k + \mathbf E_k,
    \qquad
    \mathbf M_k
    =
    (\mathbf B-i\beta_k^*\mathbf A)^{-1}
    (\mathbf B-i\beta_k^-\mathbf A),
\end{equation}
where $\mathbf E_k$ collects the finite-precision error incurred in
solving against $(\mathbf B-i\beta_k^*\mathbf A)$, so that the computed
projection is
\begin{equation}\label{eqn:owns_r_error}
    \mathbf{P}_{N}^+ \phi
    =
    \frac{1}{c+1}
    \left(\mathbf M_N+\mathbf E_N\right)
    \cdots
    \left(\mathbf M_1+\mathbf E_1\right)\phi .
\end{equation}
Each stage takes the output of the previous one, so the error
$\mathbf E_k$ committed at stage~$k$ is subsequently acted on by the
$N-k$ stages that follow it. The magnitude of $\mathbf E_k$ is set by
the conditioning of the $k$-th solve alone, exactly as $\delta\psi_k$
is in OWNS-S; what differs is that it is then propagated through the
remainder of the recursion.

Three consequences follow. First, the amplification applied to
$\mathbf E_k$ is a product of $N-k$ stage operators, so if their norms
exceed unity the amplification grows geometrically with the
approximation order, and an error committed early is amplified most.
Second, the stage operators commute, the resolvents at distinct shifts doing
so by the first resolvent identity, so the exact operator is unchanged by any
permutation of the factors and the sequence may be chosen freely.  The
amplification may not, since it depends on the position of each stage in the
sequence and not only on which auxiliary poles are used.  The ordering is
therefore a design choice that must be made, and one that has no counterpart
in the summation.  \citet{sleemanGreedyRecursionParameter2025a} order the recursion parameters by
ascending magnitude on a criterion concerning the conditioning of each stage
individually; the accumulated product considered here is a distinct quantity.
Third, the retention and removal parameters therefore carry
two roles in OWNS-R: they determine the quality of the rational filter
and, through the norms of the resulting stage operators, the stability
of the evaluation. Parameter placement and evaluation stability are
not separable.
 
When this error growth becomes appreciable it does not degrade the projection
gracefully.  The recursion returns a state dominated by the amplified error
rather than by the filtered field, so the computed projection is unusable
rather than inaccurate.  Increasing $N$ can make the situation worse.
\citet{zhuRecursiveOnewayNavierStokes2023} document the deterioration and
attribute it to the polynomial solve for $\{\beta_k^*\}_{k=1}^N$, recommending
extended precision there. \citet{sleemanGreedyRecursionParameter2025a} adopt that recommendation
and report that the deterioration persists, the error rising again beyond an
order of roughly forty even with the poles computed in quadruple precision,
and identify the residual with the application of the filter.  The
accumulation mechanism above is a candidate for that residual, and the
experiment below measures it directly.

As the amplification of an early error is set by the product of all stage
operators applied after it, the position of each pair $(\beta_k^*,\beta_k^-)$
in the sequence controls how strongly that error is amplified, and one may ask
whether some ordering bounds the growth for a fixed set of poles.  It is not
obvious how such an ordering would be found, nor that one exists at all, and
it is a difficulty OWNS-S does not face.  Of the orderings we tried, most left
the growth unchanged or increased it, and none removed it.  In the experiment
of \S\ref{sec:synthetic} the parameters are generated ascending the imaginary
axis from $\alpha_\mathrm{mid}=0$ and are applied in that order, so the
sequence is already in ascending modulus and coincides with the ordering
recommended by \citet{sleemanGreedyRecursionParameter2025a}; the growth reported there is not an
artefact of an unsorted product. 

This does not render OWNS-R ill-posed. The operator~\eqref{eqn:R_matrix}
is well defined in exact arithmetic for any $N$, and with identical
recursion parameters, auxiliary poles and weights it is the same
operator that OWNS-S evaluates; the instability is a property of the
finite-precision evaluation alone.

The additive evaluation removes the mechanism rather than mitigating
it. In OWNS-S $N$ is therefore a convergence parameter in the
ordinary sense: given a correct placement of $\beta_k^\pm$ that resolves
the relevant spectral regions, increasing $N$ improves the projection
and continues to do so.

\subsubsection{Synthetic experiment}\label{sec:synthetic}

To isolate the floating-point behaviour from any particular
calculation, we consider a fixed representative pencil
$(\mathbf B,\mathbf A)$, taken from an incompressible flat-plate
boundary layer at $R_\delta=400$, $M=0.02$, $f=86\times10^{-6}$, and
$b=0$, discretised with $n_y=251$ points ($y_{\max}=600$,
$y_{\mathrm{half}}=30$); see \S\ref{sec:numerics}. The experiment
concerns the arithmetic of the projection operator, not the physical
evolution of a specific disturbance.

We consider two sets of retention/removal parameters. First,
$\{\beta_k^+,\beta_k^-\}_{k=1}^N$ are placed on the positive and
negative imaginary axes with mild Malik-type clustering
(Eq.~\eqref{eqn:malik_grid}) towards the origin, satisfying the
rotation symmetry $\beta_k^- = 2\alpha_{\mathrm{mid}}-\beta_k^+$ of
\S\ref{sec:implicit}, conditioning value $c=1$, and Eq.~\eqref{eq:recursion_parameters_minus} with
$\alpha_{\mathrm{mid}}=0$. Second, we use the heuristic parameters for
a subsonic disturbance field defined in \S\ref{sec:rec:sub}.

For each $N\in\{10,20,\ldots,100\}$, the auxiliary poles
$\{\beta_k^*\}$ are computed from Eq.~\eqref{eqn:poly_general} using
the same root-finding routine for both formulations. The OWNS-S
weights are computed from the residue
formula~\eqref{eqn:standard_weights}; the least-squares construction
is not used in this experiment. Thus OWNS-S and OWNS-R use identical
pole data; they differ only in how the resulting rational filter is
evaluated.

We monitor the accumulated operators generated by the two evaluations.
For OWNS-R, the diagnostic is the norm of the recursive product after
$k$ stages,
\begin{equation}\label{eqn:rhoR}
    \rho_k^{\mathrm R}
    =
    \left\|
    \prod_{j=1}^k
    (\mathbf B-i\beta_j^*\mathbf A)^{-1}
    (\mathbf B-i\beta_j^-\mathbf A)
    \right\|_{\mathrm F},
\end{equation}
where $\lVert \mathbf{M}\rVert_{\mathrm{F}} =
\tfrac{1}{n}\bigl(\sum_{i,j=1}^n\lvert M_{ij}\rvert^{2}\bigr)^{1/2}$,
with $n=5n_y$, is a normalised Frobenius norm.  For OWNS-S, the
corresponding diagnostic is the norm of the partial summation operator
after $k$ terms,
\begin{equation}\label{eqn:rhoS}
    \rho_k^{\mathrm S}
    =
    \left\|
    w_0\mathbf I
    +
    \sum_{j=1}^k
    w_j
    (\mathbf B-i\beta_j^*\mathbf A)^{-1}
    \mathbf A
    \right\|_{\mathrm F}.
\end{equation}
These quantities are not sharp error bounds. They measure the size of
the intermediate operators through which floating-point errors are
propagated.

\begin{figure}[!htbp]
    \centering
    \includegraphics[width=0.49\textwidth]{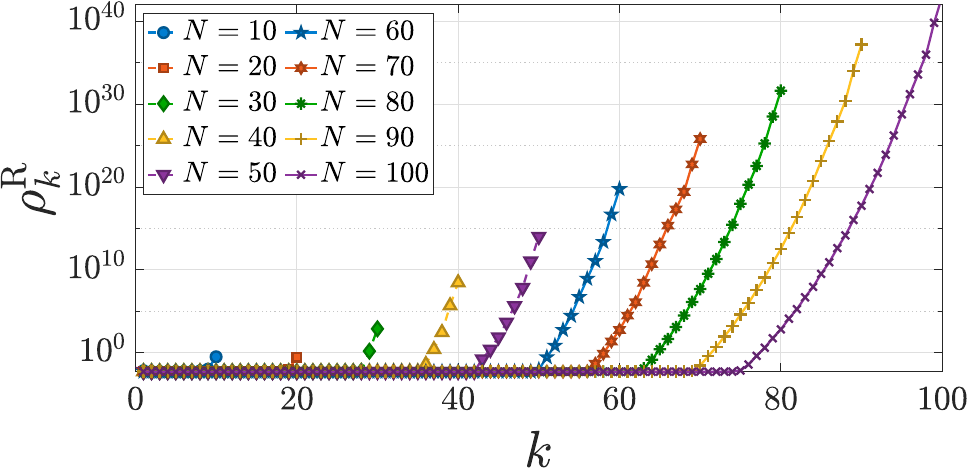}
    \includegraphics[width=0.49\textwidth]{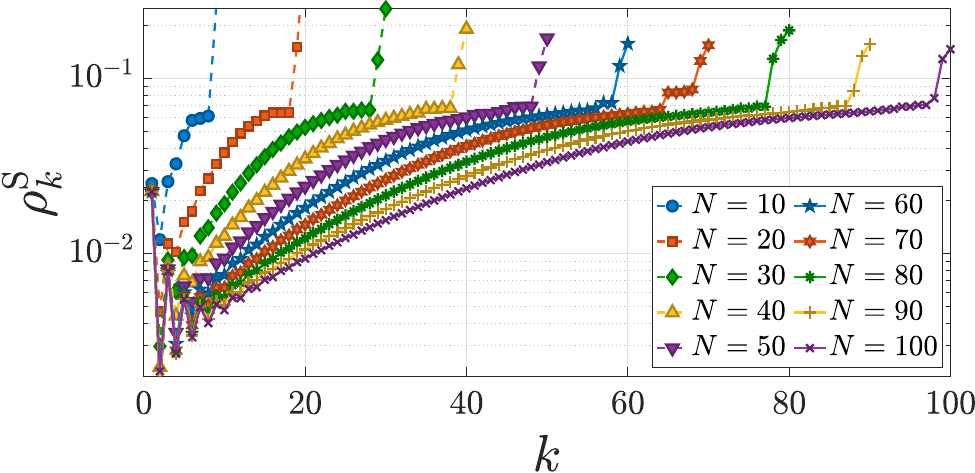}
    \caption{Accumulated-stage Frobenius norms for the recursive
    (OWNS-R, left, Eq.~\eqref{eqn:rhoR}) and summation
    (OWNS-S, right, Eq.~\eqref{eqn:rhoS}) evaluations at a fixed pencil
    $(\mathbf B,\mathbf A)$ using identical auxiliary poles
    $\{\beta_k^*\}$, for $N\in\{10,20,\ldots,100\}$ with the
    Malik-clustered retention/removal parameters. OWNS-R develops accumulated
    norms of order $10^{40}$ at $N=100$, whereas OWNS-S remains
    bounded throughout.}
    \label{fig:synthetic_norms}
\end{figure}

Figure~\ref{fig:synthetic_norms} reports $\rho_k^{\mathrm R}$ and
$\rho_k^{\mathrm S}$ for the Malik-clustered parameters.  The recursive
diagnostic grows rapidly with $k$, reaching approximately
$\mathcal O(10^{40})$ at $N=100$, far beyond the inverse of double-precision
machine accuracy.  The summation diagnostic remains bounded over the same
range, varying by less than two orders of magnitude across all $k$ and all
tested~$N$.  The absolute level of either quantity carries the $1/n$
normalisation and is not by itself meaningful; the load-bearing observation is
the differential between them, roughly forty orders of magnitude.  OWNS-S and
OWNS-S~(LS) have also been tested at higher approximation orders
($N\approx300$) without exhibiting the growth seen in OWNS-R.

The behaviour is not an artefact of the idealised parameter set.  The
heuristic subsonic parameters of \S\ref{sec:rec:sub}, which are tuned for this
configuration, produce the same result, with $\rho^{\mathrm R}_k$ diverging
and $\rho^{\mathrm S}_k$ bounded (not shown); a good placement postpones the
growth but does not remove it.  The same growth and boundedness are observed
when the pencil and pole data are transformed consistently to the
Cayley-compactified $\xi$-representation (Appendix~\ref{app:Cayley}), so the
instability is not an artefact of the unbounded $\alpha$-plane either.

Both evaluations received identical auxiliary poles from the same polynomial
solve, so any error in the pole computation is common to both and cannot
account for a difference between them.  Under that control the recursive
diagnostic grows by roughly forty orders of magnitude while the summation
diagnostic remains bounded. 
The growth is therefore a property of the recursive evaluation. This is
consistent with \citet{sleemanGreedyRecursionParameter2025a}, who computed the poles in
extended precision, observed that the deterioration survived, and were
thereby led to locate the residual error in the application of the filter
rather than in the poles. The additive evaluation removes this recursive
growth entirely.

\subsection{Solve-count comparison}\label{sec:solve_count}

The relative cost of the two evaluations, benchmarked against PSE, is
determined by the number of linear solves required per marching step, by the
auxiliary work needed to form their right-hand sides, and by the dependency
structure between them.  PSE requires one base solve and typically
$n_{\mathrm{iter}}\approx2$ to~$4$ sequential wave-carrier correction solves,
each carrying additional per-iteration overhead from the carrier correction
and convergence monitoring.  OWNS-R requires one base solve and $N$ projection
solves, and forms a distinct right-hand side at every stage: the shifted
operator $\mathbf B-i\beta_k^-\mathbf A$ must be assembled and applied to
$\phi_{k-1}$, $N$ times per station.  OWNS-S requires the same $1+N$ solves,
but every right-hand side is the same vector $\mathbf A\phi$, formed once.

Since no OWNS-S solve depends on
another, the $N$ solves may be distributed across $n_\mathrm{t}$ threads,
giving a critical path of $1+\lceil N/n_\mathrm{t}\rceil$ solves. With sufficient threads and a low enough $N$, the computational cost of OWNS-S can fall below PSE, albeit at the cost of additional memory.

OWNS-R factorisations are likewise independent and may be formed concurrently
at the same storage cost, but the back-substitutions they feed are not.  The
obstruction is a data dependency and not a question of ordering: whichever
permutation of the factors is chosen, stage~$k$ consumes the vector
$\phi_{k-1}$ produced by stage~$k-1$, so the OWNS-R substitution critical path
cannot fall below $N$ at any thread count.

PSE uses a linear solve at each of its $n_{\mathrm{iter}}$ carrier
iterations.  The OWNS-S critical-path count $1+\lceil N/n_\mathrm{t}\rceil$ does not
exceed PSE's $1+n_{\mathrm{iter}}$ once $n_\mathrm{t}\ge N/n_{\mathrm{iter}}$.  The
serial cost remains the full $1+N$ solves; parallelism is what brings the
per-station critical path of OWNS-S to PSE levels.

\begin{table}[!htbp]
\centering
\small
\renewcommand{\arraystretch}{1.1}
\begin{tabular}{@{}lcccccc@{}}
\toprule
        & \multicolumn{3}{c}{Per marching step}
        & \multicolumn{2}{c}{Critical path, $n_\mathrm{t}$ threads} \\
\cmidrule(lr){2-4}\cmidrule(lr){5-6}
Method  & Factorisations & Back-substitutions & Matrix--vector
        & Factorisations & Back-substitutions \\
\midrule
PSE     & $1 + n_{\mathrm{iter}}$ & $1 + n_{\mathrm{iter}}$ & --
        & $1 + n_{\mathrm{iter}}$ & $1 + n_{\mathrm{iter}}$ \\
OWNS-R  & $1 + N$ & $1 + N$ & $N$
        & $1 + \lceil N/n_\mathrm{t}\rceil$ & $1 + N$ \\
OWNS-S  & $1 + N$ & $1 + N$ & $1$
        & $1 + \lceil N/n_\mathrm{t}\rceil$
        & $1 + \lceil N/n_\mathrm{t}\rceil$ \\
\bottomrule
\end{tabular}
\caption{Operation counts per marching step for PSE, OWNS-R and OWNS-S, and
the corresponding critical paths on $n_\mathrm{t}$ threads. }
\label{tab:solve_count}
\end{table}

\section{Retention and Removal Parameter Selection}\label{sec:parameter_selection}
 
The retention and removal parameters $\{\beta_k^+,\beta_k^-\}_{k=1}^N$
determine the auxiliary poles $\{\beta_k^*\}_{k=1}^N$ through the
polynomial identity~\eqref{eqn:poly_general}, and therefore control how
accurately the rational filter retains downstream modes and suppresses
upstream ones.  Their placement is the principal design choice in
OWNS-S and OWNS-R computations.
 
For OWNS-R the placement carries two demands at once: it must separate
the upstream and downstream spectral subsets, and it must keep the sequential product from amplifying
(\S\ref{sec:forward_error}).  Because the second demand cannot be
assessed in advance, a placement that proves inadequate cannot reliably
be repaired by raising~$N$, since raising $N$ is itself what
destabilises the product. 

In OWNS-S, high-$N$ stability is 
a property of the summation structure, so an imperfect placement costs accuracy alone and is recovered by
increasing~$N$.  Placement therefore governs efficiency, meaning how few
parameters are needed for a converged projection, rather than whether
the march survives at all.
 
Two strategies are used.  A physically motivated \emph{heuristic
placement}, developed for OWNS-R
in~\citet{badcocktcfd2026,IUTAM2024_ejbmug} and
adapted here, provides a baseline.  A \emph{greedy
refinement} selects parameters from an oversampled candidate pool by
minimising a scalar filter-error functional, following the framework
of~\citet{sleemanGreedyRecursionParameter2025a} with two structural modifications.  Both
are placements and are in principle independent of how the filter is
evaluated; in this work the greedy refinement is applied only to OWNS-S,
and all OWNS-R results use the heuristic set.

Combining the greedy selection with OWNS-R is admissible and would very
likely improve it at low order, since the two are independent choices.  We do
not report it, for a reason that is itself part of the argument.  The benefit
of a better placement is that fewer parameters are needed, but a placement
cannot be validated without raising~$N$ to check that the result has
converged, and raising~$N$ is what OWNS-R cannot survive.  The greedy
selection would therefore be constrained to operate inside a window whose
extent is unknown in advance, which is not a setting in which convergence can be established.
 
\subsection{Heuristic Placement}\label{sec:heuristic}
 
The heuristic strategy distributes the retention parameters along
analytical estimates of the spectral
branches~\citep{schmidStabilityTransitionShear2001}, with each
$\beta_k^+$ placed near a downstream spectral region and its removal
partner generated by the involution~\eqref{eqn:involution}.  Explicit
placement formulas are collected in Appendix~\ref{app:recursion}; the
features relevant to what follows are summarised here.
 
For subsonic disturbances, the upstream
and downstream acoustic branches are rotations of each
other by~$\pi$ about the pinch-point $\alpha_{\mathrm{mid}}$.  Choosing
the retention and removal parameters to respect this symmetry, that is
taking $\beta_k^-$ as the image of~$\beta_k^+$ under this rotation,
forces the auxiliary poles onto a balanced curve ($\Gamma^+$ contour) that passes
symmetrically through the pinch region.  For supersonic disturbances no
such rotational symmetry exists, but the upstream and downstream
spectra generally lie on opposite sides of a line parallel to and
slightly below the real axis, and complex conjugation provides an
analogous pairing.  In both regimes the removal parameters are
generated from the retention parameters by an involution
$\mathrm T:\mathbb C\to\mathbb C$,
\begin{equation}\label{eqn:involution}
    \beta_k^- = \mathrm T(\beta_k^+),
    \qquad
    \mathrm T(\beta) =
    \begin{cases}
        2\alpha_{\mathrm{mid}} - \beta, & \text{subsonic behaviour},\\[2pt]
        \overline{\beta}, & \text{supersonic behaviour},
    \end{cases}
\end{equation} where $\overline{\cdot}$ denotes complex conjugate. This involution is used throughout this work: in the present section to justify
the placement of the auxiliary poles, and in \S\ref{sec:greedy} as a
constraint on the admissible parameter sets.

The transonic regime uses both pairings at once.  In the SWiFT
configuration of \S\ref{sec:swift} the continuous branches are paired by
rotation about $\alpha_\mathrm{mid}$, following the subsonic scheme,
while the upstream discrete acoustic modes are paired by conjugation,
following the supersonic one.  Rotation is unsuitable for those modes,
since it would place their retention partners inside the region that
must be removed, whereas conjugation keeps both members of the pair near
the origin and independent of $\alpha_\mathrm{mid}$.  Different groups
therefore carry different involutions; what is preserved in every case
is that each removal parameter is generated as the image of its
retention partner. 

This differs from the construction
of~\citet{towneOnewaySpatialIntegration2015}, who distribute the removal
parameters along the paired acoustic branch, staggered relative to the
retention parameters, rather than taking each as the exact image of its
partner, and who rotate the convective parameters by $-\pi/2$ about the
same centre.  Taking exact images is what makes the pairing relation
hold exactly, and is what allows \S\ref{sec:greedy} to treat a parameter
pair as one indivisible
object~\citep{badcocktcfd2026}.
 
See Appendix~\ref{sec:C}, Fig.~\ref{fig:c1}, for an example in which
Eq.~\eqref{eqn:involution} with $c=1$ produces a $\beta^*$~contour
tracing the unit circle in Cayley space; the parameters there lie on a
single line through $\alpha_\mathrm{mid}$, and the figure is an
illustration of the pairing rather than a general property of it.
 
The heuristic is a physically motivated default, not an optimised
construction.  Its accuracy depends on how closely the analytical branch
estimates approximate the discretised spectrum, and on the quality of
the estimates for the discrete-mode locations, which are not analytic.
At moderate orders it is generally sufficient.  Near pinch-points, or
through rapid spectral reorganisation, it can leave residual projection
error that accumulates over a long march.

It is worth being explicit about what the heuristic requires, since the greedy
refinement of \S\ref{sec:greedy} inherits the same inputs.  Two things are
needed.  The first is the analytical continuous branches of
Appendix~\ref{sec:continuous_spectra}, which follow from the freestream
profile at each station and cost nothing to evaluate.  The second is an
approximate location for the upstream discrete acoustic modes, which we do not yet
obtain analytically.  These modes are trapped between the wall and the sonic
waveguide \citep{gushchinExcitationDevelopmentUnstable1990b}, and a WKB treatment of that region may well furnish their locations from
the profile alone; we have not pursued it, and take the locations instead from a computed spectrum at a single station.  Where they sit varies with the
configuration.  In the transonic case of \S\ref{sec:swift} they lie on the
imaginary axis near the origin; in the hypersonic case of \S\ref{sec:rec:sup}
they sit well away from it, at a real part that must be estimated and that
shifts with frequency.  Everything downstream of these two inputs, including
the greedy selection, is automatic.

\subsection{Greedy Refinement}\label{sec:greedy}

The ideal parameter set would make the scalar
filter~\eqref{eqn:quadrature_scalar} equal to one throughout $\Omega^+$
and zero throughout $\Omega^-$.  A greedy procedure approaches this by
adding parameters one at a time, each chosen to reduce the departure
from those values, and is defined by three choices: a finite proxy for
$\Omega^\pm$ on which the filter is evaluated, the structure of the
admissible parameter sets, in particular whether $\beta^+$ and
$\beta^-$ are selected independently or as linked pairs, and the norm in
which the departure is measured.  We first describe the choices made
by~\citet{sleemanGreedyRecursionParameter2025a}, then those made here.

The first choice is forced.  $\Omega^+$ and $\Omega^-$ are uncountable
subsets of the complex plane, so imposing the conditions on them exactly
is a minimax problem over uncountable sets; and in any case most of each
region contains no spectrum, so enforcing the conditions there would be
wasted effort.  Both algorithms therefore evaluate the filter on a
finite set of test points $\mathcal Z = \mathcal Z^+\cup\mathcal Z^-$
standing in for the regions, with
$\mathcal Z^\pm\subset\Omega^\pm$.  That set is where knowledge of the
spectrum enters, and the two algorithms differ mainly in where they
obtain it.

Write $\mathcal S^{(n)} = \{(\beta_k^+,\beta_k^-)\}_{k=1}^{n}$ for a set
of $n$ parameter pairs, $n=1,\ldots,N$, with $\mathcal S^{(0)}=\emptyset$
and the empty product taken as unity, and
$\mathcal P^+(\,\cdot\,;\mathcal S^{(n)})$ for the filter it induces.
Throughout the selection the filter is evaluated in the
product form~\eqref{eqn:product},
\begin{equation}\label{eqn:filter_product_form}
    \mathcal P^+(\alpha;\mathcal S^{(n)})
    = \bigl(1+cR(\alpha;\mathcal S^{(n)})\bigr)^{-1},
    \qquad
    R(\alpha;\mathcal S^{(n)})
    = \prod_{(\beta^+_k,\beta^-_k)\in\mathcal{S}^{(n)
    }}\frac{\alpha-\beta_k^+}{\alpha-\beta_k^-},
\end{equation}
which depends only on $c$ and the retention/removal parameters.  The
partial-fraction form~\eqref{eqn:quadrature_scalar} is not available at
this stage, since the auxiliary poles follow from the polynomial
identity~\eqref{eqn:poly_general} and are computed once the parameter
set is fixed.  The two forms represent the same function
(\S\ref{sec:owns_r}), so the selection is independent of which
evaluation is subsequently used.

The quantity to be reduced is the departure of the filter from its
target value at a test point,
\begin{equation}\label{eqn:residual}
    e(\zeta;\mathcal S^{(n)}) =
    \begin{cases}
        \bigl|\mathcal P^+(\zeta;\mathcal S^{(n)})-1\bigr|,
            & \zeta\in\mathcal Z^+,\\[2pt]
        \bigl|\mathcal P^+(\zeta;\mathcal S^{(n)})\bigr|,
            & \zeta\in\mathcal Z^- .
    \end{cases}
\end{equation}
The two algorithms minimise different norms of the same residual, and
select different objects in doing so.

\subsubsection{Algorithm of Sleeman and Colonius}\label{sec:sleeman_greedy}

\citet{sleemanGreedyRecursionParameter2025a} compute the full local eigenspectrum of the
two-way operator and partition it by Briggs' criterion, taking
$\mathcal Z^+$ and $\mathcal Z^-$ to be the downstream- and
upstream-propagating eigenvalues.  The same sets serve as the candidate
pool, so the parameters are themselves eigenvalues.

What is selected at each iteration is a single test point in each
subset, taken independently: the downstream eigenvalue at which the
current residual is largest becomes the next $\beta^+$, and the upstream
eigenvalue at which it is largest becomes the next $\beta^-$.  Because
the candidates are the test points, placing a parameter on one sets the
corresponding factor of~$R$ exactly to zero, so no trial evaluation is
needed.  No relation is imposed between the two selections.  The
iteration is initialised from a randomly chosen eigenvalue in each
subset, in practice from several such initialisations, retaining the set
with the lowest objective.  During marching the selected parameters are
tracked by local sparse eigen-solves rather than a full decomposition at
every station, and the full spectrum is recomputed and the selection
re-run when the count of upstream and downstream characteristics
changes.

Their norm is the worst case, which in the present notation reads
\begin{equation}\label{eqn:J_inf}
    \mathcal J_\infty(\mathcal S^{(n)})
    = \max_{\zeta\in\mathcal Z} e(\zeta;\mathcal S^{(n)}),
\end{equation}
so that the selection removes the largest current departure at each
selection iteration.  They evaluate it through $|R|$ on $\mathcal Z^+$ and $|R|^{-1}$ on
$\mathcal Z^-$, which for their choice $c=1$ are the leading-order forms
of~\eqref{eqn:residual} and are the quantities appearing in their error bound
for the projected operator.  The form quoted is their OWNS-R objective, which
takes the larger of the two worst cases; their OWNS-P objective takes their
product, following the corresponding bound.

Two aspects bear on the present application.  Selecting $\beta^+$ and
$\beta^-$ independently imposes no geometric relationship between them,
so the paired structure that places the auxiliary poles on a separating
contour (\S\ref{sec:implicit}) is not preserved.  Additionally, the number of
distinct removal parameters is bounded by $|\mathcal Z^-|$, which in
supersonic configurations is a few discrete modes. 

\subsubsection{Paired-symmetric greedy refinement}\label{sec:greedy_ours}

We retain the iterative error-reduction principle
of~\citet{sleemanGreedyRecursionParameter2025a} and change the candidate pool, from which
the paired structure follows.  The objective also differs in form.

\paragraph*{(i) Analytical candidate pool.}
Candidates are drawn from an oversampled version of the heuristic
placement of \S\ref{sec:heuristic} rather than from eigenvalues.  The
heuristic distributions, comprising the continuous-branch estimates, the
discrete-mode placements, and any regime-specific groups, are sampled at
$N_\mathrm{c}\gg N$ points, typically $5$ to $20$ times~$N$.  The
resulting set serves as both the candidate pool and the test set
$\mathcal Z$, partitioned as $\mathcal Z^\pm$ by construction, and no
eigendecomposition is required during the numerical march.

The pool is not free of spectral input: it inherits the branch estimates
and discrete-mode locations of the heuristic.  What the oversampling
buys is tolerance, in that the individual placements need only be
approximately right for the selection to find a good subset, and the
constants of Appendix~\ref{app:recursion} matter much less than the
support of the distribution.

\paragraph*{(ii) Paired candidates.}
Because the heuristic generates each removal parameter from its
retention partner through the involution~\eqref{eqn:involution}, every
element of the pool is already a pair $(\beta^+,\mathrm T(\beta^+))$.
The selection therefore operates on pairs rather than on individual
parameters, and the symmetry that \S\ref{sec:implicit} invokes to
justify the placement of the auxiliary poles is carried through the
greedy iteration at no additional cost.  The same dependence holds in
reverse for the eigenvalue pool of \S\ref{sec:sleeman_greedy}: computed
eigenvalues carry no pairing, so selection from them is necessarily
independent.  In each case the structure of the pool determines the
structure of the selection.

Two consequences follow.  The auxiliary poles retain the contour
properties of \S\ref{sec:implicit} for every intermediate set
$\mathcal S^{(n)}$, not only for the final one.  And because each
removal parameter is generated rather than selected, the number of
admissible pairs is fixed by the sampling of the candidate distribution
and not by the size of the upstream spectrum, which removes the
cardinality bound of \S\ref{sec:sleeman_greedy}.

The second of these matters most where the upstream spectrum is smallest.
In a supersonic disturbance the continuous acoustic branches are entirely
downstream-propagating and the upstream content reduces to a handful of
discrete trapped modes, so a pool drawn from computed eigenvalues admits only
that handful of distinct removal parameters.  Under the analysis
of~\citet{sleemanGreedyRecursionParameter2025a} this is sufficient, since parameters placed
exactly on the upstream eigenvalues annihilate them; it becomes a constraint
only when $N$ is to be raised past that count, either to improve the retention
of the downstream branches or because the upstream modes have moved and the
placement is no longer exact.  Both occur here: the hypersonic case of
\S\ref{sec:hypersonic} requires $N>30$ because a discrete upstream mode
migrates during the numerical march, and the transonic case of \S\ref{sec:swift}
requires the parameter set to track modes that detach from the continuum as
the flow accelerates.  Generating the removal parameters from a continuous
candidate distribution keeps $N$ free in both.

\paragraph*{Selection procedure.}
What is selected here is a whole pair, and the norm is the aggregate,
\begin{equation}\label{eqn:J_2}
    \mathcal J_2(\mathcal S^{(n)})
    = \sum_{\zeta\in\mathcal Z} e(\zeta;\mathcal S^{(n)})^2 .
\end{equation}
Selection proceeds one pair at a time from
$\mathcal S^{(0)}=\emptyset$.  Given the $n$ pairs already chosen, every
candidate pair not yet selected is trialled by forming
$\mathcal S^{(n)}\cup\{(\beta_j^+,\beta_j^-)\}$ and evaluating the
objective, and the minimiser is retained,
\begin{equation}\label{eqn:greedy_step}
    \mathcal S^{(n+1)}
    = \mathcal S^{(n)}\cup\bigl\{(\beta_{j^\ast}^+,\beta_{j^\ast}^-)\bigr\},
    \qquad
    j^\ast = \arg\min_{j}\;
    \mathcal J_2\bigl(\mathcal S^{(n)}\cup\{(\beta_j^+,\beta_j^-)\}\bigr).
\end{equation}
The procedure terminates at $n=N$, or earlier if the reduction in
$\mathcal J_2$ between iterations falls below a prescribed threshold \emph{e.g.} $10^{-14}$.
Because an aggregate measure is not reduced to zero at any single point,
the direct rule of \S\ref{sec:sleeman_greedy} is unavailable and the
trial evaluation in~\eqref{eqn:greedy_step} is required.  The two norms
target different things, the least well separated test point against the
bulk fit, and we do not claim an advantage for either.

No weights enter the selection: the product form depends only on the
retention and removal parameters.  They are computed once the set is
fixed and the auxiliary poles have been obtained from
Eq.~\eqref{eqn:poly_general}. 

Where least-squares weights are used, the separation conditions are
enforced on the full candidate pool $\mathcal Z$ of $N_\mathrm{c}$
points rather than only at the selected parameters; this is the
enlarged test set anticipated in \S\ref{sec:weights}.

The product form corresponds to the
residue weights~\eqref{eqn:standard_weights}, so where least-squares
weights are used the filter finally deployed differs slightly from the
one selected; the two constructions agree closely
(\S\ref{sec:weights}), and the distinction is immaterial here.

The greedy selection is not free of spectral input;  it requires analytical estimates of the continuous
branches, which follow from the freestream profile, and an approximate
location for the upstream discrete acoustic modes, which is here taken from a
computed spectrum at one station.  What the method removes is the
per-configuration tuning of the individual placements.

\paragraph*{Greedy-spectrum benchmark.}
To test whether the analytical pool is a faithful surrogate for the true
spectrum, we also run a variant that applies the same paired selection
and the same objective but takes $\mathcal Z$ from the computed
eigenvalues, obtained by QZ decomposition of the
pencil~$(\mathbf B,\mathbf A)$.  This requires a full eigendecomposition
at every station and is not used for routine analysis.  It is not the
algorithm of \S\ref{sec:sleeman_greedy}, since the pairing constraint is
retained; it isolates the effect of the pool alone.  As reported in
\S\ref{sec:inc}, the two selections produce closely matching parameters,
indicating that the analytical pool captures the spectral features
relevant to placement.

\begin{table}[!htbp]
\centering
\footnotesize
\renewcommand{\arraystretch}{1.25}
\begin{tabular}{@{}lp{5cm}lcp{3cm}@{}}
\toprule
Configuration & Candidate and test set $\mathcal Z$
    & Pair structure & Norm & Eigendecomposition \\
\midrule
Heuristic (\S\ref{sec:heuristic})
    & the $N$ pairs themselves; no selection
    & paired by $\mathrm T$ & -- & not required \\
Sleeman \& Colonius (\S\ref{sec:sleeman_greedy})
    & computed eigenvalues
    & independent & $\mathcal J_\infty$ & required, with downstream tracking \\
Greedy (\S\ref{sec:greedy_ours})
    & heuristic placement oversampled,
      $N_\mathrm{c}=5$--$20\,N$
    & paired by $\mathrm T$ & $\mathcal J_2$ & not required \\
Greedy-spectrum benchmark
    & computed eigenvalues
    & paired by $\mathrm T$ & $\mathcal J_2$ & required at every station \\
\bottomrule
\end{tabular}
\caption{Candidate and test sets.  In every configuration one set
supplies both the candidates and the test points, so the two are not
distinguished.  The
first row is not a greedy method and is listed because it is the special
case $N_\mathrm{c}=N$, in which every pair is retained; it is also the
default test set for the weight fit of \S\ref{sec:weights}.  The final
row is not the algorithm of \citet{sleemanGreedyRecursionParameter2025a}: it retains the
pairing and the objective of \S\ref{sec:greedy_ours} and changes only
the pool, so that the effect of the pool is isolated from the effect of
the pairing.}
\label{tab:greedy}
\end{table}

\section{Results}\label{sec:results}

The OWNS-S framework is assessed across three configurations that together
span the primary spectral challenges of boundary-layer stability analysis. An
incompressible boundary layer ($M = 0.02$) tests projection accuracy near the
acoustic pinch-point, where the upstream and downstream branches coalesce and
separation is most challenging. A hypersonic boundary layer ($M = 4.5$) tests
the removal of discrete upstream-propagating trapped acoustic modes that
coexist with the downstream fast and slow acoustic, vorticity, and entropy
branches supporting Mack's second-mode growth. A transonic swept-wing boundary
layer (SWiFT,~\citealp{coppinSWiFTModelDevelopment2025}) presents a strongly
non-parallel, three-dimensional developing flow whose disturbance spectrum
reorganises topologically during the numerical march, passing continuously through
subsonic, transonic, and supersonic regimes.

The formulations compared below differ along three independent choices:
how the rational filter is \emph{evaluated}, how the \emph{weights} are
obtained, and how the \emph{parameters} are placed.  Only the first is a
structural difference between OWNS-R and OWNS-S; the remaining two apply
to the summation form alone, which is itself a consequence of the
decoupling described in \S\ref{sec:forward_error}.  Each configuration is
included to isolate one of these choices, as summarised in
Table~\ref{tab:methods}.

\begin{table}[!htbp]
\centering
\footnotesize
\renewcommand{\arraystretch}{1.0}

\begin{tabularx}{\linewidth}{@{}
l
l
l
>{\raggedright\arraybackslash}p{1.8cm}
c
>{\raggedright\arraybackslash}X
@{}}
\toprule
Method & Evaluation & Weights & Placement & Large $N$ & Isolates \\
\midrule
PSE
    & --            & --            & --
    & --
    & External reference; independent of the projector \\
OWNS-R
    & product       & --            & heuristic
    & no$^{\dagger}$
    & Baseline; establishes the failure mode \\
OWNS-S
    & sum           & residue       & heuristic
    & yes
    & Evaluation structure: same operator as OWNS-R in exact arithmetic \\
OWNS-S~(LS)
    & sum           & least squares & heuristic
    & yes
    & Pole--weight consistency at fixed poles \\
Greedy OWNS-S~(LS)
    & sum           & least squares & greedy (oversampled heuristic pool)
    & yes
    & Parameter placement, with the weights refitted to the selected poles \\
\bottomrule
\end{tabularx}
\caption{Configurations compared in \S\ref{sec:results}.  Evaluation is
either the recursive product~\eqref{eqn:R_matrix} or the
partial-fraction sum~\eqref{eqn:owns_s_apply}. \\[2pt]
$^{\dagger}$~OWNS-R is well defined at any $N$ in exact arithmetic.  The
entry records that its finite-precision evaluation fails once $N$ exceeds
an empirical threshold which is not predictable in advance
(\S\ref{sec:forward_error}).}
\label{tab:methods}
\end{table}

All computations use an in-house compressible spatial-marching solver
implementing LST, PSE~\citep{mughalactive}, OWNS-R
\citep{zhuRecursiveOnewayNavierStokes2023}, and the new OWNS-S formulations in
both wave-carrier-factored and non-factored form
\citep{badcocktcfd2026, IUTAM2024_ejbmug}. Except for
the projection operator, the numerical procedure is identical across OWNS-R and
OWNS-S, so any differences arise solely from the projector. All steady base
flows are computed from the fully compressible non-similar boundary-layer
equations using a second-order Keller-box method
\citep{thomasetal2016, kaupscebeci1977}.

\subsection{Numerical Implementation}\label{sec:numerics}

The wall-normal direction is discretised with fourth-order finite differences
on the stretched grid of \citet{malikNumericalMethodsHypersonic1990},
\begin{equation}\label{eqn:malik_grid}
    y_{\mathrm{Malik}}(\eta) = \frac{a_\mathrm{M}\eta}{b_\mathrm{M}-\eta},
\end{equation}
with $\eta \in [0,1]$, far-field truncation at $y_{\max}$, and more than half
the points placed in the near-wall boundary-layer region $0 < y < y_{\mathrm{half}}$,
\begin{equation}\label{eqn:malik_ab}
    a_\mathrm{M} = \frac{y_{\max}\,y_{\mathrm{half}}}{\,y_{\max}-2y_{\mathrm{half}}\,},
    \qquad
    b_\mathrm{M} = 1 + \frac{a_\mathrm{M}}{y_{\max}}.
\end{equation}
Streamwise marching uses a second-order backward-difference scheme.

No-slip is imposed at the wall with either isothermal or adiabatic thermal
conditions, and Thompson one-way conditions
\citep{thompsonTimeDependentBoundary1987a} in the far field suppress
reflections. At the inflow plane the disturbance is introduced as a local LST
eigenfunction; freestream forcing and controlled wall actuation are also
supported but are not used here. The governing equations use the orthogonal
body-fitted coordinate system of \citet{mughalactive}, whose metric
coefficients appear explicitly in the LHNS operators.

Wave-carrier factoring
\citep{badcocktcfd2026} is
applied: the disturbance is factored by a fixed carrier wave, with the carrier
wavenumber set to the vorticity/entropy branch endpoint $\alpha_{1,2,3}(0)$ (Eq.~\eqref{eq:cont_vort}).
This is a WKB-type ansatz that splits off the rapid oscillation and leaves a
slowly varying envelope, so the marching step can be enlarged without loss of
spatial resolution. Because the carrier is fixed, no PSE-type iteration is
required.  All OWNS results reported here, both OWNS-R and OWNS-S, use this
factoring, so the comparison between them is unaffected by it; the PSE
reference uses its own carrier iteration.  We record it because the marching
step sizes quoted below are not directly comparable with those of unfactored
OWNS implementations.

We adopt the standard non-dimensionalisation of compressible boundary-layer
stability analysis. Spatial variables are scaled by the inlet boundary-layer
thickness, and velocities, temperature, and pressure by their freestream
values. The inlet Reynolds number $R$ is defined on this thickness, the local
value varying downstream through $R_\delta(x)$. Disturbances are characterised
by the non-dimensional frequency and spanwise wavenumber, $f = \omega/R$ and
$b = \beta/R$.

A local diagnostic for the streamwise wavenumber $\alpha$ is given by the
Rayleigh quotient associated with the Chu energy inner product
\citep{chuEnergyTransferSmall1965, hanifiTransientGrowthCompressible1996},
\begin{equation}\label{eqn:wave}
    \alpha = -i\,\frac{(\phi,\mathbf A\phi_x)_{\mathcal{H}}}
                       {(\phi,\mathbf A\phi)_{\mathcal{H}}}.
\end{equation}
For a single mode, $\alpha$ returns the corresponding eigenvalue. It is
evaluated at each station as a post-processing diagnostic, providing the
wavenumber and growth-rate curves reported below; it does not feed back into
the march.

The projection operator $\mathbf{P}_N^+$ is constructed from the full
discretised operators $(\mathbf{B},\mathbf{A})$, including boundary-condition
rows; no deflation of the infinite-eigenvalue subspace is performed. The
projection-quality metrics of \S\ref{sec:errors}, which classify eigenvalues
via Briggs' criterion, are evaluated on the finite eigenvalues returned by the
QZ decomposition. 

Additionally, the projection is deployed on the state rather than on the equations.  At each
station the BDF2 update of Eq.~\eqref{eqn:two-way2} is formed from the
unmodified operators, and the projector is applied once to the resulting state,
$\phi \leftarrow \mathbf P^+_N\phi$, before the march proceeds to the next
station.  \citet{zhuRecursiveOnewayNavierStokes2023} show that for an equation with no forcing and implicit
integration this per-step projection of the state is equivalent to integrating
the projected evolution equation, so the stability of the march rests on the projector alone and not on the integration scheme.  The
insertion point is identical for OWNS-R and OWNS-S, so the comparison between
them is unaffected by it. Alternatively, one can numerically march the projected problem defined by the one-way pencil Eq.~\eqref{eqn:project_pencil}.

In this work we use double precision throughout. 

\subsection{Projection-Quality Metrics}\label{sec:errors}

Three metrics are used throughout this section to assess the approximate
projector $\mathbf P^+_N$ at a single station, independently of the marching
procedure.  They test internal consistency, fidelity of the retained
downstream modes, and suppression of the upstream modes.  All are computed
from the eigenvalues of the original pencil $(\mathbf B,\mathbf A)$ and of the
projected pencil $(\mathbf{B}\mathbf{P}^+_N,\,\mathbf A)$, obtained by QZ
decomposition and reported in standard $\alpha$-space.

\paragraph*{Idempotency error.}
An exact projector satisfies $(\mathbf P^+)^2=\mathbf P^+$, so the departure
\begin{equation}\label{eqn:idem_metric}
    \mathcal E_\mathrm{idem}
    = \bigl\lVert(\mathbf P^+_N)^2 - \mathbf P^+_N\bigr\rVert_\mathrm{F}
\end{equation}
measures internal inconsistency of the computed operator.  Here
$\lVert\cdot\rVert_\mathrm{F}$ is the normalised Frobenius norm of
\S\ref{sec:synthetic}.

\paragraph*{Retention error.}
The projection should leave the downstream eigenvalues where it found them.
Each projected downstream eigenvalue $\alpha_k^+\in\sigma(\mathbf \Pi^+)$ is paired with the eigenvalue $\alpha_k\in\sigma\left(\mathbf \Pi\right)$ it most
plausibly originated from, and the displacements are summed,
\begin{equation}\label{eqn:retain_metric}
    \mathcal E_\mathrm{retain}
    = \sum_{\alpha_k^+\in\,\sigma(\mathbf \Pi^+)}\lvert\alpha_k^+-\alpha_k\rvert .
\end{equation}
The pairing is not done by nearest neighbour, which can assign two original
eigenvalues to the same image and leave another unmatched.  Instead every
original downstream eigenvalue is assigned to a distinct projected one, and
among all such one-to-one assignments the one with the smallest total
displacement is taken.  This is the classical assignment problem and is solved
here by the Hungarian algorithm~\citep{kuhnHungarianMethodAssignment1955}.

\paragraph*{Leakage error.}
Modes the projection should have removed are mapped to zero by the exact
projector, so the residual magnitude of their images,
\begin{equation}\label{eqn:leak_metric}
    \mathcal E_\mathrm{leak}
    = \sum_{\alpha^-_k\in\,\sigma (\mathbf \Pi^-)}\lvert\alpha^-_k\rvert ,
\end{equation}
measures how much upstream content survives.  Ideal one-way behaviour is
$\mathcal E_\mathrm{leak}\to0$.

Both $\mathcal E_\mathrm{retain}$ and $\mathcal E_\mathrm{leak}$ are sums over
the discretised spectrum, so their absolute values depend
on $n_y$ and are meaningful only in comparison between methods on a common
grid.  The two also floor at different levels, and for a reason worth stating:
$\mathcal E_\mathrm{leak}$ measures distance from zero and is limited only by
round-off on small numbers, whereas $\mathcal E_\mathrm{retain}$ measures
distance between two large numbers and inherits the absolute accuracy of the
QZ decomposition on the far reaches of the continuous branches, where
$\lvert\alpha\rvert$ is large.  A retention floor several orders above the
leakage floor is therefore expected and does not indicate a defect in the
projection.

\paragraph*{Per-station and marching thresholds.}
With the metrics defined, the order at which OWNS-R ceases to be usable can be
stated in the two senses in which it appears below, which differ and are
quoted repeatedly.  The \emph{per-station} threshold is the order at which the
metrics above first degrade.  It is a property of the projector alone,
measured on the whole discretised spectrum, and it is the earlier of the two.

The \emph{marching} threshold is higher, and the reason is that a march never
sees the whole spectrum.  What is marched is a disturbance introduced at the
inlet and advanced by a scheme of finite order at a finite step, so the
solution carries only the part of the spectrum that the inlet condition
excites and the discretisation resolves.  Leakage into modes outside that
range degrades the projector as the metrics record, but has little to act on
in the marched solution, and the march continues to produce the right answer
for some way past the point at which the operator has stopped being a
projector.  How far depends on the scheme, the step size and the disturbance,
which is why the two thresholds are reported separately throughout.

\subsection{Incompressible Boundary Layer ($M = 0.02$)}\label{sec:inc}

The incompressible flat-plate boundary layer is demanding for projection-based
formulations because the upstream and downstream acoustic branches coalesce at
the pinch-point $\alpha_{\mathrm{mid}}$ (Fig.~\ref{fig:spec_cartoon1}a), where
the spectra become inseparable within any fixed closed contour. This case
therefore tests the numerical robustness of the projection operator rather
than physical complexity.

The base flow follows \citet{bertolottiAnalysisLinearStability1991}, with
freestream Reynolds number $R_\infty = 10^6\,\mathrm{m}^{-1}$, temperature
$T_\infty = 298\,\mathrm{K}$, a two-dimensional disturbance of frequency
$f = 86 \times 10^{-6}$, and spanwise wavenumber $b = 0$. The wall-normal
domain uses Malik stretching with $y_\mathrm{max} = 600$, $y_\mathrm{half} = 30$,
and $n_y = 251$ points. For the spatial-evolution tests an LST T--S wave is
marched from the inlet at $R_\delta = 400$ to $R_\delta = 1100$ with
$n_x = 1000$ streamwise points; the PSE reference uses $n_x = 110$.

\subsubsection{Projection Quality}

\begin{figure}[!htbp]
    \centering
    \begin{subfigure}[b]{0.49\textwidth}
        \centering
        \includegraphics[width=\textwidth]{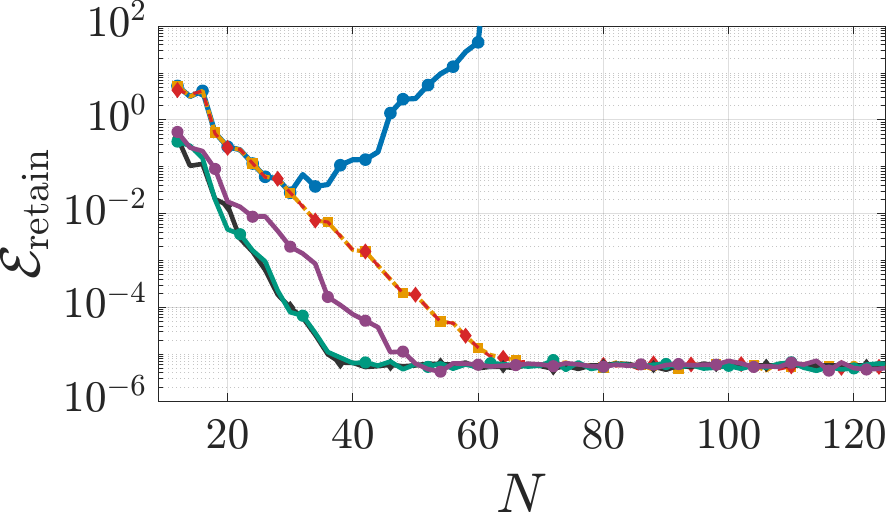}
        \caption{Downstream retention.}
        \label{fig:e1}
    \end{subfigure}
    \hfill
    \begin{subfigure}[b]{0.49\textwidth}
        \centering
        \includegraphics[width=\textwidth]{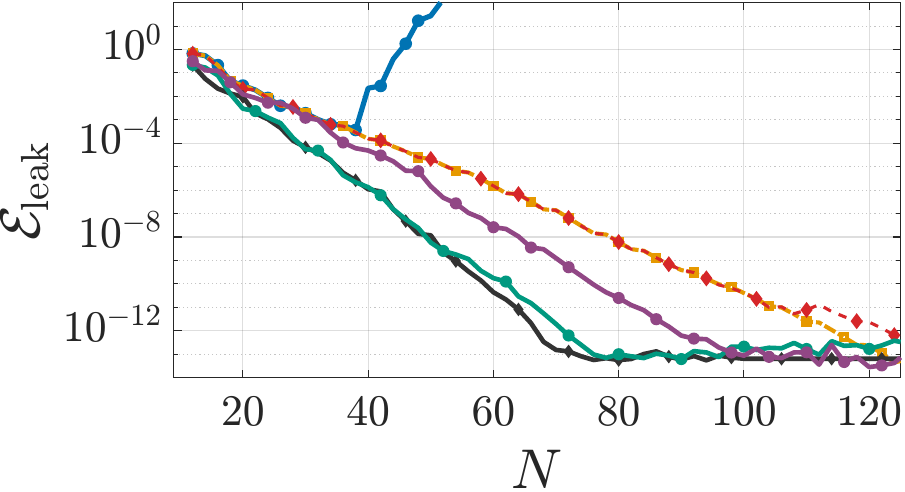}
        \caption{Upstream removal.}
        \label{fig:e2}
    \end{subfigure}

    \begin{subfigure}[b]{0.49\textwidth}
        \centering
        \includegraphics[width=\textwidth]{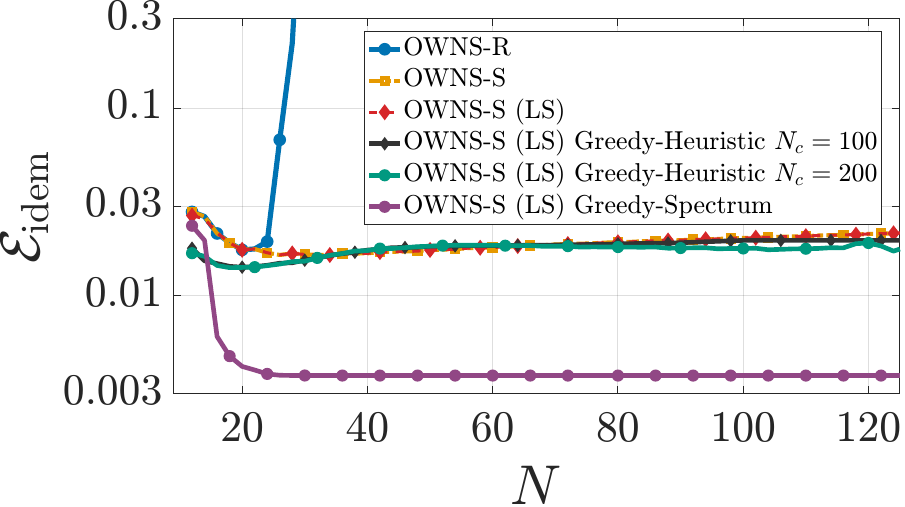}
        \caption{Idempotency.}
        \label{fig:e3}
    \end{subfigure}
    \caption{Projection-quality metrics for the 2D incompressible boundary
    layer, comparing OWNS-R, OWNS-S, and OWNS-S~(LS) with greedy and non-greedy
    selection, against approximation order $N$: (a) downstream retention
    $\mathcal{E}_{\mathrm{retain}}$, (b) upstream leakage
    $\mathcal{E}_{\mathrm{leak}}$, (c) idempotency $\mathcal{E}_{\mathrm{idem}}$.
    OWNS-R diverges for $N > 36$ for retention and removal metrics, and  $N>28$ for idempotency; all OWNS-S variants remain stable and improve
    until numerical saturation.}
    \label{fig:e}
\end{figure}

Figure~\ref{fig:e} reports the projection-quality metrics against approximation
order $N$. OWNS-R deteriorates rapidly beyond $N \approx 36$ in retention and leakage metrics, and earlier ($N>28$) for idempotency , all indicating loss of projection quality,
while OWNS-S and OWNS-S~(LS) remain stable for all orders tested. The
downstream-retention error falls to a numerical floor of
$\mathcal{E}_{\mathrm{retain}} \sim 10^{-5}$, the upstream-leakage error
continues to $\mathcal{E}_{\mathrm{leak}} \sim 10^{-12}$ for the best-resolved
cases, and the idempotency error decreases and then saturates. The continued
convergence of $\mathcal{E}_{\mathrm{retain}}$ and
$\mathcal{E}_{\mathrm{leak}}$ shows the saturation does not reflect loss of
fidelity in the dynamically relevant spectrum; this is confirmed by the
subsequent spatial marching in every test case (\S\ref{sec:marching}).

For the non-greedy pole distribution, least-squares recalibration has only a
minor effect, indicating the residue weights are already well conditioned.
Greedy selection is more influential at low and moderate orders: concentrating
parameters near the pinch-point, where the upstream and downstream spectra are
closest, reduces the number of parameters needed for accurate separation.
Greedy selection from the full eigenvalue spectrum, used here only as a
benchmark, gives the lowest idempotency error, while greedy selection from the
heuristic candidate pool achieves comparable retention and leakage without
requiring the full spectrum. The corresponding placements are shown in
Fig.~\ref{fig:spec1} (\S\ref{sec:rec:sub}).

\subsubsection{Spatial Marching}\label{sec:marching}

\begin{figure}[!htbp]
    \centering
    \includegraphics[width=0.6\textwidth]{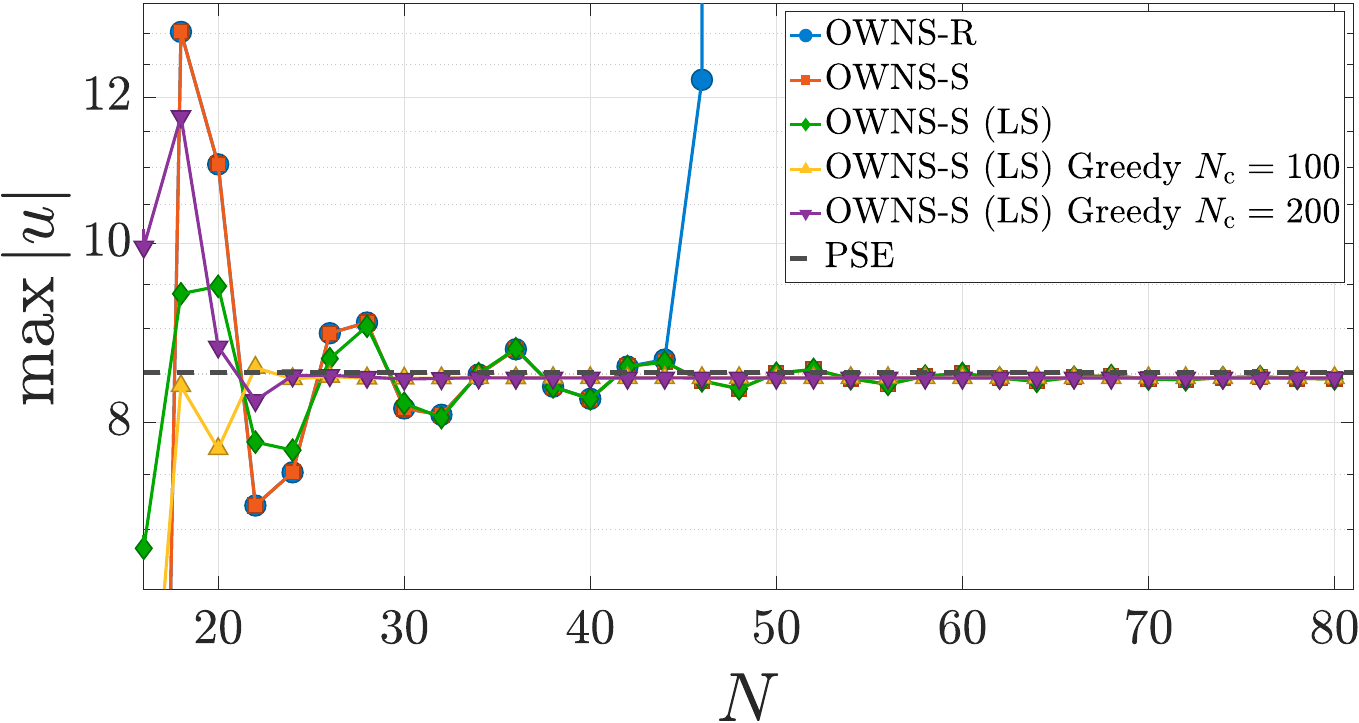}
    \caption{Convergence of $\max|u|$ at $R_\delta \approx 815$ against
    approximation order $N$. OWNS-R and standard OWNS-S coincide for $N \le 44$ at
    identical retention/removal parameters (OWNS-R is hidden beneath OWNS-S over this
    range), but OWNS-R fails drastically for $N > 44$. The greedy variants reach
    converged amplitudes at the smallest orders.}
    \label{fig:umax_convergence}
\end{figure}

Figure~\ref{fig:umax_convergence} shows the convergence of the disturbance
amplitude $\max|u|$ at the peak-growth location ($R_\delta \approx 815$) against
approximation order $N$. OWNS-R and standard OWNS-S give nearly identical amplitudes
for $N \le 44$, confirming that the two evaluate the same rational projector
if the recursive product remains numerically stable. Standard OWNS-S reaches
the converged amplitude at $N > 60$, but OWNS-R fails sharply for
$N > 44$. The march survives several orders beyond the
single-station breakdown of Fig.~\ref{fig:e} ($N \approx 36$) because the
projection error must accumulate over the $n_x$ stations before it corrupts
the disturbance.

Least-squares recalibration improves the behaviour at lower orders, giving
accurate amplitudes for $N > 50$. The greedy variants (candidate pools
$N_\mathrm{c} = 100$ and $200$) converge at $N<30$, again reflecting
the concentration of parameters near the pinch-point. For sufficiently large
$N$, all OWNS-S variants collapse onto the PSE reference, the greedy variant
requiring the fewest retention/removal parameters.

\subsection{Hypersonic Boundary Layer ($M = 4.5$)}\label{sec:hypersonic}

The hypersonic flat-plate boundary layer at $M = 4.5$ poses a qualitatively
different challenge. The upstream spectrum consists of a few discrete trapped
acoustic modes beneath the sonic line \citep{gushchinExcitationDevelopmentUnstable1990b}, whose removal is essential for the
stability of any one-way formulation.

The flow conditions follow \citet{maNumericalSimulationReceptivity2001,maReceptivitySupersonicBoundary2003a} and
match \citet{zhuRecursiveOnewayNavierStokes2023}:
$R_\infty = 7.2 \times 10^6\,\mathrm{m}^{-1}$, $T_\infty = 65.15\,\mathrm{K}$,
$p_\infty = 728.44\,\mathrm{Pa}$, Prandtl number $\sigma = 0.72$, and
disturbance frequency $f = 2.2 \times 10^{-4}$. Two wall-normal grids are used:
the eigenvalue spectra are computed with $n_y = 121$ ($y_\mathrm{half} = 30$,
$y_\mathrm{max} = 1500$), and the spatial march with $n_y = 801$
($y_\mathrm{half} = 200$, $y_\mathrm{max} = 1500$). The streamwise march spans
$R_\delta = 400$--$1280$ with $n_x = 3000$, using a slow LST mode as the inlet
condition. Parameter placements are given in \S\ref{sec:rec:sup}.

\subsubsection{Projection Quality}

\begin{figure}[!htbp]
    \centering
    \begin{subfigure}[b]{0.49\textwidth}
        \centering
        \includegraphics[width=\textwidth]{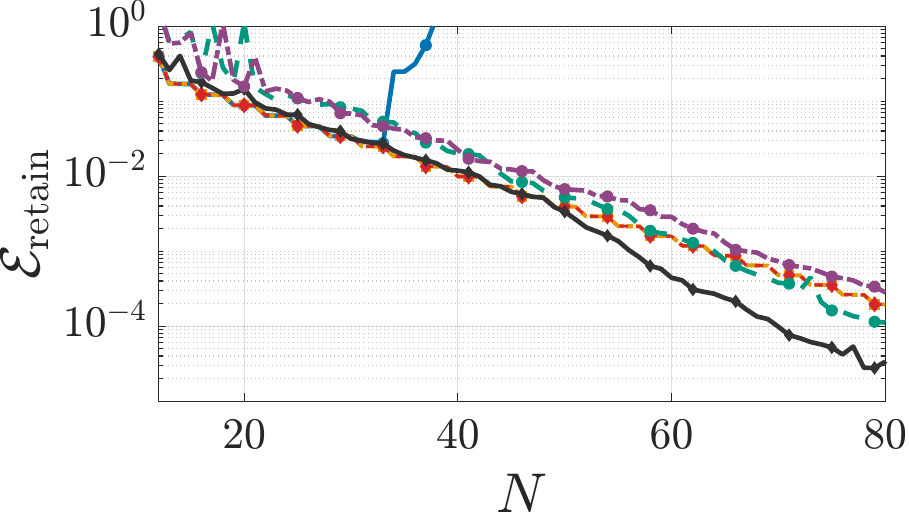}
        \caption{Downstream retention.}
        \label{fig:m4p5:e1}
    \end{subfigure}
    \hfill
    \begin{subfigure}[b]{0.49\textwidth}
        \centering
        \includegraphics[width=\textwidth]{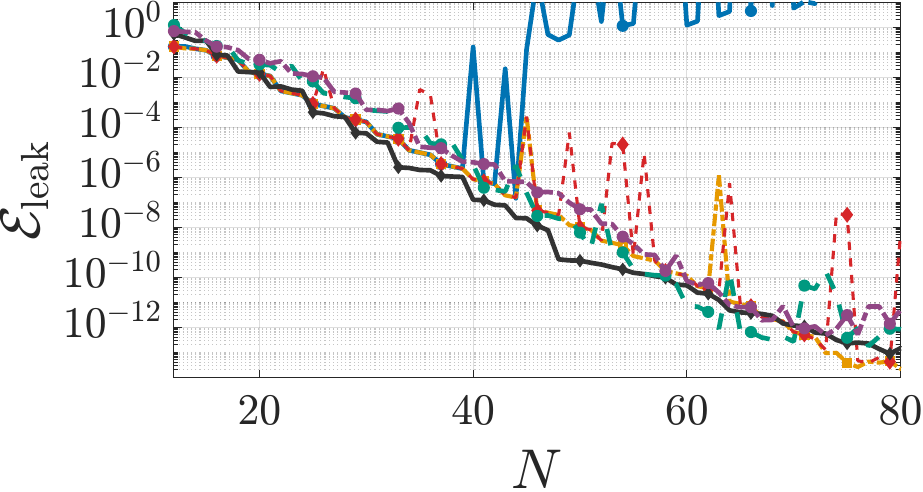}
        \caption{Upstream removal.}
        \label{fig:m4p5:e2}
    \end{subfigure}

    \begin{subfigure}[b]{0.49\textwidth}
        \centering
        \includegraphics[width=\textwidth]{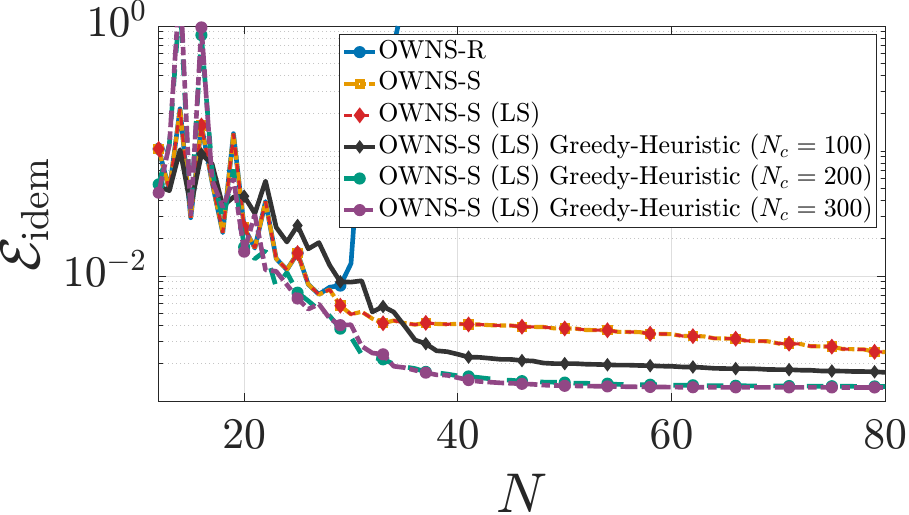}
        \caption{Idempotency.}
        \label{fig:m4p5:e3}
    \end{subfigure}
    \caption{Projection-quality metrics for the hypersonic boundary layer
    ($M = 4.5$). OWNS-R diverges for $N \ge 36$; OWNS-S and OWNS-S~(LS) remain
    stable. Greedy LS selection gives the lowest retention and leakage errors.
    Isolated leakage spikes in (b) are mode-pairing artefacts of the diagnostic
    and do not affect convergence.}
    \label{fig:m4p5:e}
\end{figure}

The metrics in Fig.~\ref{fig:m4p5:e} confirm the same structural instability as
in the incompressible case: OWNS-R diverges for $N \ge 36$, while OWNS-S and
OWNS-S~(LS) remain stable for all $N$. Greedy LS recalibration achieves the
lowest errors across all three metrics, and is insensitive to the candidate
pool over $N_\mathrm{c} = 100$, $200$, and $300$, which all perform well.
The isolated leakage spikes in Fig.~\ref{fig:m4p5:e2} are mode-pairing
artefacts of the diagnostic rather than failures of the projection; they appear
proportionally large because the upstream spectrum contains very few
eigenvalues, and were verified by inspecting the projected spectra at the
affected $N$.

\subsubsection{Spatial Marching: Mack's Second Mode}

\begin{figure}[!htbp]
    \centering
    \begin{subfigure}[b]{0.49\textwidth}
        \centering
        \includegraphics[width=\textwidth]{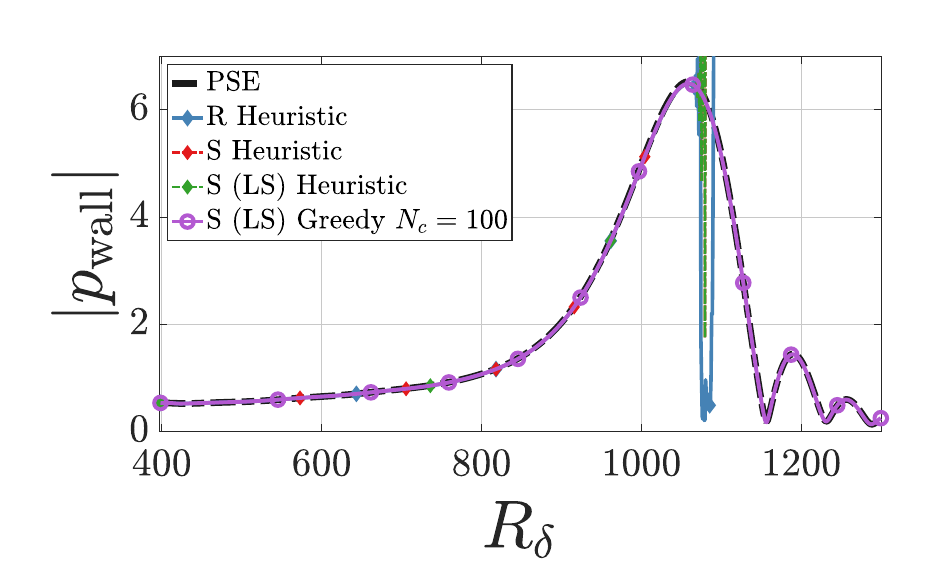}
        \caption{$N = 30$.}
        \label{fig:m4p5:marching_wave}
    \end{subfigure}
    \hfill
    \begin{subfigure}[b]{0.49\textwidth}
        \centering
        \includegraphics[width=\textwidth]{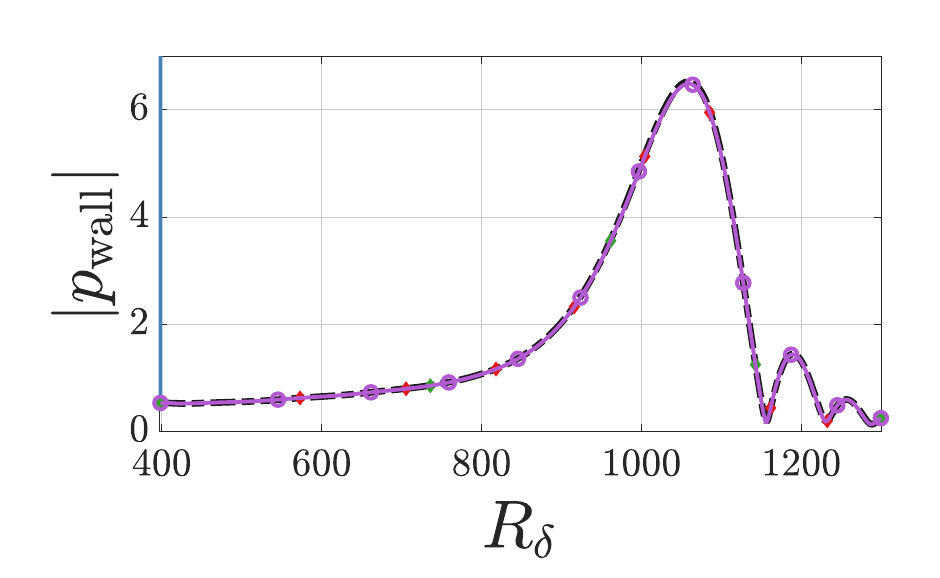}
        \caption{$N = 60$.}
        \label{fig:m4p5:marching_pwall}
    \end{subfigure}
    \caption{Wall-pressure amplitude $|p_\mathrm{wall}|$ for the hypersonic
    boundary layer ($M = 4.5$) at (a) $N = 30$ and (b) $N = 60$. At $N = 30$ the
    three heuristic formulations (OWNS-R, OWNS-S, OWNS-S~(LS)) diverge at the
    same $R_\delta \approx 1080$, indicating a common cause: a moving or
    emerging upstream discrete mode falls outside the heuristic coverage. At
    $N = 60$ the larger parameter count covers this mode and OWNS-S and
    OWNS-S~(LS) recover; OWNS-R instead diverges immediately from the inlet through the
    recursive-product instability. Greedy OWNS-S~(LS) ($N_\mathrm{c} = 100$)
    tracks PSE at both $N$ by re-optimising the placement at each station.}
    \label{fig:m4p5:marching}
\end{figure}

Figure~\ref{fig:m4p5:marching} compares wall-pressure amplitudes at $N = 30$ and
$N = 60$. The way the $N = 30$ case fails identifies its cause. The three
heuristic formulations (OWNS-R, OWNS-S, and OWNS-S~(LS), all using the
placement of \S\ref{sec:rec:sup}) share an identical recursion-parameter set and
differ only in how the projection is built from it, yet all three diverge at the
same station, $R_\delta \approx 1080$. A failure common to three independent
projection evaluations at one location cannot belong to any one evaluation; it
must originate in the parameter set. The cause is spectral migration: discrete
upstream acoustic modes shift or emerge. The heuristic placement does not yet track
this, and by $R_\delta \approx 1080$ an upstream discrete mode lies outside its
coverage, so the resulting leakage corrupts the march from that point on. The
greedy OWNS-S~(LS) has a well chosen placement at every station and stays
coincident with PSE throughout.

At $N = 60$ (Fig.~\ref{fig:m4p5:marching}b) the additional parameters extend the
heuristic coverage, the missed upstream mode is suppressed, and both OWNS-S and
OWNS-S~(LS) reproduce PSE to plotting accuracy. OWNS-R, by contrast, now
diverges from the inlet through the multiplicative amplification of
\S\ref{sec:forward_error}. The two failure modes are structurally distinct: the
$N = 30$ failure is a parameter-set deficit shared by all three heuristic
evaluations and is cured in OWNS-S by raising $N$; the $N = 60$ failure is
exclusive to OWNS-R and is incurable, since raising $N$ is what destabilises the
recursive product. Grid-refinement tests confirm OWNS-S~(LS) remains stable for
streamwise resolutions up to $n_x = 3 \times 10^4$ (about 200 points per
wavelength).

\begin{figure}[!htbp]
    \centering
    \includegraphics[width=0.7\textwidth]{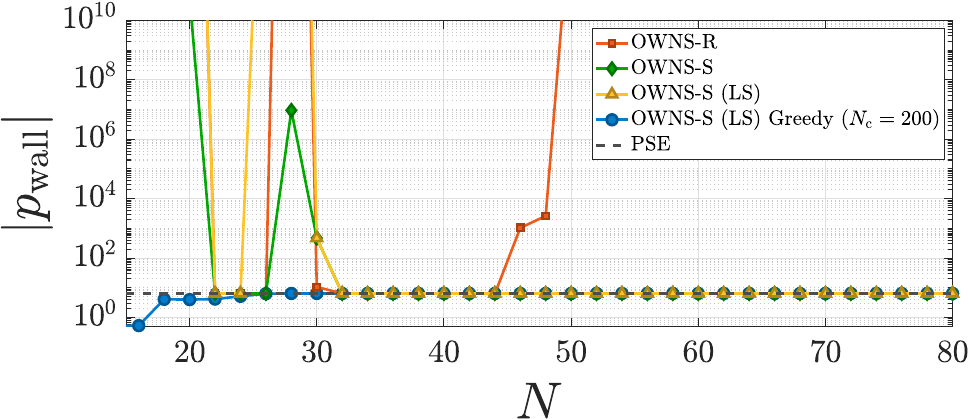}
    \caption{Convergence of $\max|p_\mathrm{wall}|$ at the amplification peak $R_\delta \approx 1050$.
    OWNS-R diverges for $N \le 30$ and $N \ge 46$; OWNS-S and OWNS-S~(LS)
    converge for $N > 30$. Greedy OWNS-S~(LS) ($N_\mathrm{c} = 200$) converges
    smoothly by $N \approx 26$.}
    \label{fig:m4p5:conv}
\end{figure}

The convergence of $\max|p_\mathrm{wall}|$ at the amplification peak
(Fig.~\ref{fig:m4p5:conv}) quantifies these observations. OWNS-R shows a narrow
window of apparent convergence, $30 < N < 46$, bounded below by insufficient
coverage and above by the finite-precision error accumulation of the recursive
product (\S\ref{sec:forward_error}).  This is the two-sided window of
\S\ref{sec:forward_error}, and its upper edge again sits above the per-station
breakdown at $N \ge 36$ (Fig.~\ref{fig:m4p5:e}) for the same reason as in the
incompressible case. OWNS-S and OWNS-S~(LS) stabilise for $N > 30$,
and greedy OWNS-S~(LS) matches PSE for $N \ge 26$, insensitive to the candidate
pool over $N_\mathrm{c} = 100$, $200$, $300$. Greedy selection yields better
parameter sets than the heuristic at the same $N$, even though it acts on a
scalar projection functional and cannot account for discretisation effects
directly.

\subsection{SWiFT Transonic Boundary Layer}\label{sec:swift}

The Swept Wing (SWiFT) configuration~\citep{coppinSWiFTModelDevelopment2025}
is the most demanding test of recursion-parameter selection considered in
this work. It is a fully three-dimensional boundary layer developing over a
wing surface with variable curvature, non-zero pressure gradients, and
transonic flow conditions. The experimental model was tested in the
transonic wind-tunnel of the Aircraft Research Association (ARA), with the
laminar-turbulent transition characterised by
\citet{eppinkTransitionMeasurementsSWiFT2025} using infra-red thermography.
A single dataset is used here, not for physical analysis but as a stress
test of the selection strategy in a flow whose spectrum passes through mixed
subsonic and supersonic states during the numerical march. For a discussion on the exact body-fitted coordinate system see \citet{badcocktcfd2026, ejb2025}.

\begin{figure}[!htbp]
    \centering
    \begin{subfigure}[b]{0.49\textwidth}
        \centering
        \includegraphics[height=4cm]{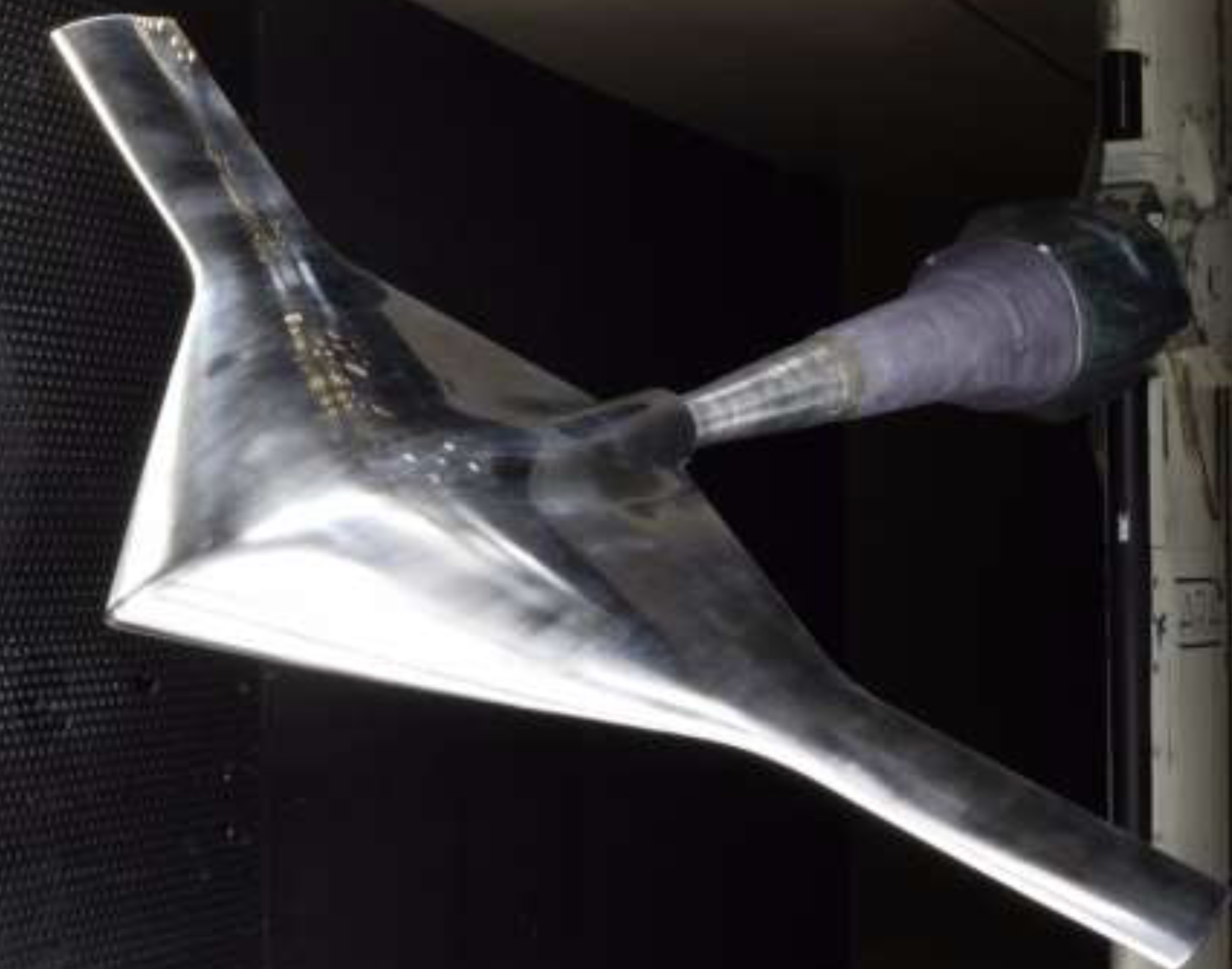}
        \caption{SWiFT model}
        \label{fig:SWiFT:Model}
    \end{subfigure}
    \hfill
    \begin{subfigure}[b]{0.49\textwidth}
        \centering
        \includegraphics[height=4cm]{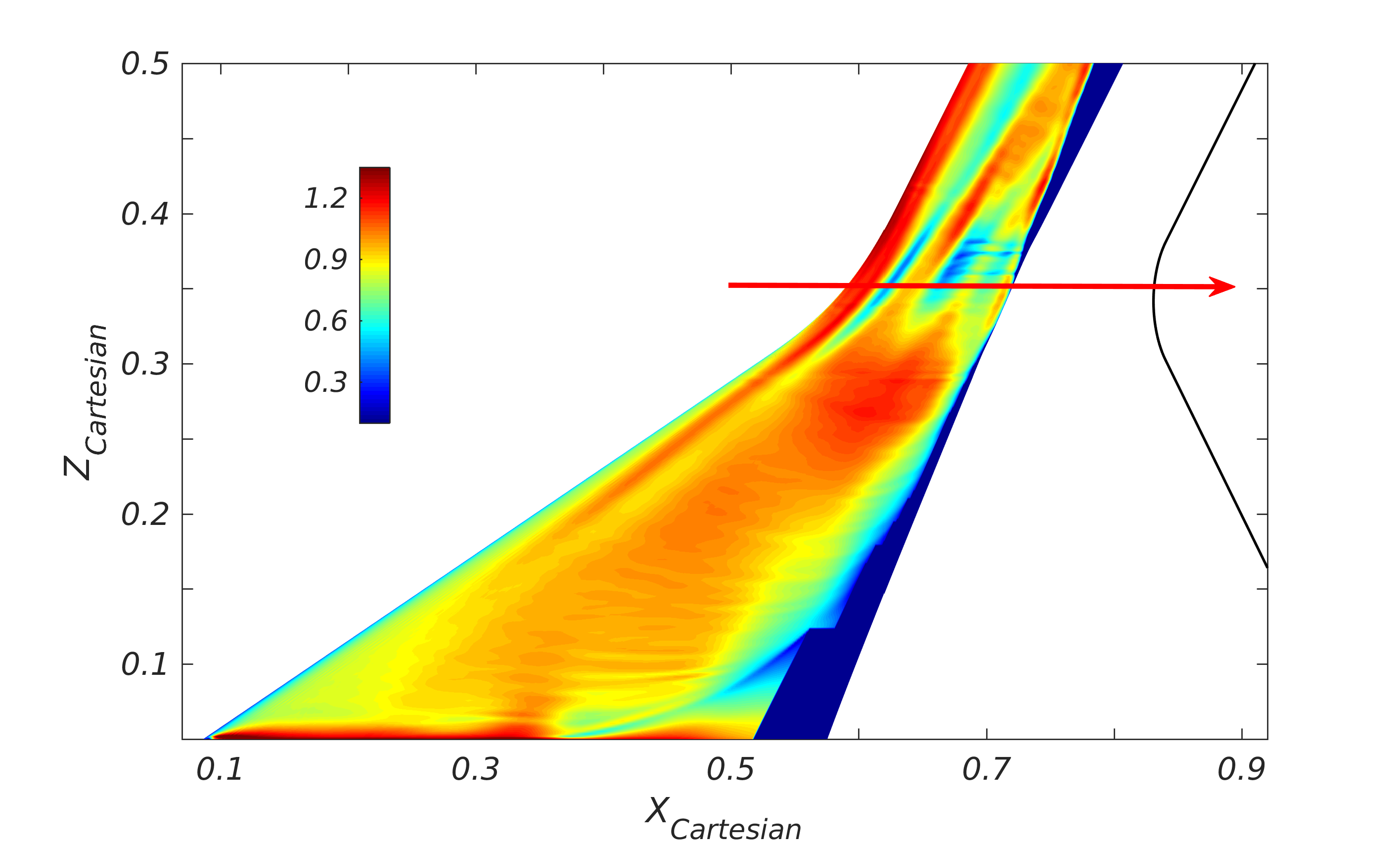}
        \caption{Streamwise skin friction}
        \label{fig:SWiFT:SF}
    \end{subfigure}
    \caption{(a) SWiFT model in the ARA wind-tunnel
    facility~\citep{coppinSWiFTModelDevelopment2025}.
    (b) Streamwise skin-friction coefficient showing the 2.5D slice used for
    the present stability analysis (red arrow).}
    \label{fig:swift:geo2}
\end{figure}

\begin{figure}[!htbp]
    \centering
    \begin{subfigure}[b]{0.49\textwidth}
        \centering
        \includegraphics[width=\textwidth]{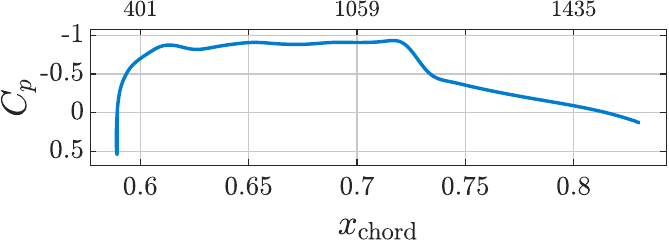}
        \caption{Surface pressure coefficient $C_p$.}
        \label{fig:SWiFT:Cp}
    \end{subfigure}
    \hfill
    \begin{subfigure}[b]{0.49\textwidth}
        \centering
        \includegraphics[width=\textwidth]{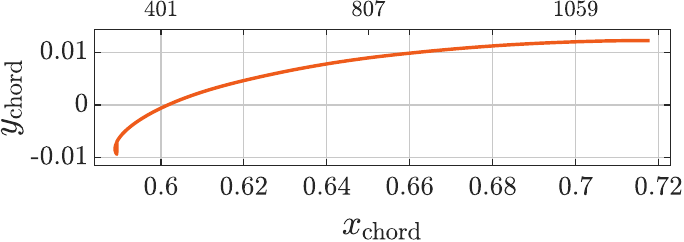}
        \caption{Local camber-line geometry.}
        \label{fig:SWiFT:Geo}
    \end{subfigure}
    \caption{Surface pressure distribution and camber-line geometry of the
    SWiFT model. The chordwise coordinate (bottom axis) is mapped to
    $R_\delta$ (top axis) for comparison with the marching domain.}
    \label{fig:swift:Cp2}
\end{figure}

An infinite-swept ``2.5D'' analysis is used: a sectional slice is extracted
and the steady boundary-layer development assumes spanwise invariance of the
external pressure field. The slice location is marked in
Fig.~\ref{fig:SWiFT:SF}. Along it the flow experiences a mild favourable
pressure gradient followed by a shock-limited, strongly adverse region at
$x_\mathrm{chord} \approx 0.72$, evident in the surface-pressure coefficient
and camber-line geometry of Fig.~\ref{fig:swift:Cp2}. The boundary-layer
solver is shock-limited and fails past $R_\delta = 1139$, which sets the
downstream limit of the marching domain. 

The instability analysis spans $145 \le R_\delta \le 1133$ at a local sweep
angle of $45.12^\circ$, with inlet freestream Mach number $M = 0.5645$, unit
Reynolds number $R_\infty = 6.748 \times 10^{6} \,\mathrm{m}^{-1}$, temperature
$T_\infty = 261.545\,\mathrm{K}$, and disturbance parameters
$f = 84.49 \times 10^{-6}$ and $b = 6.76 \times 10^{-4}$. Note that $\beta$ here denotes the spanwise wavenumber throughout.

For the spatial
marching the wall-normal direction is discretised with Malik stretching
($y_\mathrm{max} = 100$, $y_\mathrm{half} = 20$, $n_y = 321$); the eigenvalue
spectra are computed on a separate grid ($y_\mathrm{max} = 100$,
$y_\mathrm{half} = 7$, $n_y = 100$).
A vanishing wall temperature-fluctuation (isothermal) condition is imposed on
the disturbance.

\begin{figure}[!htbp]
    \centering
    \begin{subfigure}[b]{0.49\textwidth}
        \centering
        \includegraphics[width=\textwidth]{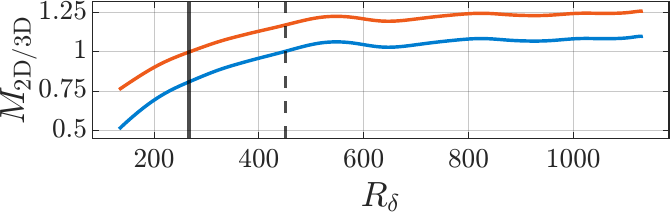}
        \caption{$M_\mathrm{3D}$ (red) and $M_\mathrm{2D}$ (blue).}
        \label{fig:SWiFT:Ma}
    \end{subfigure}
    \hfill
    \begin{subfigure}[b]{0.49\textwidth}
        \centering
        \includegraphics[width=\textwidth]{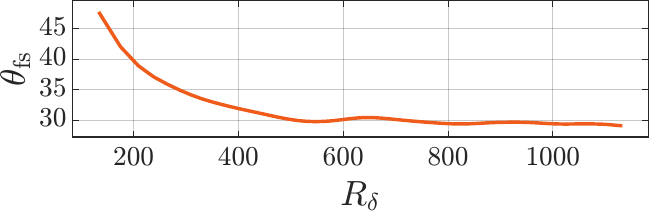}
        \caption{Flow-deflection angle $\theta_\mathrm{fs}$ (deg).}
        \label{fig:SWiFT:Fs}
    \end{subfigure}
    \caption{Streamwise variation of (a) the base-flow and disturbance Mach
    numbers $M_\mathrm{3D}$ and $M_\mathrm{2D}$, and (b) the freestream
    flow-deflection angle $\theta_\mathrm{fs}$. In (a) the solid line marks
    where the base flow becomes supersonic ($M_\mathrm{3D}=1$,
    $R_\delta\approx270$) and the dashed line where the continuous acoustic
    spectrum becomes fully downstream-propagating ($M_\mathrm{2D}=1$,
    $R_\delta\approx450$).}
    \label{fig:swift:MA_FS}
\end{figure}

The local streamwise-edge and total-edge Mach numbers $M_\mathrm{2D}$ and
$M_\mathrm{3D}$ (Eq.~\eqref{eq:mach2d3d}) and the freestream flow-deflection
angle $\theta_\mathrm{fs}$ (Eq.~\eqref{eq:deflection}) are shown in
Fig.~\ref{fig:swift:MA_FS}. Since $M_\mathrm{3D} \ge M_\mathrm{2D}$, the base
flow goes supersonic before the continuous acoustic spectrum: $M_\mathrm{3D}$
crosses unity at $R_\delta \approx 270$, the acoustic pinch-point forms at
$R_\delta \approx 300$, and the acoustic continuous branches become entirely
downstream-propagating ($M_\mathrm{2D} > 1$) at $R_\delta \approx 450$. This
offset between the base-flow and continuous-spectrum sonic transitions,
spanning approximately $180$ units in $R_\delta$, is a defining feature of
the transonic regime and the source of the spectral reorganisation that the
selection strategy must track.

\subsubsection{Spectral Evolution and Transonic Recursion-Parameter Strategy}

\begin{figure}[!htbp]
    \centering
    \begin{subfigure}[b]{0.19\textwidth}
        \centering
        \resizebox{\linewidth}{!}{
            \input{latex/super_trans0}
        }
        \caption{}
        \label{fig:swift:spec_trans0}
    \end{subfigure}
    \begin{subfigure}[b]{0.19\textwidth}
        \centering
        \resizebox{\linewidth}{!}{
            \input{latex/super_trans1}
        }
        \caption{}
        \label{fig:swift:spec_trans1}
    \end{subfigure}
        \begin{subfigure}[b]{0.19\textwidth}
        \centering
        \resizebox{\linewidth}{!}{
            \input{latex/super_trans2}
        }
        \caption{}
        \label{fig:swift:spec_trans2}
    \end{subfigure}
    \begin{subfigure}[b]{0.19\textwidth}
        \centering
        \resizebox{\linewidth}{!}{
            \input{latex/super_trans3}
        }
        \caption{}
        \label{fig:swift:spec_trans3}
    \end{subfigure}
    \begin{subfigure}[b]{0.19\textwidth}
        \centering
        \resizebox{\linewidth}{!}{
            \input{latex/super_trans4}
        }
        \caption{}
        \label{fig:swift:spec_trans4}
    \end{subfigure}

    \caption{Schematic of the spectral reorganisation through the transonic
    transition, in standard $\alpha$-space, at five representative stages:
    (a) subsonic inlet, the acoustic continua separated and all remaining
    modes near the origin; (b) the least-damped upstream discrete mode
    detaches near $\alpha=-i\beta$; (c) the continua meet at
    the pinch-point ($R_\delta\approx300$); (d) $\alpha_\mathrm{mid}$
    migrates outward, dragging the continua along the real axis as further
    discrete modes peel off; (e) supersonic regime, both continua on the
    real axis with the discrete ladder established with real parts approximately $\beta$. Retention/removal parameters are
    overlaid.}
    \label{fig:swift:cartoon}
\end{figure}
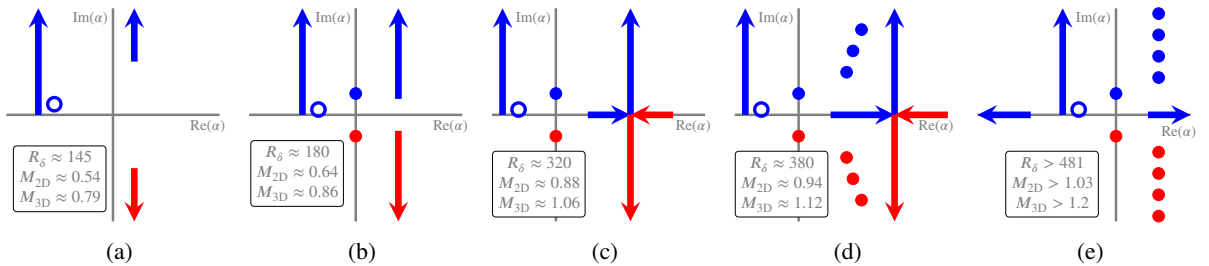

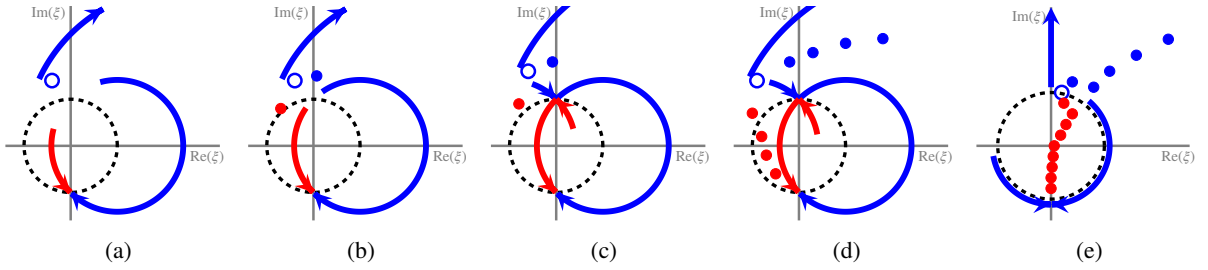
\begin{figure}[!htbp]
    \centering
       \begin{subfigure}[b]{0.19\textwidth}
        \centering
        \resizebox{\linewidth}{!}{
            \input{latex/super_trans0_c}
        }
        \caption{}
        \label{fig:swift:cay_trans0}
    \end{subfigure}
    \begin{subfigure}[b]{0.19\textwidth}
        \centering
        \resizebox{\linewidth}{!}{
            \input{latex/super_trans1_c}
        }
        \caption{}
        \label{fig:swift:cay_trans1}
    \end{subfigure}
        \begin{subfigure}[b]{0.19\textwidth}
        \centering
        \resizebox{\linewidth}{!}{
            \input{latex/super_trans2_c}
        }
        \caption{}
        \label{fig:swift:cay_trans2}
    \end{subfigure}
    \begin{subfigure}[b]{0.19\textwidth}
        \centering
        \resizebox{\linewidth}{!}{
            \input{latex/super_trans3_c}
        }
        \caption{}
        \label{fig:swift:cay_trans3}
    \end{subfigure}
    \begin{subfigure}[b]{0.19\textwidth}
        \centering
        \resizebox{\linewidth}{!}{
            \input{latex/super_trans4_c}
        }
        \caption{}
        \label{fig:swift:cay_trans4}
    \end{subfigure}

    \caption{The same five stages as Fig.~\ref{fig:swift:cartoon}, in the
    Cayley $\xi$-plane.  The unit circle is an estimate of a separating contour and depends on the choice of rotation in the transform; the discrete upstream modes may lie outside it.  The
    excursion of $\alpha_\mathrm{mid}$ through infinity between (d) and (e),
    which takes the acoustic branches off the visible part of the
    $\alpha$-plane, is a bounded and continuous motion here.}
    \label{fig:swift:cartoon_cayley}
\end{figure}

\begin{figure}[!htbp]
    \centering
    \includegraphics[width=0.8\linewidth]{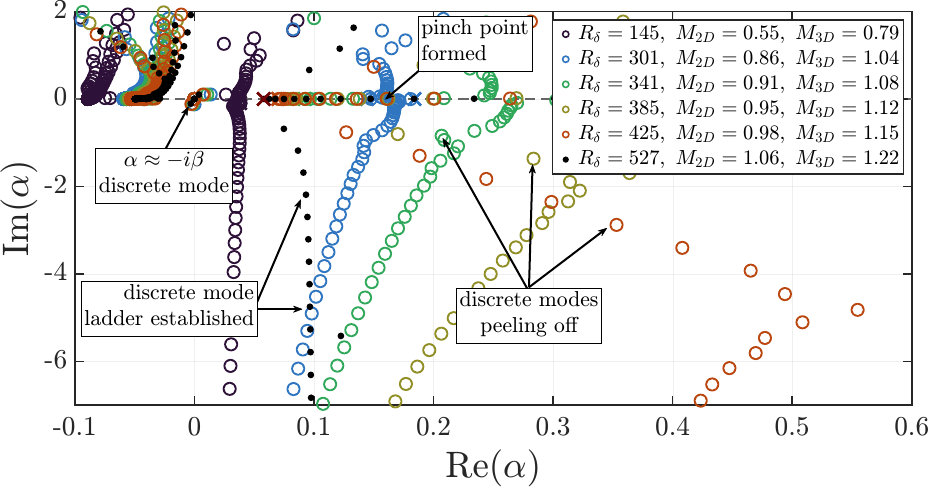}
    \caption{Computed spectra at successive stations
    ($R_\delta = 145$--$527$), standard $\alpha$-space, showing the
    continuous rightward migration of the acoustic branches and the
    progressive peeling of discrete modes as the flow accelerates through the
    transition. The near-origin cluster is where the least-damped upstream
    discrete mode ($\alpha=-i\beta$) lies; the descending
    column of open circles is the discrete-mode ladder.}
    \label{fig:swift:trajectory}
\end{figure}

As the flow accelerates downstream the disturbance spectrum reorganises
continuously, and the retention/removal parameters must follow it.
Figure~\ref{fig:swift:cartoon} sketches the sequence in standard
$\alpha$-space and Fig.~\ref{fig:swift:cartoon_cayley} its Cayley-plane
image; Fig.~\ref{fig:swift:trajectory} shows the same behaviour in the
computed $\alpha$-spectrum, sampled at successive stations. In the fully subsonic
inlet region the two acoustic continuous branches are clearly separated,
with endpoints sitting symmetrically about the midpoint $\alpha_\mathrm{mid}$. The vorticity and entropy branches, the viscous
(Tollmien--Schlichting or crossflow-type) discrete mode, and a pair of trapped discrete acoustic modes (associated with $\alpha^2 + \beta^2=0$) all sit near the origin.

Downstream, the acoustic endpoints migrate toward $\alpha_\mathrm{mid}$ and
the branches meet on the real axis at $R_\delta \approx 300$, forming the
pinch-point. Beyond it $\alpha_\mathrm{mid}$ grows and drags both continua
outward along the real axis: the upstream continuum migrates entirely toward
$+\infty$, while the downstream continuum stretches, its far endpoint
following $\alpha_\mathrm{mid}$ and its near endpoint staying close to the
origin. Discrete acoustic modes peel off the migrating branches as this
proceeds, first singly and then in growing number. The vorticity and entropy
branches and the near-origin discrete modes move only weakly during the
migration, remaining near the origin throughout. When the continuous spectrum
goes supersonic ($M_\mathrm{2D} > 1$), $\alpha_\mathrm{mid}$ passes through
infinity: the real parts of the acoustic branches diverge, and the upstream
continuum re-emerges from $-\infty$ as a second downstream branch, leaving
both continua on the real axis. This continuous migration, and in particular
the excursion of $\alpha_\mathrm{mid}$ through infinity, is what the
recursion-parameter placement must track through the march.

\begin{table}[!htbp]
\centering
\caption{Key streamwise events in the SWiFT transonic march. The local
Reynolds number $R_\delta$ and edge Mach numbers $M_\mathrm{2D}$,
$M_\mathrm{3D}$ are listed at each. The base-flow sonic transition
($M_\mathrm{3D}=1$), the pinch-point, and the continuous-spectrum sonic
transition ($M_\mathrm{2D}=1$) bound the transonic window over which the
spectrum reorganises. These three transitions are sharp events fixed by the
base-flow profile; detachment of discrete modes from the continuous spectrum
is gradual in the discretised system, so the quoted $R_\delta$ marks where
each is first clearly resolved as distinct.}
\label{tab:swift:events}
\begin{tabular}{@{}p{4.5cm}ccc p{7cm}@{}}
\toprule
Event & $R_\delta$ & $M_\mathrm{2D}$ & $M_\mathrm{3D}$ & Spectral consequence \\
\midrule
March inlet
    & $145$ & $0.54$ & $0.79$
    & Acoustic continua separated; all discrete modes near the origin \\[3pt]
Least-damped upstream mode first resolving
    & $\approx 162$ & $\approx0.60$ & $\approx0.82$
    & Upstream discrete acoustic mode separates from the continuum and moves
      near $\alpha = -i\beta$, while the flow is still fully
      subsonic \\[3pt]
Base flow sonic
    & $269$ & $\approx 0.80$ & $1.00$
    & Base flow supersonic while the disturbance remains subsonic; no change
      in the upstream/downstream partition \\[3pt]
Pinch-point
    & $301$ & $0.86$ & $1.04$
    & Acoustic continua meet at $\alpha_\mathrm{mid}$ on the real axis \\[3pt]
Discrete ladder first resolving
    & $\approx 360$ & $\approx 0.93$  & $\approx 1.10$
    & Further modes separate onto $\mathrm{Re}(\alpha)\approx\beta$, quantised
      in $\mathrm{Im}(\alpha)$ \\[3pt]
Continuous spectrum sonic
    & $447$ & $1.00$ & $\approx 1.18$
    & $\alpha_\mathrm{mid}$ passes through $\infty$; branches all
      downstream-propagating, ladder established \\[3pt]
Base-flow solver limit
    & $1139$ & $-$ & $-$
    & Boundary-layer solver shock-limited; instability analysis
      terminated at $R_\delta = 1133$ \\
\bottomrule
\end{tabular}
\end{table}

\paragraph{Upstream discrete modes.}
Beyond the continuous branches, the upstream content that the projection must
remove includes a set of discrete acoustic modes. The least-damped sits on
the imaginary axis near $\alpha = -{i}\beta$; closest to the
origin, it is the upstream constituent most able to corrupt the downstream
march if retained, and is the most vital to remove. A ladder of progressively
more damped discrete acoustic modes descends from it
\citep{gushchinExcitationDevelopmentUnstable1990b}; 
these must additionally be removed to obtain a stable march. The location of
these modes is here identified empirically from the computed spectrum: for the
present disturbance they lie conveniently on the imaginary axis. This is not
general; in the hypersonic case of \S\ref{sec:rec:sup} the corresponding modes
lie further from the origin and shift with frequency. Determining these
locations analytically, from the profile rather than the computed spectrum, is
a natural continuation of this work, left for the future.

\paragraph{Recursion-parameter placement.}
The SWiFT placement combines the two schemes of
Appendix~\ref{app:recursion}. The continuous branches are paired by rotation
about $\alpha_\mathrm{mid}$, following the subsonic placement of
\S\ref{sec:rec:sub}. The upstream discrete modes are removed following the
supersonic treatment of \S\ref{sec:rec:sup}: a branch of removal parameters
$\beta^-$ is placed near $\alpha = -{i}\beta$ and descending the
imaginary axis, covering the least-damped mode and the ladder beneath it, with
the retention partners $\beta^+$ set by complex conjugation. Rotation about
the pinch is unsuitable for these modes, since it would place the retention
parameters into the upstream region that must be removed; conjugation keeps
both near the origin, independent of $\alpha_\mathrm{mid}$, and clear of the
vorticity branch.

\begin{figure}[!htbp]
    \centering
    \begin{subfigure}[b]{0.85\textwidth}
        \centering
        \includegraphics[width=\linewidth]{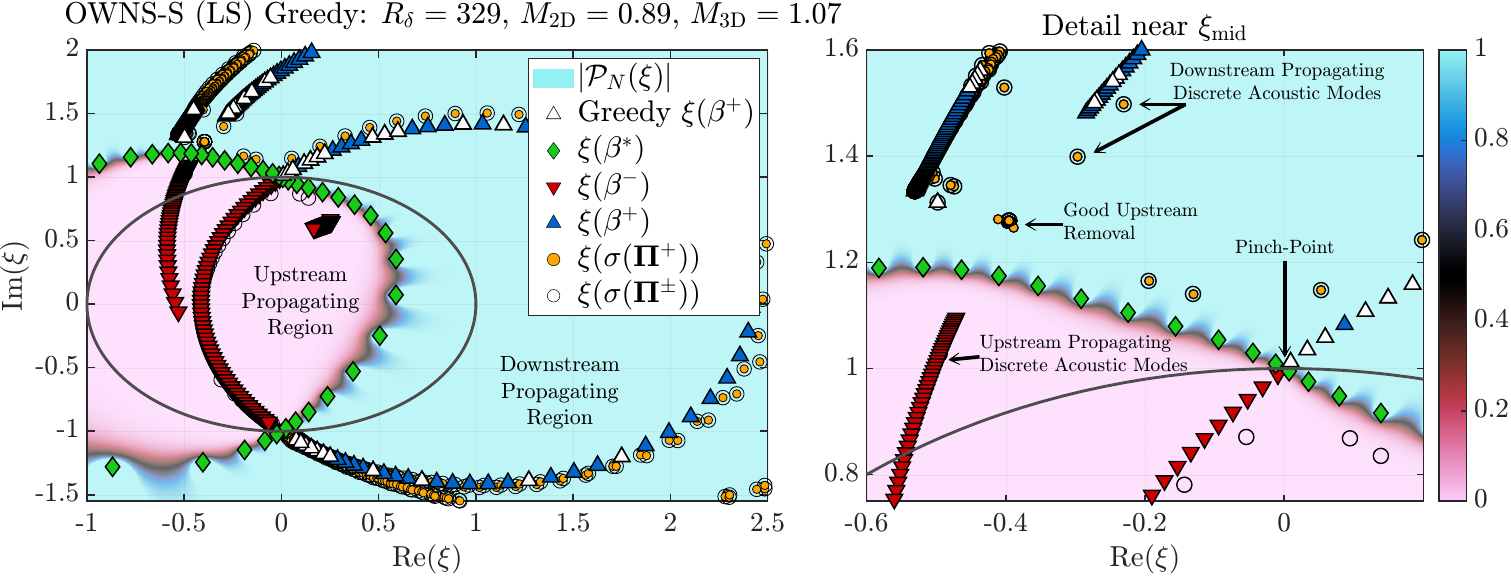}
        \caption{Greedy OWNS-S~(LS) placement.}
        \label{fig:swift:spec2b}
    \end{subfigure}

    \vspace{0.2cm}

    \begin{subfigure}[b]{0.85\textwidth}
        \centering
        \includegraphics[width=\linewidth]{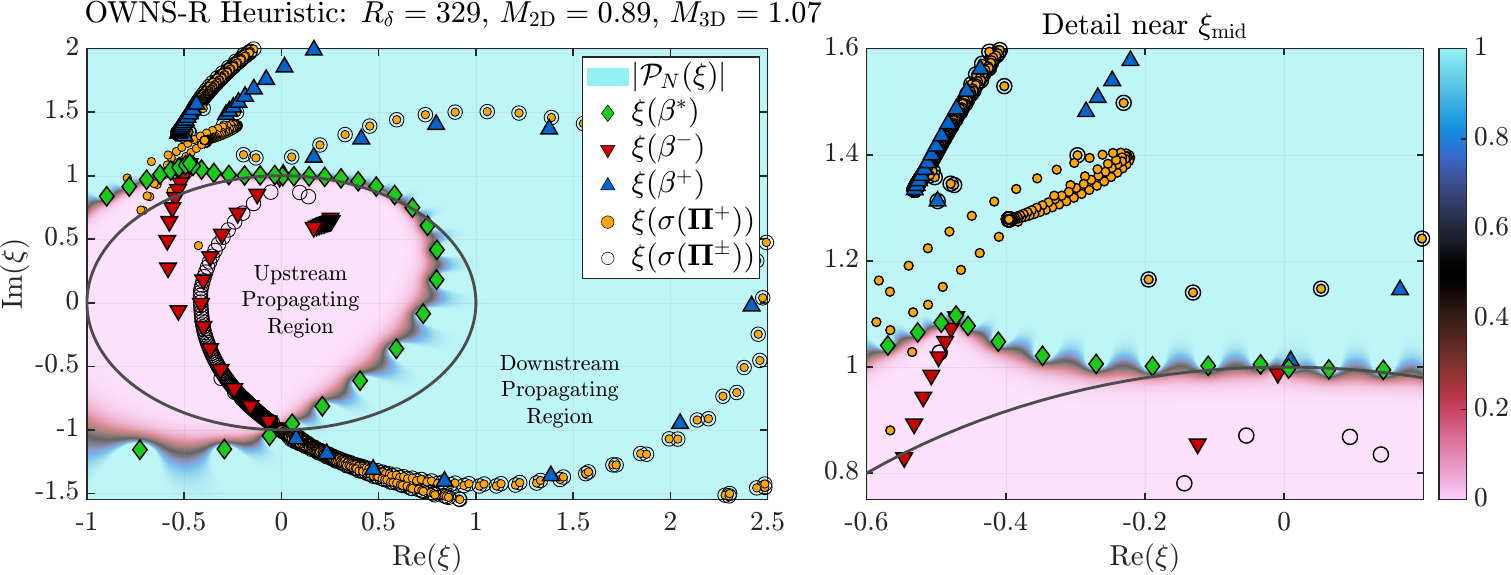}
        \caption{OWNS-R heuristic placement.}
        \label{fig:swift:spec2a}
    \end{subfigure} 
    \caption{Computed spectra and recursion-parameter placement at a mixed
    subsonic station ($R_\delta = 329$), in the Cayley $\xi$-plane, for
    (a) greedy OWNS-S~(LS) and (b) OWNS-R heuristic placement at $N = 40$.
    The freestream is supersonic but the disturbance is firmly subsonic; the
    pinch-point has formed and upstream discrete modes are beginning to
    detach, some mapped outside the unit circle by the transform.}
    \label{fig:swift:spec2}
\end{figure}

\begin{figure}[!htbp]
    \centering
    \begin{subfigure}[b]{0.85\textwidth}
        \centering
        \includegraphics[width=\linewidth]{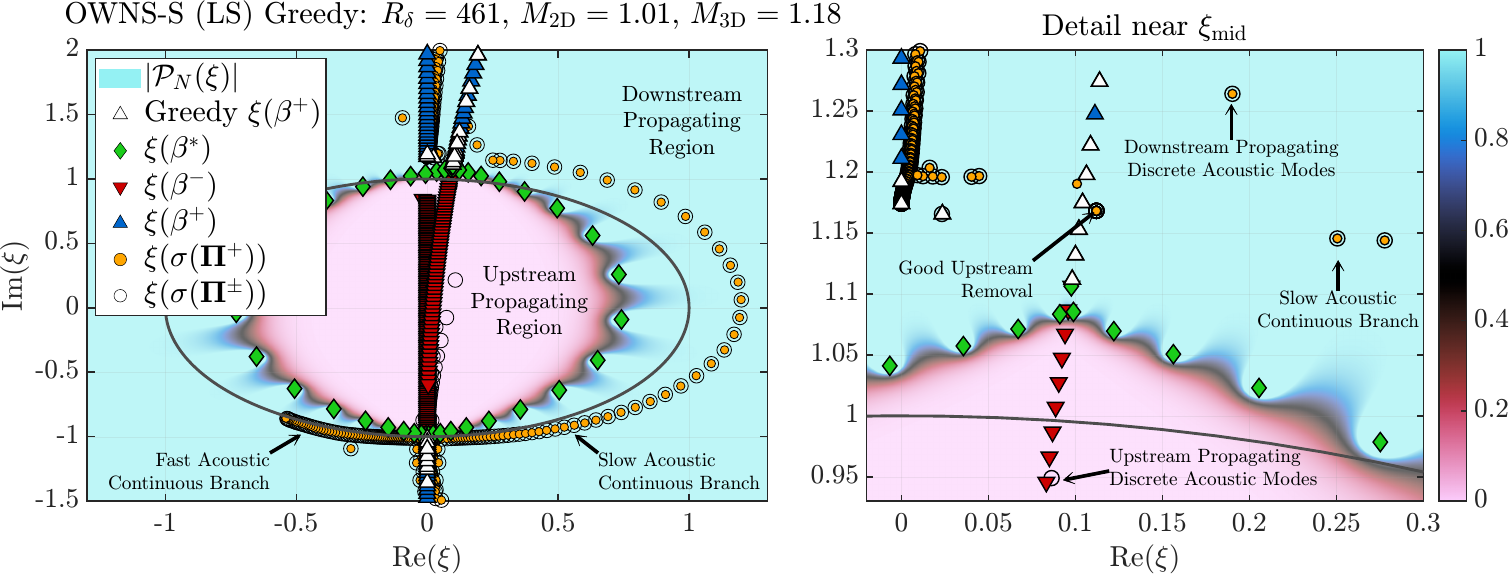}
        \caption{Greedy OWNS-S~(LS) placement.}
        \label{fig:swift:spec3c}
    \end{subfigure}

    \vspace{0.2cm}

    \begin{subfigure}[b]{0.85\textwidth}
        \centering
        \includegraphics[width=\linewidth]{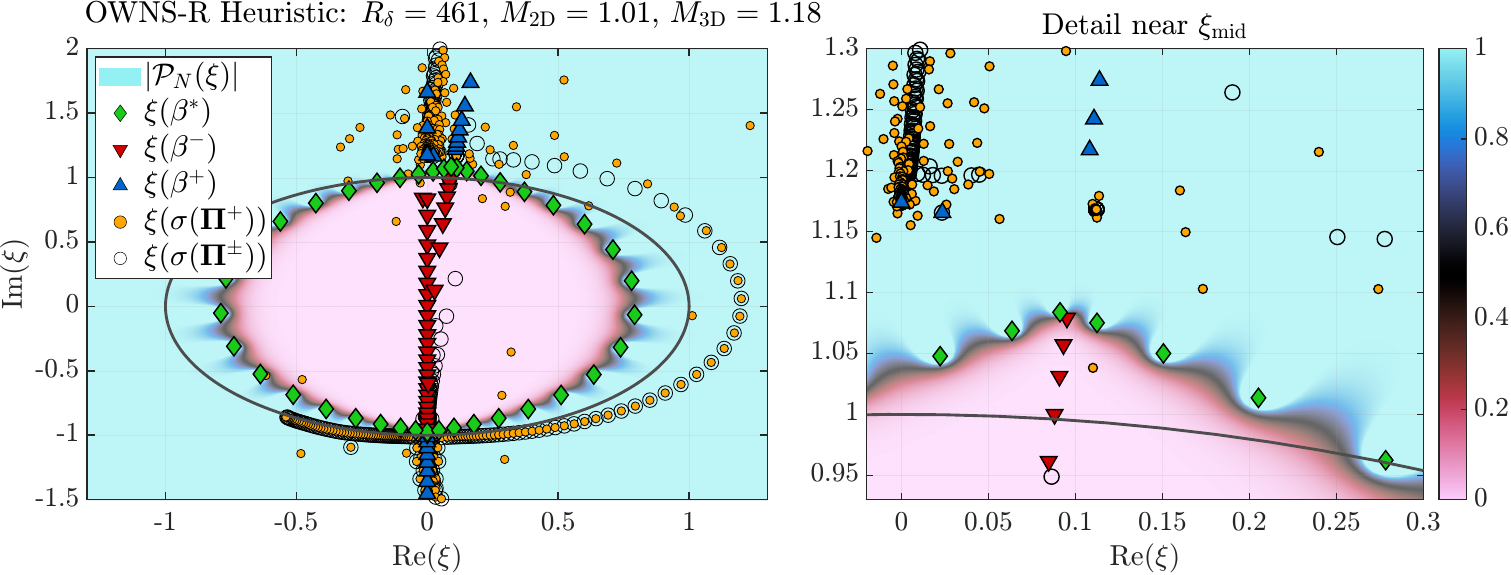}
        \caption{OWNS-R heuristic placement.}
        \label{fig:swift:spec3a}
    \end{subfigure}  
    \caption{Computed spectra and recursion-parameter placement at a
    supersonic station ($R_\delta = 461$), in the Cayley $\xi$-plane, for
    (a) greedy OWNS-S~(LS) and (b) OWNS-R heuristic placement at $N = 40$.
    The disturbance is supersonic: upstream discrete modes are fully
    separated and the acoustic branches map around the unit circle. OWNS-R
    breaks down near $\xi_\mathrm{mid}$, whereas greedy OWNS-S~(LS) achieves
    clean upstream removal.}
    \label{fig:swift:spec3}
\end{figure}

Figures~\ref{fig:swift:spec2} and~\ref{fig:swift:spec3} show the computed
placements in the Cayley plane ($\xi$ denoting the transform of $\alpha$) at
a mixed subsonic station ($R_\delta = 329$) and a supersonic station
($R_\delta = 461$), comparing greedy OWNS-S~(LS) with OWNS-R heuristic
placement at $N = 40$. At the mixed station the freestream is supersonic but
the disturbance is firmly subsonic; the pinch-point has formed and the
upstream discrete modes are detaching, several mapped outside the unit circle
by the transform. At the supersonic station the acoustic continua lie on the
real axis; a small tuning constant $c = 0.01$ further favours downstream
retention, reflecting the strong downstream/upstream imbalance, with
parameters otherwise as in \S\ref{sec:rec:sup}. OWNS-R breaks down near
$\xi_\mathrm{mid}$, whereas greedy OWNS-S~(LS) removes the upstream content
cleanly while retaining the downstream discrete and continuous acoustic modes.

\subsubsection{Spatial Marching}\label{sec:swift_runs}

\begin{figure}[!htbp]
    \centering
    \includegraphics[width=0.65\textwidth]{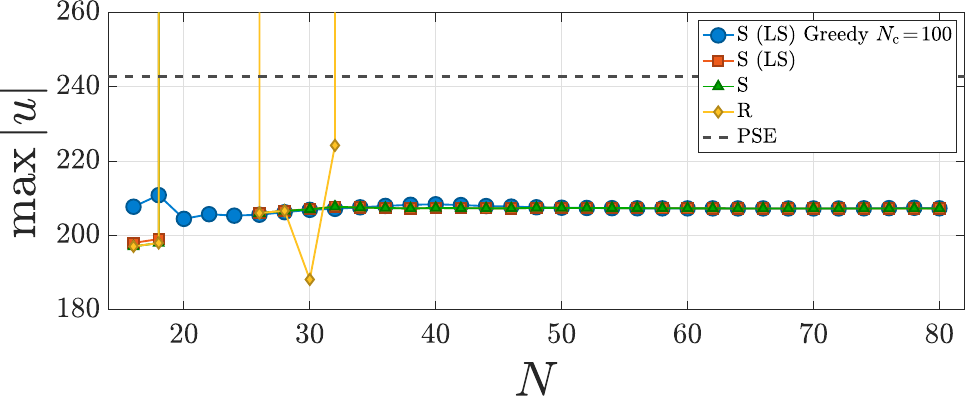}
\caption{
 Convergence of the peak $\max|u|$ with approximation order $N$. OWNS-R converges
 only within a narrow window ($N \approx 26$--$28$) and diverges outside it;
 all OWNS-S variants remain stable, the greedy variant converging earliest.
 Dashed line: PSE reference ($n_x = 150$).
}
    \label{fig:swift:marchconv}
\end{figure}

Figure~\ref{fig:swift:marchconv} shows the convergence of $\max|u|$ at the
amplification peak with approximation order $N$. OWNS-R, heuristic OWNS-S, and
heuristic OWNS-S~(LS) all use the same retention/removal parameters, finely tuned for
this configuration; the heuristic OWNS-S variants converge for $N \ge 26$,
while OWNS-R converges only within a narrow window near $N = 26$--$28$ and
diverges outside it. There is no a priori indicator that a given $N$ lies
inside that window, so OWNS-R is not usable in practice here.

The greedy variant requires no configuration-specific tuning of the individual
placements: selecting from an oversampled version of the same candidate
distribution (shown for $N_\mathrm{c} = 100$, with $N_\mathrm{c} = 200$ in
agreement), it reaches comparable accuracy by $N \approx 20$.  It is not
input-free, and the distinction matters here.  It draws on the same two
spectral inputs as the heuristic, the analytical branch estimates and the
approximate location of the upstream discrete modes (\S\ref{sec:heuristic});
what it removes is the per-configuration choice of how many parameters to
allocate to each group and where within each group to put them, which for this
flow changes continuously along the march.

The greedy result shown here uses the residue weights~\eqref{eqn:standard_weights}.
The least-squares variant was run at every~$N$ and is indistinguishable from
it, which is a further instance of the agreement reported in
\S\ref{sec:weights}: with the poles well placed, refitting the weights to them
changes nothing.
 
\begin{figure}[!htbp]
    \centering
    \begin{subfigure}[b]{0.49\textwidth}
        \centering
        \includegraphics[width=\textwidth]{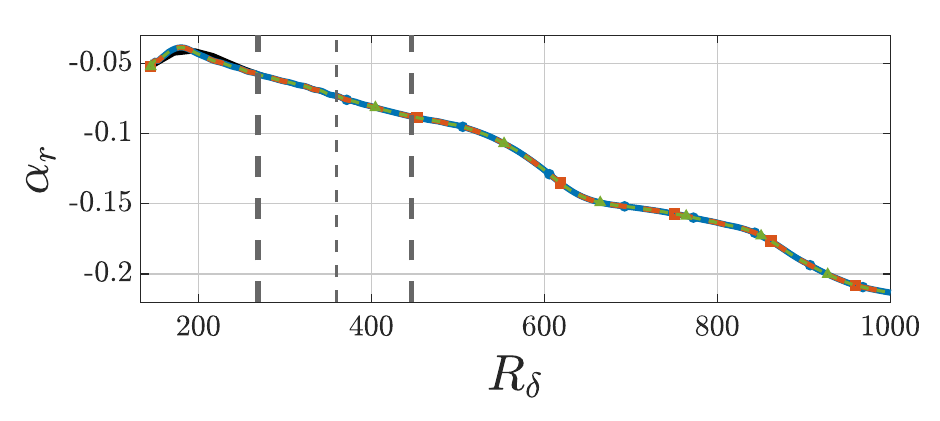}
        \caption{Wavenumber evolution $\mathrm{Re}\left(\alpha\right)$.}
    \end{subfigure}
    \hfill
    \begin{subfigure}[b]{0.49\textwidth}
        \centering
        \includegraphics[width=\textwidth]{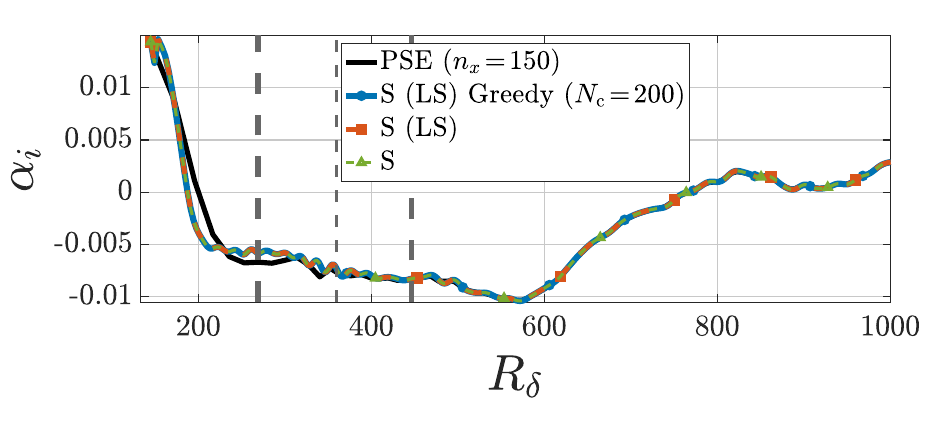}
        \caption{Growth-rate evolution $\mathrm{Im}\left(\alpha\right)$.}
    \end{subfigure}

    \begin{subfigure}[b]{0.49\textwidth}
        \centering
        \includegraphics[width=\textwidth]{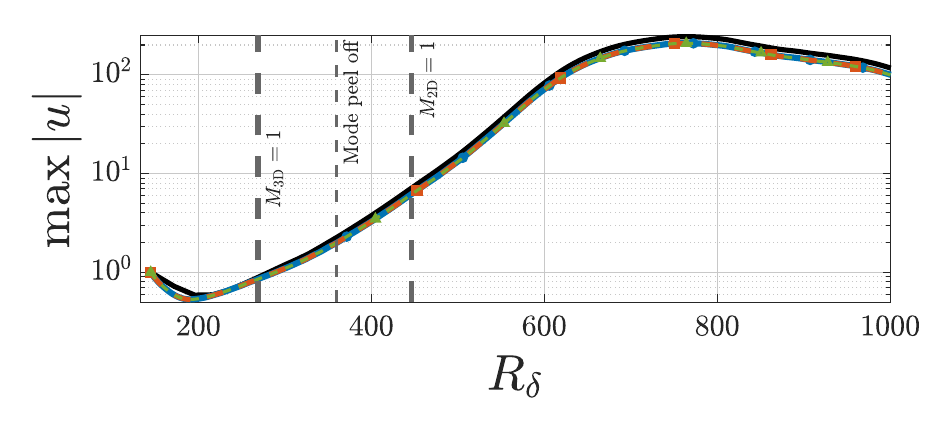}
        \caption{Maximum streamwise velocity amplitude $\max|u|$.}
    \end{subfigure}
    \hfill
    \begin{subfigure}[b]{0.49\textwidth}
        \centering
        \includegraphics[width=\textwidth]{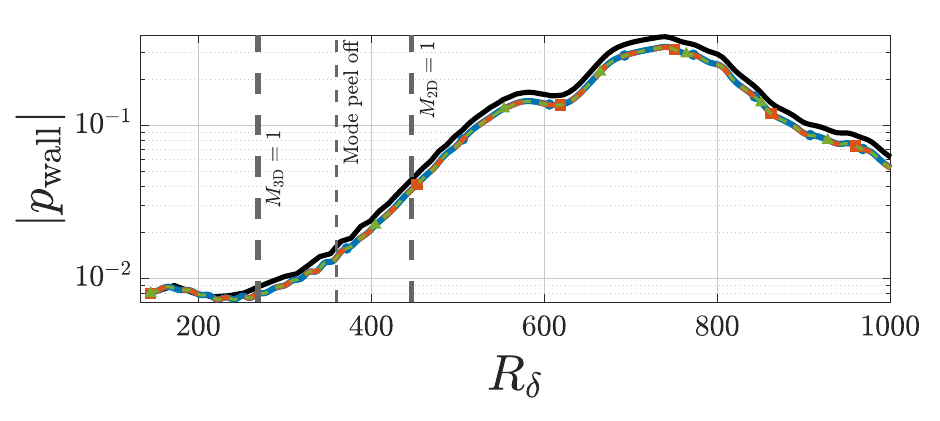}
        \caption{Wall pressure amplitude.}
    \end{subfigure}

    \caption{
 Spatial evolution at $N = 60$: (a) wavenumber $\mathrm{Re}(\alpha)$,
 (b) growth rate $\mathrm{Im}(\alpha)$, (c) maximum streamwise velocity amplitude
 $\max|u|$, (d) wall pressure amplitude. All OWNS-S variants coincide.
 Vertical lines mark the base-flow ($M_\mathrm{3D}=1$) and continuous-spectrum
 ($M_\mathrm{2D}=1$) sonic transitions and the onset of laddering-modes
 peel-off. The PSE reference ($n_x = 150$) lies above the OWNS-S amplitudes;
 the offset is a PSE resolution artefact (text).}
    \label{fig:swift:marching}
\end{figure}

Figure~\ref{fig:swift:marching} shows the converged march at $N = 60$. All
OWNS-S variants coincide, and the wavenumber, growth rate, and both amplitude
measures are given in full, these being, to the authors' knowledge, the first
one-way results reported for this configuration and the first carried
continuously through a subsonic-to-supersonic reorganisation of the
disturbance spectrum. The vertical lines mark where the base flow
($M_\mathrm{3D} = 1$) and the continuous spectrum ($M_\mathrm{2D} = 1$) become
supersonic, and where the laddering modes begin to peel off the acoustic
spectrum. The OWNS-S amplitudes sit below the PSE reference, with wavenumber
and growth-rate offsets near the inlet; the evolution is otherwise similar.
 
\begin{figure}[ht!]
\centering
\begin{subfigure}{0.9\textwidth}
    \includegraphics[width=\linewidth]{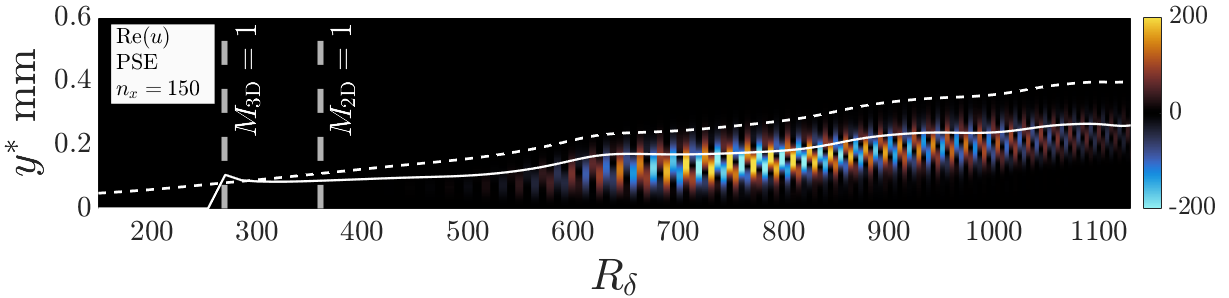}
\end{subfigure}

\begin{subfigure}{0.9\textwidth}
    \includegraphics[width=\linewidth]{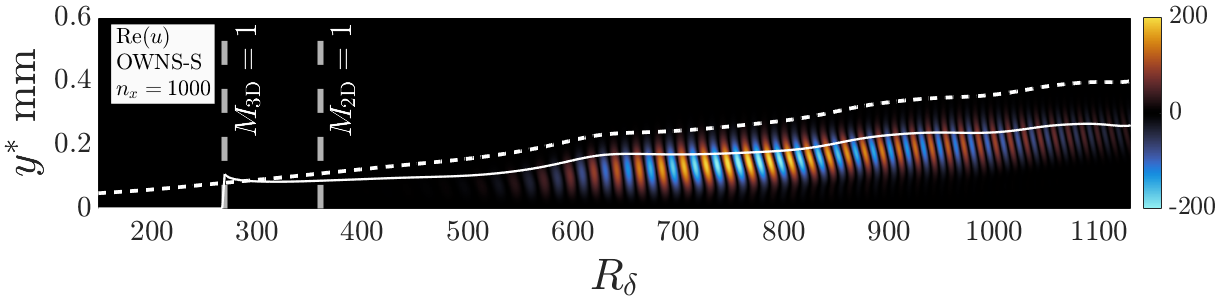}
\end{subfigure}

\caption{
Real part of streamwise velocity $\mathrm{Re}(u)$ for the SWiFT
transonic boundary layer, comparing PSE ($n_x = 150$, top) and greedy
OWNS-S~(LS) ($N=60$, $N_\mathrm{c}=200$, $n_x=1000$, bottom). The PSE solution is under-resolved. }
\label{fig:fa_profile}
\end{figure}

The offset is not a convergence artefact. The OWNS-S result is step-converged,
with no change up to $n_x = 5000$, and running OWNS-S at the coarse PSE step
($n_x = 150$) reproduces the PSE result to plotting accuracy. It reflects a
structural limitation of PSE. PSE factors the solution into a streamwise wave
carrier and a shape function assumed to vary slowly in $x$; here the base flow
and geometry vary rapidly enough that the factored shape function varies too
fast for this assumption, and resolving it demands a fine step. PSE cannot
supply one: upstream-propagating modes remain in its system and destabilise
the march below a minimum step size $\Delta x_\mathrm{min}$
\citep{liNaturePSEApproximation, liSpectralAnalysisParabolized1997}, forcing a
coarse step that underresolves the rapid variation and overpredicts the
amplitude.  A projection-based formulation removes upstream content
explicitly rather than by step-size damping, so no stability-driven lower
bound on $\Delta x$ arises from that mechanism.  We demonstrate this for
OWNS-S in the present configuration, where the solution is step-converged and
resolves the disturbance on scales well below $\Delta x_\mathrm{min}$; we do
not claim it has been established for other one-way variants in the transonic
regime.
The velocity
fields of Fig.~\ref{fig:fa_profile} agree in eigenfunction shape and differ
only in magnitude.
 
\begin{figure}[!htbp]
    \centering

    \begin{subfigure}[b]{0.49\textwidth}
        \centering
        \includegraphics[width=\textwidth]{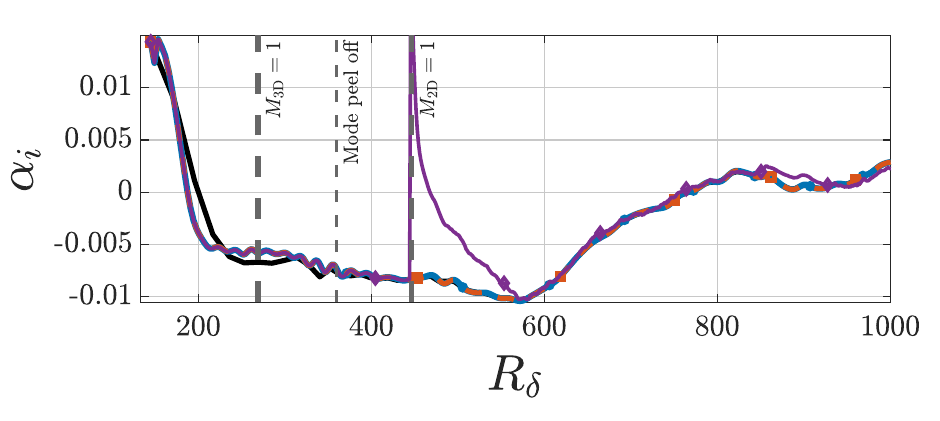}
        \caption{Growth-rate evolution $\mathrm{Im}\left(\alpha\right)$.}
    \end{subfigure}    \hfill
    \begin{subfigure}[b]{0.49\textwidth}
        \centering
        \includegraphics[width=\textwidth]{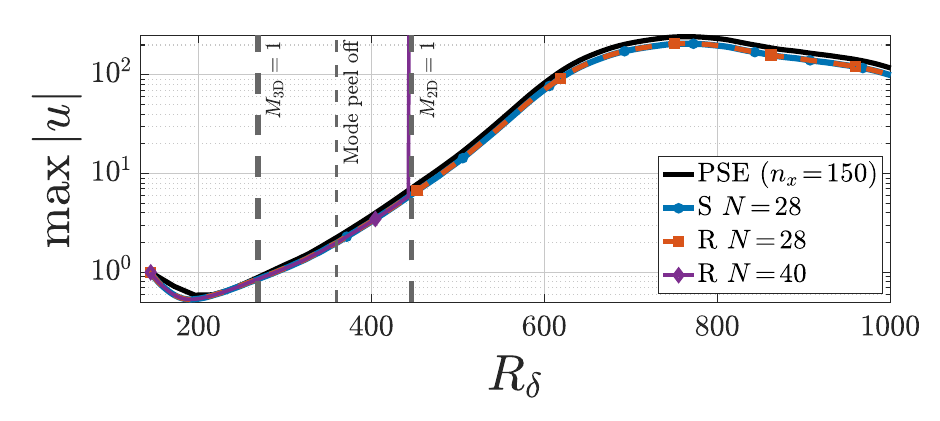}
        \caption{Maximum streamwise velocity amplitude $\max|u|$.}
    \end{subfigure}

    \caption{
 OWNS-R march: (a) growth rate, (b) maximum streamwise velocity amplitude. At $N = 28$ (within the
 stable window) OWNS-R tracks the converged result; at $N = 40$ it diverges at
 $R_\delta \approx 450$, just after the $M_\mathrm{2D}=1$ transition (vertical
 line) where the continuous spectrum goes supersonic.}
    \label{fig:swift:marching2}
\end{figure}

Figure~\ref{fig:swift:marching2} shows the OWNS-R march. At $N = 28$, within
the stable window, it tracks the converged OWNS-S result. At $N = 40$ it
diverges abruptly at $R_\delta \approx 450$, just after the continuous
spectrum goes supersonic ($M_\mathrm{2D} = 1$ at $R_\delta \approx 447$). We
attribute the narrow window to competing bounds across the two regimes: the
subsonic region requires a comparatively high $N$ to remove all of its
upstream modes, setting a lower bound, while the supersonic region tolerates a
smaller $N$ before the recursion becomes unstable, setting an upper bound.
OWNS-R is stable only where the two overlap, near $N = 26$--$28$; at $N = 40$
the supersonic bound is exceeded and the march diverges as soon as the flow
becomes supersonic.
 
\begin{figure}[!htbp]
    \centering

    \begin{subfigure}[b]{0.49\textwidth}
        \centering
        \includegraphics[width=\textwidth]{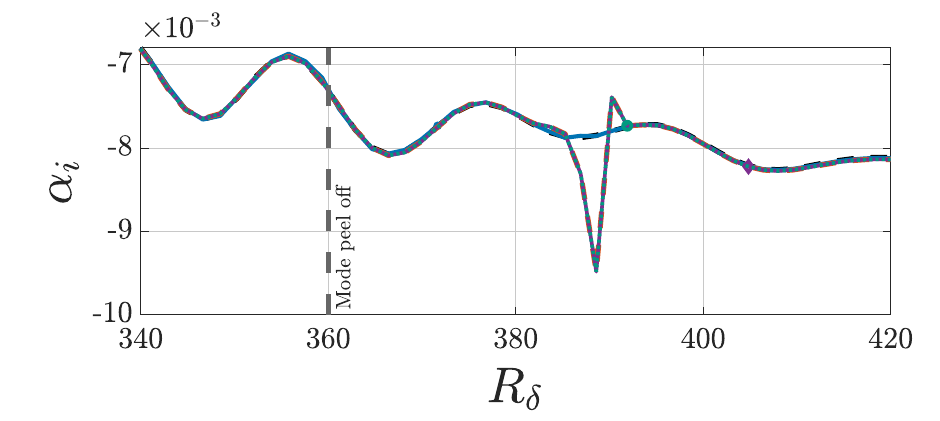}
        \caption{Growth-rate evolution $\mathrm{Im}\left(\alpha\right)$.}
    \end{subfigure}    \hfill
    \begin{subfigure}[b]{0.49\textwidth}
        \centering
        \includegraphics[width=\textwidth]{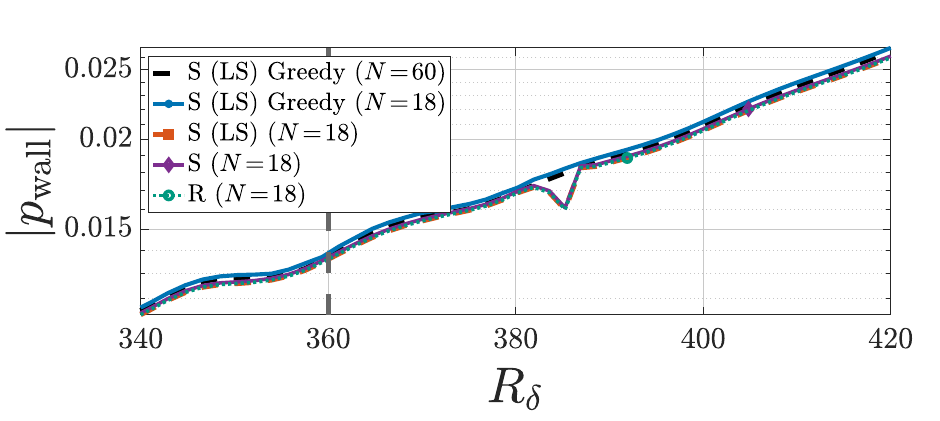}
        \caption{Wall pressure amplitude.}
    \end{subfigure}

    \caption{
 Low-$N$ ($N = 18$) march zoomed on the upstream laddering peel-off: (a) growth rate,
 (b) wall pressure amplitude. Non-greedy OWNS-S, OWNS-S~(LS), and OWNS-R, which
 share the same tuned parameters, carry a localised defect ($R_\delta \approx 385$) just after the upstream modes detach ($R_\delta \approx 360$, vertical line), absent from $\max|u|$ and
 hence acoustic. Greedy OWNS-S~(LS) at the same $N$ is defect-free (about
 $1.5\%$ from the converged high $N$ $\max|u|$).
 }
    \label{fig:swift:marching3}
\end{figure}

At low $N$ the placement quality becomes visible.
Figure~\ref{fig:swift:marching3} shows the $N = 18$ march near the upstream
peel-off. The non-greedy OWNS-S, OWNS-S~(LS), and OWNS-R results, which share
the same tuned parameters, carry a small, localised defect ($R_\delta \approx 385$) at the station
near where the upstream discrete modes detach ($R_\delta \approx 360$). It appears
in the growth rate and wall pressure but not in $\max|u|$, marking it as
acoustic, and it is the signature of the tuned parameters momentarily
mistracking the peeling mode. The greedy variant at the same $N = 18$, having
repositioned its parameters to track the peeling mode, is free of the defect
and follows the converged shape, though its peak amplitude is still about
$1.5\%$ from the $N = 60$ result. The greedy advantage here is therefore not
only earlier amplitude convergence but the removal of a genuine placement
deficiency. That the greedy march is smooth and step-converged certifies clean
upstream removal, since a retained upstream mode amplifies under the downstream
march: the low-$N$ defect is its mild form and the OWNS-R divergence its
severe one.

\subsubsection{Parameter Sweep}\label{sec:swift:sweep}

\begin{figure}[!htbp]
    \centering
    \begin{subfigure}[b]{0.49\textwidth}
        \centering
        \includegraphics[width=\textwidth]{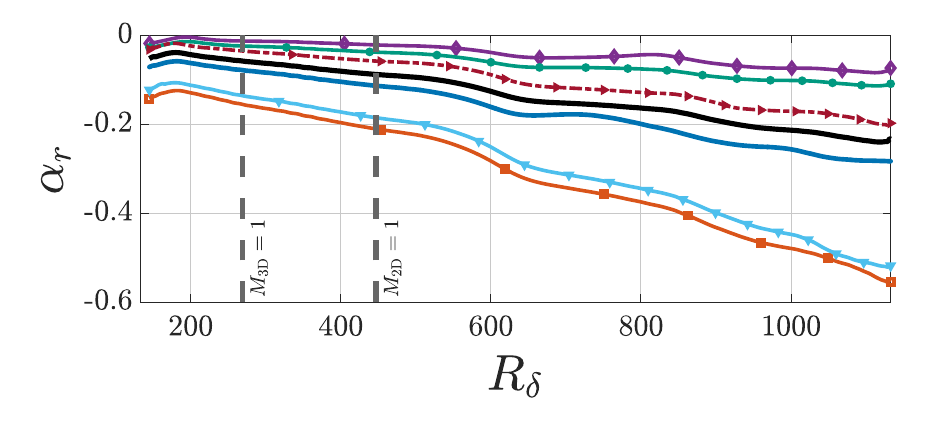}
        \caption{Wavenumber evolution $\mathrm{Re}\left(\alpha\right)$.}
    \end{subfigure}
    \hfill
    \begin{subfigure}[b]{0.49\textwidth}
        \centering
        \includegraphics[width=\textwidth]{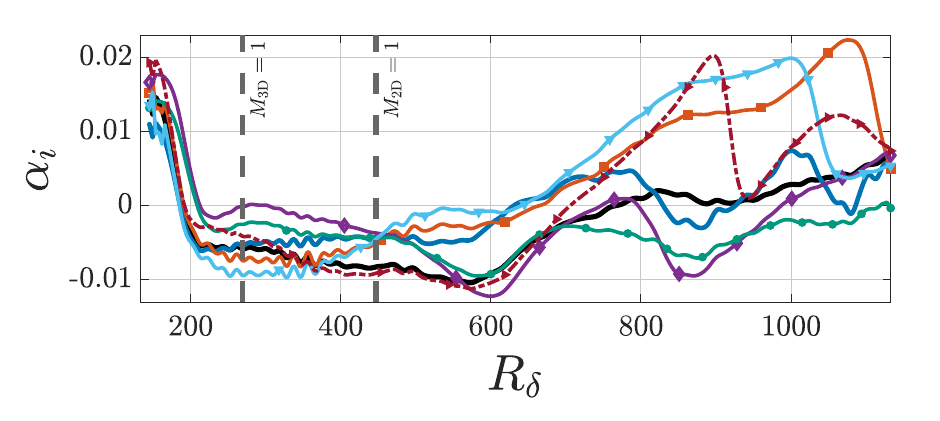}
        \caption{Growth-rate evolution $\mathrm{Im}\left(\alpha\right)$.}
    \end{subfigure}

    \begin{subfigure}[b]{0.49\textwidth}
        \centering
        \includegraphics[width=\textwidth]{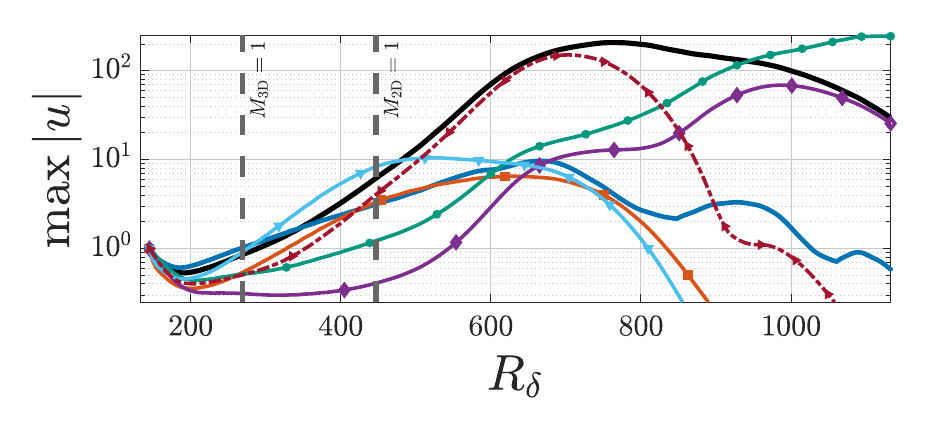}
        \caption{Maximum streamwise velocity amplitude $\max|u|$.}
    \end{subfigure}
    \hfill
    \begin{subfigure}[b]{0.49\textwidth}
        \centering
        \includegraphics[width=\textwidth]{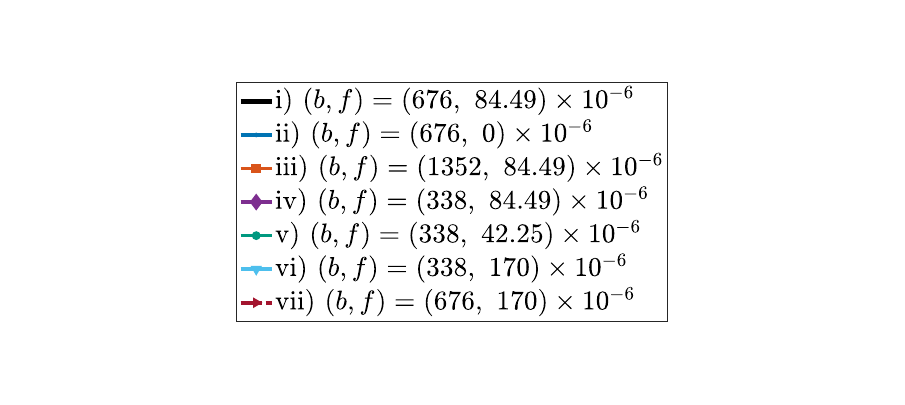}
        \caption{Legend.}
    \end{subfigure}

    \caption{Seven SWiFT marches over spanwise wavenumber and frequency, all
    with greedy OWNS-S~(LS) at $N=50$, $N_\mathrm{c}=200$, and all converged
    in~$N$: (a) wavenumber, (b) growth rate, (c) maximum streamwise velocity
    amplitude, (d) legend.  Case~(i), black, is the disturbance of
    \S\ref{sec:swift_runs}; case~(ii) is stationary ($f=0$).  The placement
    functions of Appendix~\ref{app:recursion} are identical across all seven;
    only $\omega$ and $\beta$ differ, and the parameters they return differ
    accordingly.  Vertical lines mark the base-flow ($M_\mathrm{3D}=1$) and
    continuous-spectrum ($M_\mathrm{2D}=1$) sonic transitions, common to all
    cases.}
    \label{fig:swift:marching_many}
\end{figure}

The results above establish that the greedy selection works for one
disturbance.  A more useful question for a practitioner is whether it continues
to work across the range of disturbances a wing-section analysis requires, since
such an analysis is a sweep over frequency and spanwise wavenumber rather than a
single computation.  We therefore repeated the analysis for seven combinations
of spanwise wavenumber and frequency $(b,\;f)$, including one stationary case
$f=0$.  All use greedy OWNS-S~(LS) at $N=50$ with a candidate pool of
$N_\mathrm{c}=200$, and all seven are converged in~$N$.
Fig.~\ref{fig:swift:marching_many} collects the results; the black curve is the
case analysed above.

The placement functions of Appendix~\ref{app:recursion} are unchanged across all
seven runs; only their arguments differ.  Those functions and their constants,
the pairing~\eqref{eqn:involution}, the order $N=50$ and the pool size
$N_\mathrm{c}=200$ are fixed throughout.  Their arguments are the local base
flow and the disturbance parameters $\omega$ and $\beta$.  Since
$\alpha_\mathrm{mid}$, the branch endpoints (Eqs.~\eqref{eq:cont_vort}
and~\eqref{eq:cont_acoustic}) and the pinch location all move with
$\omega-\beta W$, both the candidate pool and the subset greedy selection
returns from it differ at every station of every run.  The base flow is common
to all seven, so each crosses $M_\mathrm{3D}=1$ and $M_\mathrm{2D}=1$ at the
same $R_\delta$, but each sees a different spectral geometry on either side of
those stations: seven spectral geometries rather than seven repetitions of one.
Nothing was adjusted after inspecting a result, and all seven marches completed
without intervention.

Three remarks on the setting.  First, $N=50$ is significantly more than the order at which
the greedy selection has converged for this configuration
(\S\ref{sec:swift_runs}).  It was chosen for headroom rather than from
necessity, the summation imposing no penalty for the excess beyond the
additional solves.

Second, $N=50$ lies well outside the range $N\approx26$--$28$ within which
OWNS-R was usable for the disturbance of \S\ref{sec:swift_runs}.  That range was
identified retrospectively, by comparison against a converged summation
solution, and for that disturbance alone.

Third, case~(ii) has $f=0$, and is projected on the same footing as the rest.
The placement construction depends on the vorticity/entropy branch endpoint $\alpha_{1-3}(0)$ being distinct from the pinch point $\alpha_\mathrm{mid}$. For $\beta\neq0$ and $\omega=0$, $\omega-\beta W$ vanishes only when $W=0$: this collapses
$\alpha_{1-3}(0)$ onto $\alpha_\mathrm{mid}$.  The disturbance is then a
mean-flow distortion, carrying no propagation direction, and there is no
upstream- and downstream-propagating content for a projection to separate.  

\section{Conclusions}\label{sec:conclusion}

This paper introduced OWNS-Summation (OWNS-S), which evaluates the one-way
Navier--Stokes projection operator as an additive partial-fraction sum rather
than the recursive product of OWNS-Recursion (OWNS-R).  The contribution is
not a different projection, but a different evaluation of the same projection.
In exact arithmetic the two return the same operator, at the same number of
linear solves per station; in floating point they do not. Supplied with identical auxiliary poles from the same polynomial solve, the
summation construction remained accurate while the recursive variant diverged
for sufficiently large $N$.  Holding the poles fixed in this way isolates the
evaluation, and identifies the multiplicative propagation of solve error
through the recursion as the mechanism of the large-$N$ breakdown reported in
previous work.

Two consequences follow.  The approximation order in OWNS-S becomes a genuine
convergence parameter; $N$ may be raised until the computed solution ceases
to change. OWNS-R is limited to a window of usable orders that cannot
be located in advance, since raising $N$ is the operation that destabilises
the product.  
Additionally, in OWNS-S no solve depends on another and so
the $N$ solves
distribute across threads, reducing the per-station critical path from $1+N$
solves to $1+\lceil N/n_\mathrm{t}\rceil$. 

The formulation was validated on incompressible, hypersonic and transonic
boundary layers.  The incompressible case tests separation at the acoustic
pinch-point, where the upstream and downstream spectra approach arbitrarily
closely.  The hypersonic case tests removal of the discrete trapped acoustic
modes.  The transonic SWiFT case combines both, and adds a continuous
reorganisation of the disturbance spectrum from a subsonic to a supersonic
topology.  The march was carried continuously through that reorganisation, to
our knowledge for the first time by a one-way projection method, at a streamwise
step size below the floor imposed on the PSE by
their own stability requirement.  In every case OWNS-S remained accurate at
approximation orders where OWNS-R became unstable.

Alongside the summation, a paired greedy parameter-selection procedure was
developed.  Its candidates are drawn from an oversampled version of the
analytic heuristic placement, so no eigen-decomposition is required at any
station.  Because each candidate is already a retention and removal pair, the
symmetry that places the auxiliary poles on a separating contour is preserved
at every intermediate order.  The selection could equally be applied to
OWNS-R.  It is reported only for OWNS-S, because a placement cannot be shown
to be converged without raising $N$, which is the operation the recursive
evaluation does not survive.

Two directions remain.  The projector is currently constructed from the
first-order pencil $(\mathbf{B},\mathbf{A})$, with the streamwise-viscous
term~$\mathbf{C}$ retained only in the marching discretisation
(\S\ref{sec:pencil}).  Extending the projection to the full quadratic pencil
would provide an exact spectral splitting of the second-order system, but
requires handling the additional parasitic branches of the quadratic
eigenvalue problem.  Separately, the locations of the trapped upstream
acoustic modes are here taken from a computed spectrum at a single station.
A WKB treatment of the region between the wall and the sonic
line~\citep{gushchinExcitationDevelopmentUnstable1990b} may furnish them from
the profile alone, and would remove the one input the parameter selection
currently takes from a computed spectrum.

Alternative rational approximations were also investigated, notably
AAA-based barycentric representations \citep{nakatsukasaAAAAlgorithmRational2018}. These are
effective for constructing the projected operator explicitly, and hence
for parabolising the equations, but the barycentric form does not reduce
to independent resolvent solves acting on the marched state, so no
action-only application of the projector is easily available.
For one-way marching this made them less practical than the implicit-pole
partial-fraction evaluation used here.

In summary, the step that limits the numerical stability of earlier OWNS-R
treatments has been identified and removed. OWNS-S is a more robust
evaluation of the same projection and, with its solves distributed across
threads, brings the per-station critical path to PSE-like levels.

\subsubsection*{Credit authorship contribution statement}
\noindent\textbf{Elliot J. Badcock:} Conceptualisation, Methodology, Software,
 Investigation, Writing -- original draft, review \& editing.\\ 
 \textbf{Shahid Mughal:}
 Supervision, Project administration, Writing -- review \& editing.
 
\subsubsection*{Declaration of competing interest}
The authors declare that they have no known competing financial interests
or personal relationships that could have appeared to influence the work
reported in this paper.
 
\subsubsection*{Data availability}
The SWiFT data set used in this study and the software are not publicly available due to contractual
restrictions. All other data and results can be made available from the
corresponding author on reasonable request.

\subsection*{Declaration of AI assistance}
Claude Code was used to assist with some aspects of the numerical implementation and efficiency. It was also used during the final stages of manuscript preparation to provide grammar and proofreading suggestions. All content generated or edited with the help of generative AI was carefully reviewed and verified by the authors, who assume full responsibility for the accuracy, originality and integrity of the manuscript.

\subsubsection*{Acknowledgements \& Funding}
This work was supported by the United Kingdom Defence Science and
Technology Laboratory (DSTL), under contract DSTLX1000152974. We thank
Dr J. Coppin of the DSTL for interest in this research and arranging
funding support.

\setcounter{equation}{0}
\renewcommand{\theequation}{A.\arabic{equation}}

\appendix


\section{Continuous Spectra Theory}\label{sec:continuous_spectra}

This appendix derives the analytical continuous branches of the pencil and the
two edge Mach numbers that classify them.  It is not background material: the
branch formulae \eqref{eq:cont_vort} and \eqref{eq:cont_acoustic} are what the
heuristic placement of Appendix~\ref{app:recursion} samples to build the
retention and removal parameters, and what the greedy selection of
\S\ref{sec:greedy_ours} oversamples to build its candidate pool, so they are
evaluated at every station of every computation reported here.  The reference
point $\alpha_\mathrm{mid}$ of \S\ref{sec:spec_structure} and the
sonic-transition markers of \S\ref{sec:swift} are likewise defined from them.

The continuous spectrum of the pencil $(\mathbf{B},\mathbf{A})$ is set entirely
by the freestream: the base-flow coefficients are bounded and decay to their
freestream limits. As such, the
continuous branches are identified with the essential spectrum of the freestream
pencil \citep{katoPerturbationTheoryLinear1966} with the boundary layer contributing discrete or isolated modes. Its constant coefficients then permit direct
evaluation by frozen-coefficient analysis, substituting
$\partial_x\rightarrow i\alpha$ and $\partial_y\rightarrow i\eta$
into the linearised operator of Eq.~\eqref{eqn:LHNS}. This yields a dispersion
relation that is a fifth-order polynomial in $\alpha$, parameterised by the real
wall-normal wavenumber $\eta$. Its roots are the five continuous branches. Three
correspond to vorticity and entropy dynamics,
\begin{equation}\label{eq:cont_vort}
    \alpha_{1,2,3}(\eta)
    = \frac{\omega - \beta W}{U}
      + \frac{i\bigl(\beta^2 + \eta^2\bigr)}{U R},
\end{equation}
which are convected by the mean flow and viscously damped, so they are always
downstream-propagating. The remaining two are acoustic,
\begin{equation}\label{eq:cont_acoustic}
    \alpha_{4,5}(\eta)
    = \frac{M^2 U\,(\omega - \beta W) \pm \sqrt{\mu(\eta)}}{M^2 U^2 - T},
\end{equation}
with discriminant
\begin{equation}\label{eq:cont_mu}
    \mu(\eta)
    = T\Bigl[\,(M^2 U^2 - T)\bigl(\beta^2 + \eta^2\bigr)
              + M^2\,(\omega - \beta W)^2\,\Bigr].
\end{equation}
Here $U$, $W$ and $T$ are the freestream streamwise velocity, spanwise velocity
and temperature, $M$ is the freestream Mach number, and $R$ is the Reynolds
number based on the inlet boundary-layer thickness.  


We define the freestream flow-deflection angle as ${\theta}_{\mathrm{fs}}$ given from
\begin{equation}\label{eq:deflection}
    \tan\theta_\mathrm{fs} = \frac{W}{U}.
\end{equation} 

The acoustic branches may propagate downstream or upstream, and the distinction
is governed by the sign of the leading coefficient $M^2U^2-T$ in
\eqref{eq:cont_acoustic}. It is convenient to introduce the streamwise and
total-speed freestream Mach numbers,
\begin{equation}\label{eq:mach2d3d}
    M_{\mathrm{2D}} = \frac{M U}{\sqrt{T}},
    \qquad
    M_{\mathrm{3D}} = \frac{M\sqrt{U^2 + W^2}}{\sqrt{T}},
\end{equation}
so that $M^2U^2 - T = T\,(M_{\mathrm{2D}}^2 - 1)$. The continuous acoustic
structure is controlled by $M_{\mathrm{2D}}$, the marching-direction Mach
number, since only the streamwise velocity enters the leading coefficient. When
$M_{\mathrm{2D}}>1$ the bracket in \eqref{eq:cont_mu} is a sum of non-negative
terms, so $\mu(\eta)>0$ for all $\eta$, both acoustic branches are real, and the
disturbance is \emph{supersonic}: the acoustic continuum is strictly
downstream-propagating (Fig.~\ref{fig:spec_cartoon1b}). When $M_{\mathrm{2D}}<1$
the leading coefficient is negative, $\mu$ decreases with $\eta^2$ and changes
sign at the cut-off
\begin{equation}\label{eq:pinch}
    \eta_c^2
    = \frac{M^2\,(\omega - \beta W)^2}{T\,(1 - M_{\mathrm{2D}}^2)} - \beta^2,
\end{equation}
when this is positive. At $\eta=\pm\eta_c$ the two acoustic branches coalesce, a
pinch point in the sense of
\citet{briggsElectronStreamInteractionPlasmas1964a}, and for $|\eta|>\eta_c$
they are complex, producing the upstream-propagating acoustic continuum
characteristic of a \emph{subsonic} disturbance
(Fig.~\ref{fig:spec_cartoon1a}).

The crossflow velocity $W$ enters \eqref{eq:cont_mu} only through the term $\omega-\beta W$, so it sets the cut-off location \eqref{eq:pinch} but not
the regime. The regime is fixed by $M_{\mathrm{2D}}$ alone, whereas the base
flow itself is locally supersonic when $M_{\mathrm{3D}}>1$. The two coincide
without crossflow and separate otherwise, so a station may be base-flow
supersonic, $M_{\mathrm{3D}}>1$, while its acoustic continuum remains subsonic,
$M_{\mathrm{2D}}<1$. The transonic SWiFT boundary layer passes through exactly
this regime, and $M_{\mathrm{2D}}$ and $M_{\mathrm{3D}}$ are compared along the
march in Fig.~\ref{fig:swift:MA_FS}.

Alongside the continuous branches, the pencil admits discrete eigenvalues
associated with boundary-layer instability and acoustic trapping. Subsonic
disturbances support Tollmien--Schlichting and crossflow modes; supersonic
disturbances support fast and slow acoustic modes
\citep{fedorovHighSpeedBoundaryLayerInstability2011}; and transonic and
supersonic flows may additionally exhibit upstream-trapped modes
\citep{chuvakhovSpontaneousRadiationSound2016}. These are set by the full
wall-normal profile and, being contributed by the boundary layer rather than
the freestream pencil, lie off the continuous branches.

\setcounter{equation}{0}
\renewcommand{\theequation}{B.\arabic{equation}}
\section{Cayley Transformation}\label{app:Cayley}

The Cayley transformation compactifies the spectrum of
$(\mathbf{B},\mathbf{A})$ onto a bounded domain, providing a convenient
representation for visualising the upstream--downstream separation and for
plotting the scalar filter $|\mathcal{P}^+_N|$.  Although the discretised
spectrum is bounded, the corresponding continuous branches of the
frozen-coefficient operator are unbounded and may have widely separated
endpoints, particularly in transonic and supersonic regimes.

The transformation is implemented through three elementary steps.

\paragraph{Step 1: Centre the pinch-point.}
A shift
\begin{equation}
    t_1:\; (\mathbf{B},\mathbf{A}) \mapsto
    (\mathbf{B}-i\alpha_\mathrm{mid}\mathbf{A},\,\mathbf{A})
\end{equation}
places the spectral reference point at the origin.  Here
\begin{equation}\label{eqn:mid_app}
    \alpha_{\mathrm{mid}} =
    \begin{cases}
        \tfrac{1}{2}\bigl[\alpha_{4}(0) + \alpha_{5}(0)\bigr],
            & \text{subsonic disturbances},\\[4pt]
        \alpha_{1,2,3}(0) - \delta i,
            & \text{supersonic disturbances},
    \end{cases}
\end{equation}
where $\delta \sim 10^{-2}$.  For subsonic spectra this selects the midpoint of
the acoustic pair, coinciding with the pinch-point when one exists.  For
supersonic spectra we use the endpoint of the vorticity and entropy branches,
and the small imaginary shift ensures that the continuous branches lie outside
the unit circle under the subsequent map.

\paragraph{Step 2: Optional rotation.}
In subsonic flows the acoustic branches meet at $\alpha_\mathrm{mid}$ and lie
on the real axis; a rotation
\begin{equation}
    t_2:\; (\mathbf{B}',\mathbf{A}) \mapsto
    (e^{i\theta}\mathbf{B}',\,\mathbf{A}),
    \qquad
    \theta=\mathrm{sign}\bigl(\mathrm{Re}\,[\alpha_{1\text{--}3}(0)-\alpha_\mathrm{mid}]\bigr)\,\frac{\pi}{4},
\end{equation}
lifts the oscillatory branches off the real line to allow the Cayley map to act
cleanly.  The sign selects the direction of rotation so that the vorticity and
entropy branches move away from, rather than across, the image of the real
axis.  In supersonic flows the geometry is already favourable, so $t_2$ is the
identity.

\paragraph{Step 3: Cayley map.}
The bilinear transformation
\begin{equation}
    t_3:\; (\mathbf{B}'',\mathbf{A}) \mapsto
    (\mathbf{B}''-\mathbf{A},\,\mathbf{B}''+\mathbf{A})
\end{equation}
maps the real axis of the shifted/rotated frame to the unit circle: the upper
half-plane maps outside the circle and the lower half-plane inside.

Composing the three steps gives the compactified pencil
\begin{equation}
    \mathbf{\Lambda}
    =(\mathbf{B}_\mathrm{c},\mathbf{A}_\mathrm{c})
    =(t_3\circ t_2\circ t_1)(\mathbf{B},\mathbf{A}),
\end{equation}
with the corresponding eigenvalue mapping
\begin{equation}\label{eqn:cayley_map}
    \xi(\alpha)
    =\frac{e^{i\theta}(\alpha-\alpha_\mathrm{mid})+i}
           {\,i\,e^{i\theta}(\alpha-\alpha_\mathrm{mid})+1}.
\end{equation}
Because the transformation is eigenvector-preserving, the projectors $\mathbf{P}^\pm$ constructed in
either $\alpha$ or $\xi$ coordinates are identical.

\subsection{Auxiliary Poles Under the Cayley Map}\label{sec:cayley_poles}

The retention/removal parameters and auxiliary poles transform pointwise as
$\gamma^\pm_k = \xi(\beta^\pm_k)$ and $\gamma^*_k = \xi(\beta^*_k)$.  It is
natural to ask what form the scalar filter takes in the new variables.  A
M\"obius map does not preserve differences, but it transforms them in a
simple way: under the map~\eqref{eqn:cayley_map}, any two points $\alpha$
and $\beta$ satisfy
\begin{equation}\label{eqn:cayley_factor}
    \alpha - \beta
    = \frac{2\,e^{-i\theta}\,(\xi(\alpha) - \xi(\beta))}
           {(1 - i\xi(\alpha))(1 - i\xi(\beta))}.
\end{equation}
Applying this to the numerator and denominator of a single factor of the
product form, the terms involving the evaluation point~$\xi$ cancel, while
the terms involving the parameters do not.  Each factor therefore transforms
into its Cayley counterpart multiplied by a constant, and collecting the
constants over all $N$ factors gives the filter in the compactified
variables,
\begin{equation}\label{eqn:cayley_K}
    \mathcal{P}^+_N(\xi)
    = \frac{K_N}{c+1}\prod_{k=1}^{N}
      \frac{\xi - \gamma^-_k}{\xi - \gamma^*_k},
    \qquad
    K_N = \prod_{k=1}^{N}\frac{1 - i\gamma^*_k}{1 - i\gamma^-_k}.
\end{equation}
The product itself has the same form as in standard space; only the
prefactor changes, from $1/(c+1)$ to $K_N/(c+1)$, with $K_N$ fixed by the
mapped parameters alone.

The point $\xi \to \infty$ deserves comment.  Under~\eqref{eqn:cayley_map}
it is the image of the finite point
$\alpha = \alpha_\mathrm{mid} + i\,e^{-i\theta}$, lying in the downstream
region. Each factor of the product in~\eqref{eqn:cayley_K} tends
to unity there, so the monic product, taken without its prefactor, is the
filter rescaled to equal one at this particular downstream point.  This is
the form used for plotting $|\mathcal{P}^+_N|$ in the $\xi$-plane.  Because
a converged filter is close to one throughout the downstream region, the
rescaling amounts to a constant of the same size as the filter's own
approximation error, and the plotted contours are insensitive to it.  Where
a definite normalisation is needed, for instance in comparing filter values
quantitatively, it is fixed by requiring unit mean response over the
retention parameters.

The opposite exchange is the single point $\xi = -i$, the image of
$\alpha \to \infty$, at which the filter takes its asymptotic value
$w_0 = 1/(c+1)$.

Finally, two small consequences of the constant.  The parameter $c$
influences the auxiliary-pole locations through
Eq.~\eqref{eqn:poly_general} and enters the transformed filter through
$K_N/(c+1)$, so it is present in Cayley space even though the monic product
does not display it.  The scalar projectors in the two representations are
\begin{align}
    \text{Standard space:}\quad
    \mathcal{P}^+_N(\alpha) &= \frac{1}{c+1}\prod_{k=1}^{N}
        \frac{\alpha - \beta^-_k}{\alpha - \beta^*_k},
    \label{eqn:proj_alpha}\\[6pt]
    \text{Cayley space:}\quad
    \mathcal{P}^+_N(\xi) &= \frac{K_N}{c+1}\prod_{k=1}^{N}
        \frac{\xi - \gamma^-_k}{\xi - \gamma^*_k},
    \label{eqn:proj_xi}
\end{align}
and the complementary relation $\mathcal{P}^-_N = 1 - \mathcal{P}^+_N$
holds in both representations when the prefactors are retained as written.

\setcounter{equation}{0}
\renewcommand{\theequation}{C.\arabic{equation}}
\section{Retention and Removal-Parameter Placement}\label{app:recursion}

The retention and removal parameters $\beta^\pm_k$ introduced in
\S\ref{sec:implicit} must be placed so that the approximate projector
$\mathbf{P}^+_N$ faithfully separates the downstream and upstream spectral
subspaces.  This appendix provides the explicit placement formulae for the
subsonic and supersonic configurations studied in the main text; the transonic
SWiFT case, whose spectral geometry evolves during the numerical march, is discussed in
\S\ref{sec:swift}.

The branches $\alpha_{1\text{--}3}$ represent vorticity/entropy modes;
$\alpha_{4,5}$ are acoustic.  In subsonic flows, $\alpha_{1\text{--}4}$
propagate downstream and $\alpha_5$ propagates upstream; in supersonic flows all
five propagate
downstream.
Three locations in the $\alpha$-plane organise the placements below: the
vorticity and entropy branch endpoint $\alpha_{1,2,3}(0)$, the two acoustic
branch endpoints $\alpha_{4,5}(0)$, and their midpoint, which for subsonic
disturbances is the reference point $\alpha_\mathrm{mid}$ of
Eq.~\eqref{eqn:mid_app}.  The wall-normal grid stretching produces a
non-uniform sampling of the branches: eigenvalues cluster near the endpoints
(small $|\eta|$) and become sparse further along each branch (large $|\eta|$).
The parameter distributions below follow that clustering.

Throughout this appendix $y_\mathrm{Malik}(\eta;h,L)$ denotes the stretching
function of Eq.~\eqref{eqn:malik_grid} with $a_\mathrm{M}$ and $b_\mathrm{M}$
obtained from the half-point $h$ and the extent $L$ by Eq.~\eqref{eqn:malik_ab},
exactly as for the wall-normal grid but with constants chosen independently of
it.  The argument $\eta$ runs over $[0,1]$ and the function over $[0,L]$, with
half the interval mapped into $[0,h]$; small $h/L$ therefore gives strong
clustering near the origin, which is where the eigenvalues of the branches
themselves accumulate.

The constants quoted below are those used in the computations reported here.
They are tied to the present discretisation and to the branch extents it
produces, and are not intended as optimal values; the placements are
insensitive to them over wide ranges, which is the property the greedy
oversampling of \S\ref{sec:greedy_ours} exploits.

\subsection{Subsonic Placement}\label{sec:rec:sub}

For subsonic disturbances, the dominant geometric feature of the spectrum is the
acoustic pinch-point at $\alpha_{\mathrm{mid}}$.  Eigenvalues immediately above
this point on the imaginary axis correspond to downstream-propagating acoustic
modes, whereas those immediately below correspond to upstream-propagating modes.
Accurate projection therefore requires that the $\beta_k^*$-contour produced by
the retention/removal parameters cuts cleanly through $\alpha_{\mathrm{mid}}$.  This
motivates three design conditions for the retention/removal parameters $\beta^\pm$:
(i)~$\beta_k^{-}$ is a $\pi$-rotation of $\beta_k^{+}$ about
$\alpha_{\mathrm{mid}}$, ensuring exact symmetry;
(ii)~$c=1$, so that neither side of the contour is preferentially weighted;
(iii)~clustering $\beta_k^{\pm}$ near $\alpha_{\mathrm{mid}}$, which provides
finer control of the $\beta^*$ locus where downstream--upstream separation is
most sensitive.

The $N$ parameters are divided into three groups: a single point placed at a
potential downstream propagating discrete eigenvalue $\alpha_0\in\mathbb C$, typically the discrete
eigenvalue of interest; $N_\mathrm{a}$ points along the acoustic branches; and
$N_\mathrm{v}$ points along the vorticity and entropy branches, with
\begin{equation}
    N_{\mathrm{v}} = \Big\lceil \tfrac{N}{3} \Big\rceil,
    \qquad
    N_{\mathrm{a}} = N - N_{\mathrm{v}} - 1 .
\end{equation}

The retention parameters are then
\begin{align}
\beta^{+}_k &=
\begin{cases}
\alpha_{0},
& k = 1, \\[3pt]
\alpha_\mathrm{mid} + p + i\,
    y_{\mathrm{Malik}}
    \!\left( \dfrac{k-2}{N_{\mathrm{a}}-1};\, h_\mathrm{a},\, L_\mathrm{a} \right),
& k = 2,\dots,N_{\mathrm{a}}+1, \\[7pt]
\alpha_{1,2,3}(0) + i\,s\,|\alpha_{0}|\,y_{\mathrm{Malik}}
    \!\left( \dfrac{k-N_{\mathrm{a}}-2}{N_{\mathrm{v}}-1};\, h_\mathrm{v},\, L_\mathrm{v} \right),
& k = N_{\mathrm{a}}+2,\dots,N,
\end{cases}
\label{eq:recursion_parameters_plus}
\end{align}
so that each group is distributed over $[0,1]$ in its own index.  The acoustic
group ascends the imaginary direction from the pinch-point, offset by a small
complex shift $p$ that keeps the two groups from meeting there; the
vorticity/entropy group ascends from the branch endpoint $\alpha_{1,2,3}(0)$
with extent set by $s|\alpha_0|$.  Representative values are
$L_\mathrm{a}\sim10^2$ with $h_\mathrm{a}/L_\mathrm{a}\sim10^{-2}$,
$L_\mathrm{v}=1$ with $h_\mathrm{v}\approx0.3$, and $s=2$.

The removal parameters follow from the required rotational symmetry,
\begin{equation}
    \beta_k^{-}
    = -\left( \beta_k^{+} - \alpha_{\mathrm{mid}} \right) + \alpha_{\mathrm{mid}},
    \label{eq:recursion_parameters_minus}
\end{equation}
which guarantees that upstream and downstream filtering strengths balance
exactly at $\alpha_{\mathrm{mid}}$ and that the resulting auxiliary poles
$\beta_k^{*}$ trace a contour that intersects the pinch-point cleanly.

\begin{figure}[!htbp]
    \centering
    \begin{subfigure}[b]{0.85\textwidth}
        \centering
        \includegraphics[width=\linewidth]{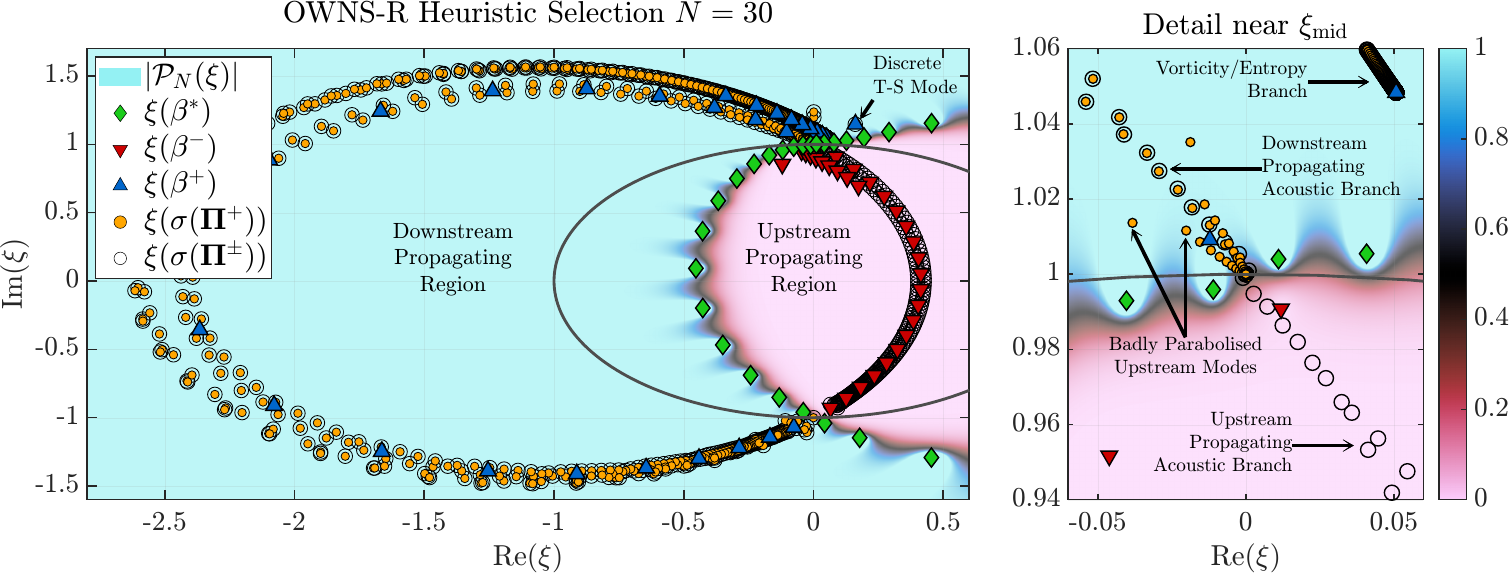}
        \caption{OWNS--R with heuristic parameter selection.}
        \label{fig:spec1a}
    \end{subfigure}
    \vspace{2mm}

    \begin{subfigure}[b]{0.85\textwidth}
        \centering
    \includegraphics[width=\linewidth]{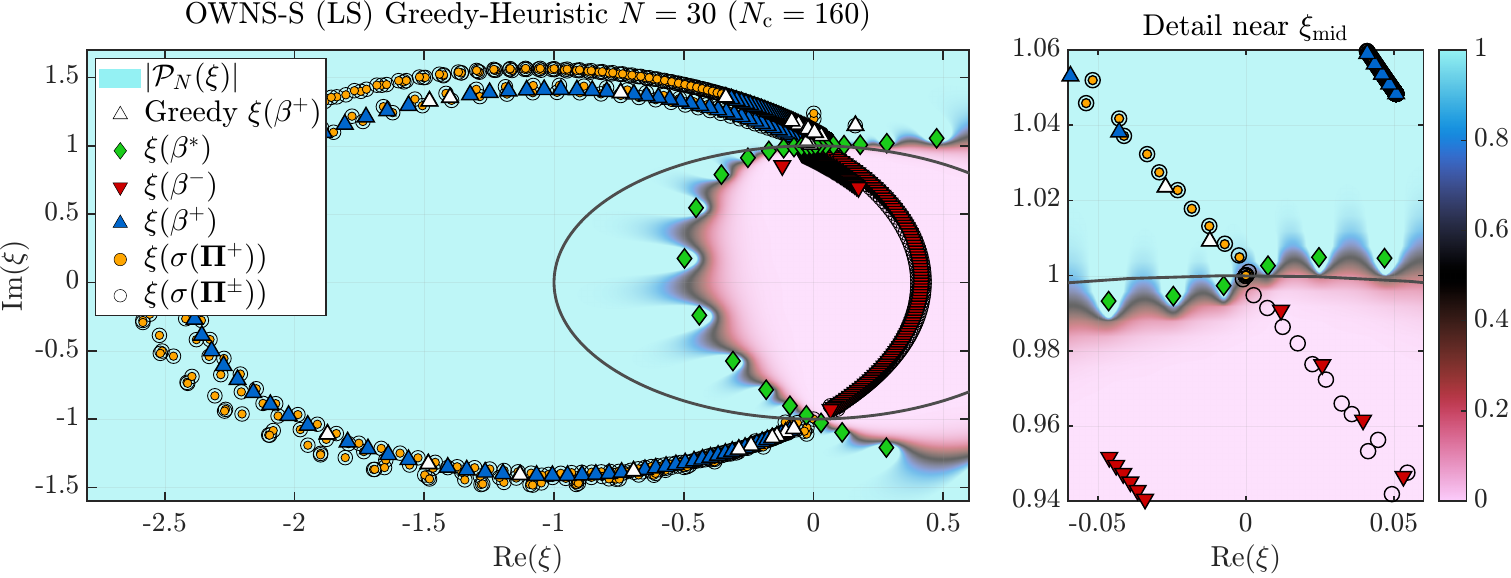}
        \caption{OWNS--S with greedy selection from 160 heuristic candidates
                 and least-squares weights.}
        \label{fig:spec1b}
    \end{subfigure}
    \vspace{2mm}

    \begin{subfigure}[b]{0.85\textwidth}
        \centering
        \includegraphics[width=\linewidth]{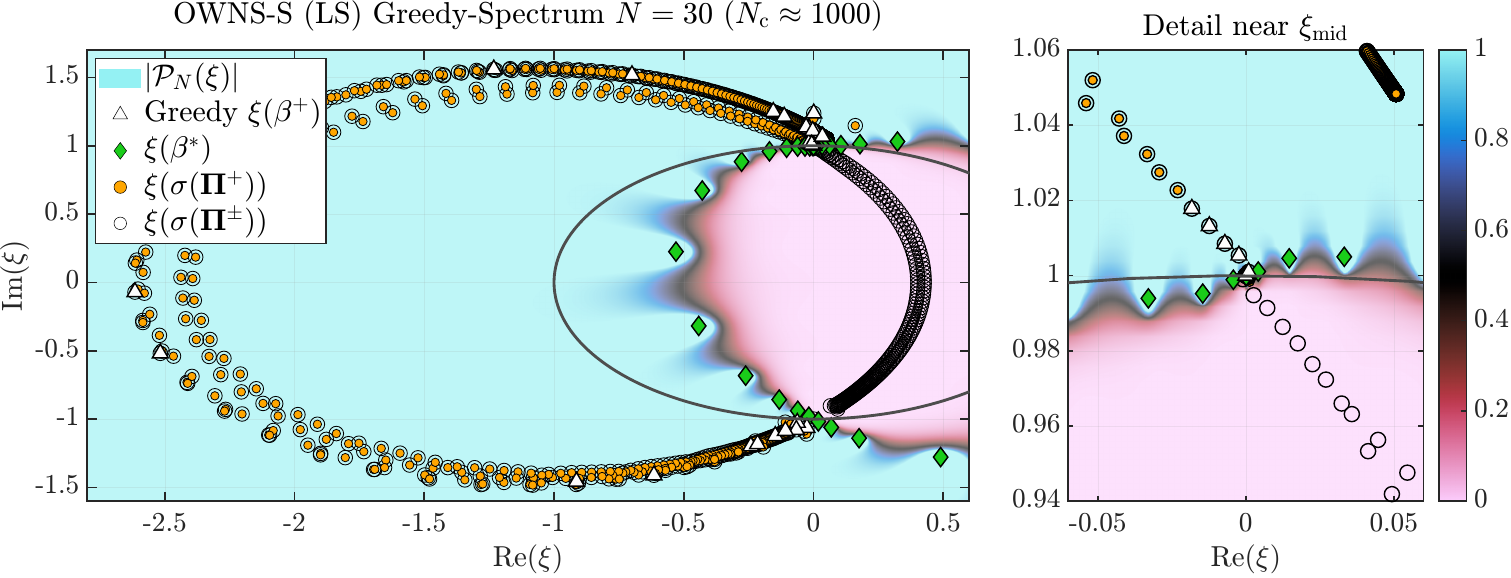}
        \caption{OWNS--S with greedy selection using the full eigenvalue spectrum
                 as possible retention/removal parameter pairs.}
        \label{fig:spec1c}
    \end{subfigure}
    \caption{%
    Numerical spectrum of the Cayley-transformed two-way system
    $\sigma(\mathbf{\Pi})$ and the parabolised downstream system
    $\sigma(\mathbf{\Pi}^+)$ for a 2D incompressible boundary layer
    ($M = 0.02$, $R_\delta = 400$, $f = 86\times10^{-6}$, $b = 0$)
    with $N = 30$ retention/removal parameters.
    Blue and red diamonds denote $\beta^+$ and $\beta^-$, green diamonds the
    auxiliary poles $\beta^*$, and contours show the modulus of the scalar
    filter $|\mathcal{P}^+_N|$ (unit circle shown).
    The heuristic selection (a) fails to remove all upstream modes,
    whereas both greedy variants (b,\,c) achieve tight clustering of upstream
    modes at $\xi_\mathrm{mid}$.}
    \label{fig:spec1}
\end{figure}

Figure~\ref{fig:spec1} illustrates these placements for the
Cayley-transformed spectrum at $M=0.02$.

\subsection{Supersonic Placement}\label{sec:rec:sup}

For the supersonic boundary layer at $M = 4.5$, the continuous acoustic branches
all propagate downstream, while upstream propagation occurs only through a small
set of trapped discrete acoustic modes whose locations vary with Mach number and
frequency and are not known analytically.  Accurate projection therefore requires
that the retention/removal parameters retain the downstream continuous spectrum while
suppressing these discrete upstream modes, without the two sets interfering with
each other.

To achieve independent control, $\beta^+$ and $\beta^-$ are constructed in two
complementary sets.  Downstream parameters $\beta_k^+$ are placed near the
downstream acoustic branches and reflected by complex conjugation to obtain the
corresponding $\beta_k^-$.  Removal parameters $\beta_k^-$ placed near the
approximate upstream-mode locations are reflected similarly.  A small tuning
constant $c<1$ reflects the imbalance between the number of downstream and
upstream modes, favouring retention.

The parameters are partitioned as
\begin{equation}
    N_{\mathrm{remove}} = \left\lfloor \frac{N}{3} \right\rfloor,
    \qquad
    N_{\mathrm{retain}} = N - N_{\mathrm{remove}},
\end{equation}
assigning $N_{\mathrm{retain}}$ pairs to the downstream continuous branches and
$N_{\mathrm{remove}}$ to the upstream trapped modes.  The larger share goes to
the branches because they are unbounded and must be covered along their length,
whereas the trapped modes are few and localised.

\begin{figure}[!htbp]
    \centering
    \begin{subfigure}[b]{0.85\textwidth}
        \centering
        \includegraphics[width=\linewidth]{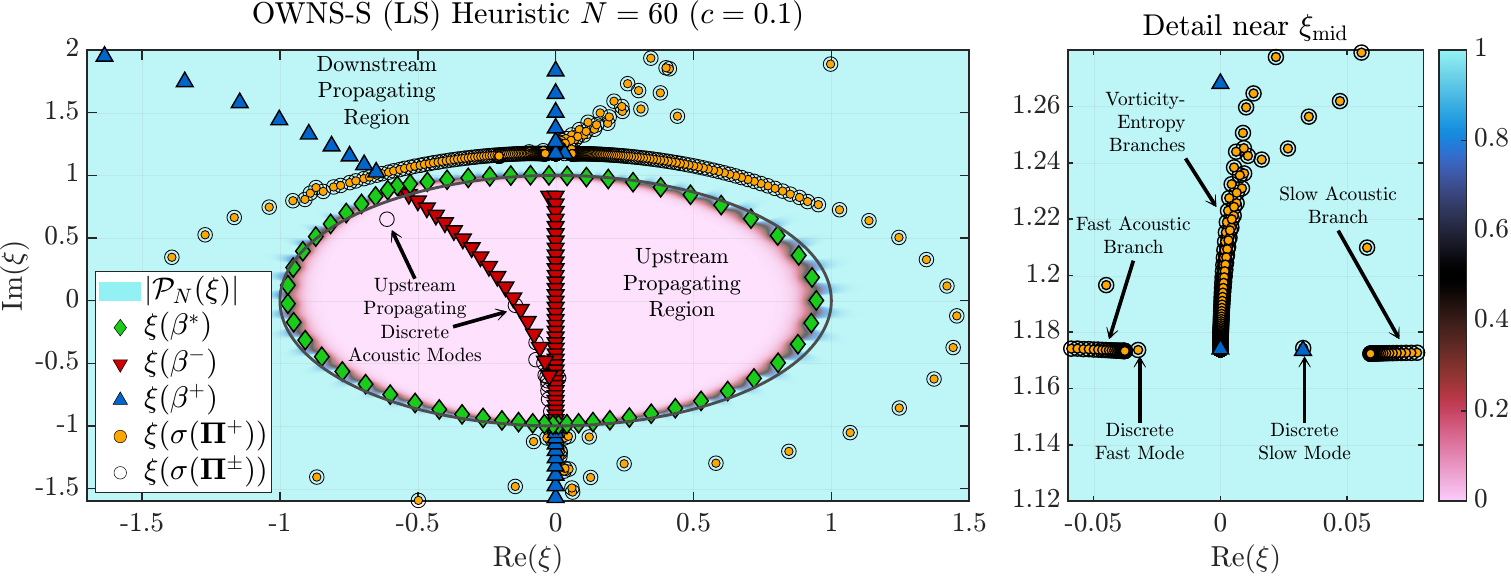}
        \caption{Heuristic placement targeting the downstream continuous branches
                 and the discrete trapped upstream modes.}
        \label{fig:spec2a}
    \end{subfigure}
    \vspace{2mm}

    \begin{subfigure}[b]{0.85\textwidth}
        \centering
        \includegraphics[width=\linewidth]{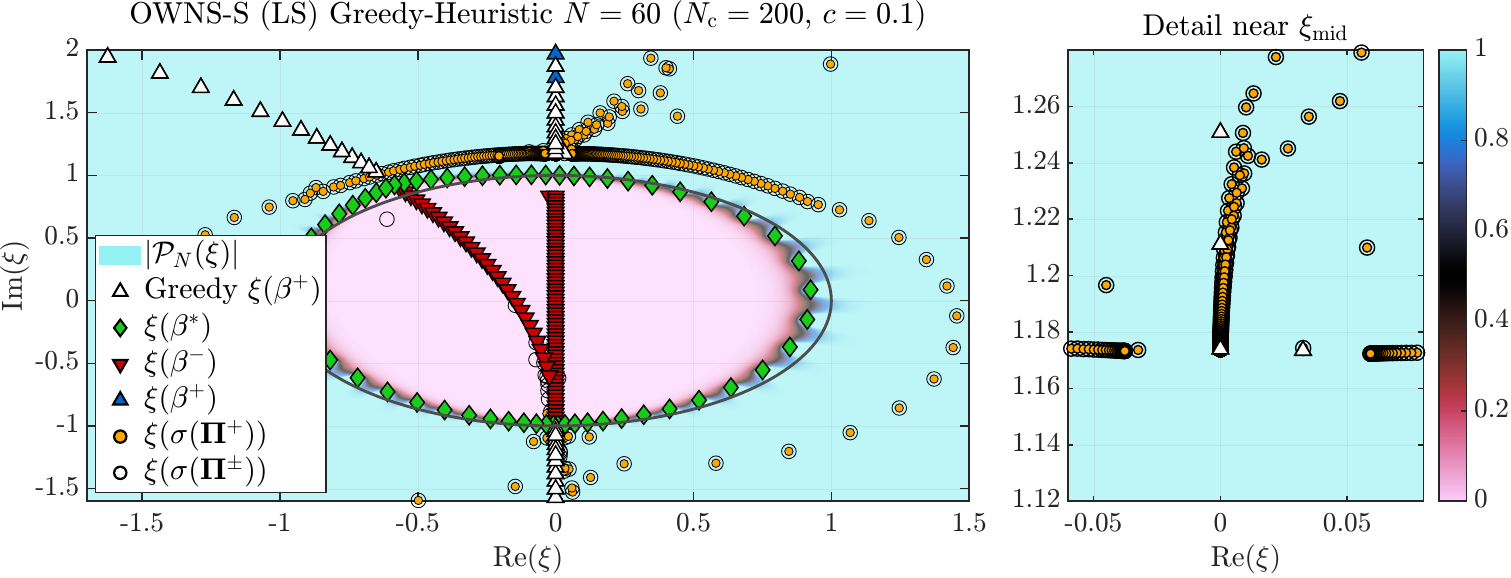}
        \caption{Greedy placement redistributes parameters across both regions.}
        \label{fig:spec2b}
    \end{subfigure}
    \caption{%
    Retention/removal parameter distributions for the hypersonic boundary layer
    ($M=4.5$) in Cayley space with $N=60$.
    Blue and red diamonds denote $\beta^+$ and $\beta^-$, green diamonds
    the auxiliary poles $\beta^*$, and contours show $|\mathcal{P}^+_N|$.
    (a)~Heuristic placement.  (b)~Greedy placement.}
    \label{fig:spec2}
\end{figure}

The downstream-retention parameters are
\begin{align}\label{eqn:recursion_parameters_sup1}
    \beta^{+}_k &=
    \alpha_{1,2,3}(0) + i\,y_{\mathrm{Malik}}
        \!\left(\dfrac{k-1}{N_{\mathrm{retain}}-1};\,h_1,L_1\right),
        && k = 1,\dots,N_{\mathrm{retain}}, \\[7pt]
    \label{eqn:recursion_parameters_sup2}
    \beta^{-}_k &=
    \alpha_{1,2,3}(0) - i\,y_{\mathrm{Malik}}
        \!\left(\dfrac{k-1}{N_{\mathrm{retain}}-1};\,h_1,L_1\right),
        && k = 1,\dots,N_{\mathrm{retain}},
\end{align}
and the upstream-removal parameters are
\begin{align}
    \beta^{-}_k &=
    \alpha_{\mathrm{remove}} - i\,y_{\mathrm{Malik}}
        \!\left(\dfrac{k-N_{\mathrm{retain}}-1}{N_{\mathrm{remove}}-1};\,h_2,L_2\right),
        && k = N_{\mathrm{retain}}+1,\dots,N,
    \label{eqn:recursion_parameters_sup3} \\[7pt]
    \beta^{+}_k &=
    \alpha_{\mathrm{remove}} + i\,y_{\mathrm{Malik}}
        \!\left(\dfrac{k-N_{\mathrm{retain}}-1}{N_{\mathrm{remove}}-1};\,h_2,L_2\right),
        && k = N_{\mathrm{retain}}+1,\dots,N,
    \label{eqn:recursion_parameters_sup4}
\end{align}
where $\alpha_{\mathrm{remove}}$ approximates the real part of the trapped
upstream modes \citep{gushchinExcitationDevelopmentUnstable1990b}, and
$L_1$, $L_2$ set the imaginary extent of the two groups with $h_1$, $h_2$ their
half-points.  Representative
values are $L_1\sim10^2$ with $h_1/L_1\sim10^{-2}$, and $L_2\approx4$ with
$h_2\approx0.5$, the second group being much the shorter because it covers a
few discrete modes rather than an unbounded branch. Fig.~\ref{fig:spec2} illustrates the resulting parameter distributions.

\subsection{Effect of \texorpdfstring{$c$}{c} on Auxiliary-Pole
Placement}\label{sec:C}

The tuning parameter $c$ in Eq.~\eqref{eqn:poly_general} controls the relative
weighting of the upstream and downstream recursion sets.  Its value determines
the shape of the transition region traced out by the auxiliary poles
$\beta_k^*$ in Cayley space and therefore the sharpness and orientation of the
scalar rational filter $\mathcal{P}^+_N$.

\begin{figure}[!htbp]
    \centering
    \begin{subfigure}[b]{0.32\textwidth}
        \centering
        \includegraphics[width=\textwidth]{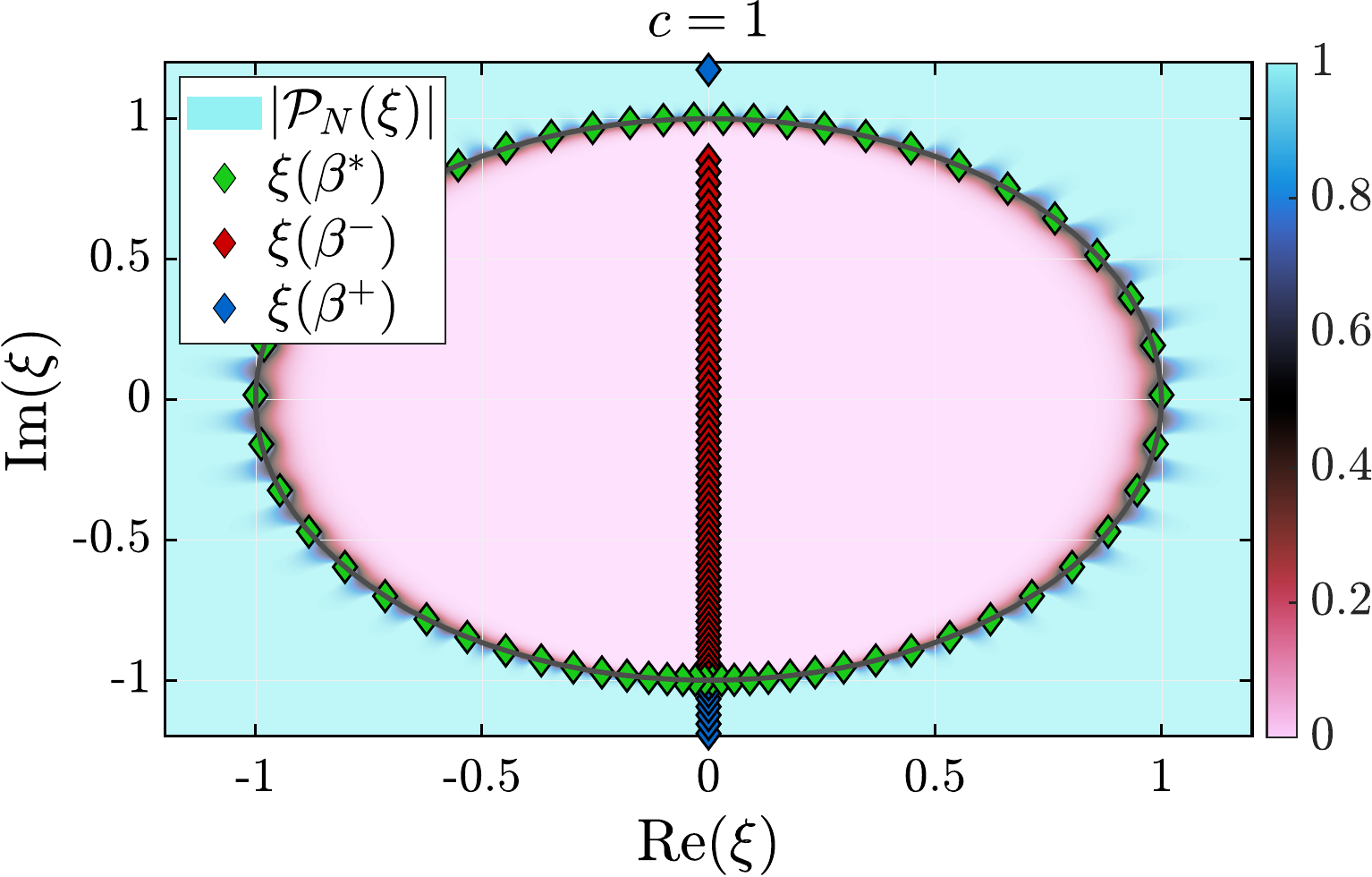}
        \caption{$c=1$.}
        \label{fig:c1}
    \end{subfigure}
    \hfill
    \begin{subfigure}[b]{0.32\textwidth}
        \centering
        \includegraphics[width=\textwidth]{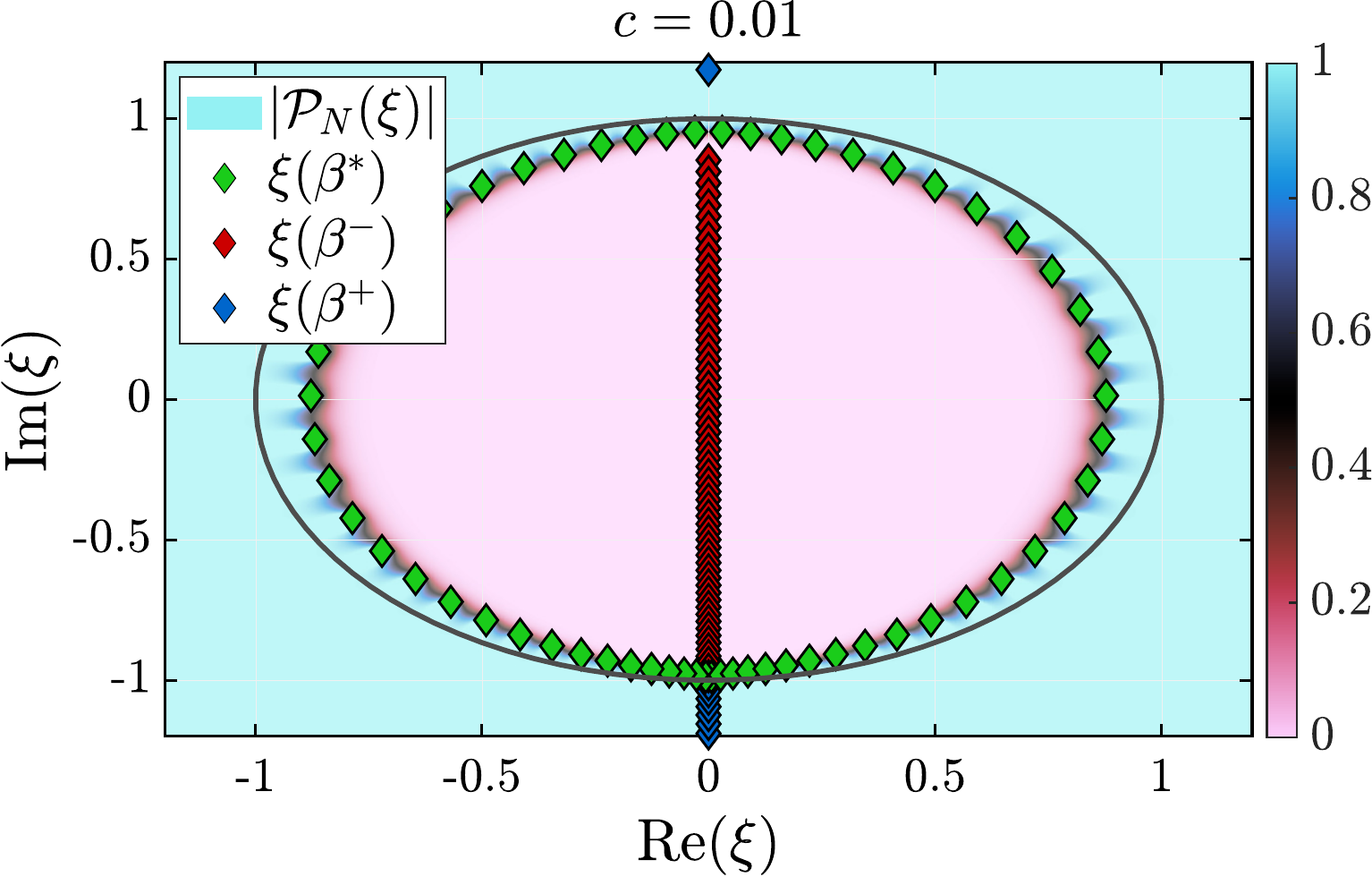} 
        \caption{$c=0.01$.} 
        \label{fig:c2}
    \end{subfigure}
    \hfill
    \begin{subfigure}[b]{0.32\textwidth}
        \centering
        \includegraphics[width=\textwidth]{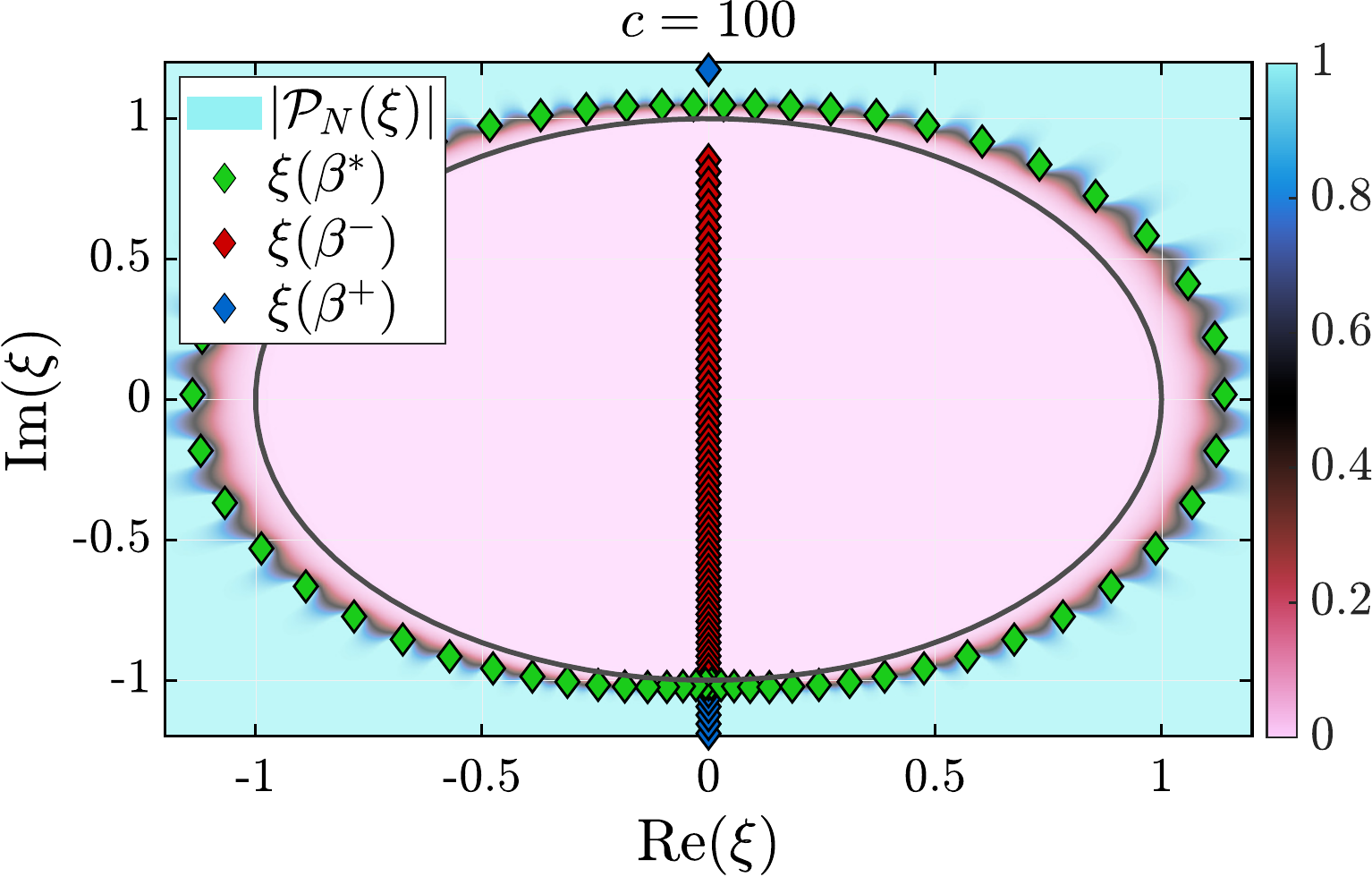}
        \caption{$c=100$.}
        \label{fig:c3}
    \end{subfigure}
    \caption{%
    Effect of the tuning parameter $c$ on auxiliary-pole placement in Cayley
    space.  Blue and red diamonds show $\beta^\pm$; green diamonds show
    $\beta^*$.  (a)~$c=1$: balanced retention and removal, $\beta^*$ contour
    near the unit circle.  (b)~$c=0.01$: contour contracted inward, favouring
    downstream retention.  (c)~$c=100$: contour pushed outward, strengthening
    upstream suppression.}
    \label{fig:c}
\end{figure}

When $c=1$ the auxiliary poles lie close to the unit circle and the transition
between retained and suppressed modes is approximately circular.  This choice is
effective in subsonic cases where the upstream and downstream recursion sets
exhibit near-symmetry (Fig.~\ref{fig:c1}).

Decreasing $c$ below unity contracts the transition region toward the unit
circle, drawing the auxiliary poles inward.  This localisation reduces unwanted
coupling from upstream retention/removal parameters and is useful when upstream discrete
modes lie far from the downstream branches, as in many supersonic configurations
(Fig.~\ref{fig:c2}).

Increasing $c$ above unity pushes the transition region outward, strengthening
suppression away from $\pm i$ without excessively tightening the region near the
acoustic-mode pinch.  This helps stabilise filtering in cases with broad spectral
separation (Fig.~\ref{fig:c3}).

In all cases, the clustering of poles near $\pm i$ reflects the non-uniform
distribution of the retention/removal parameters in standard $\alpha$-space rather than
any pathology of the projector.

\newcommand{\grouprow}[1]{\multicolumn{3}{@{}l}{\emph{#1}}\\[1pt]}

\begin{table}[!htbp]
\centering
\scriptsize

\noindent
\begin{minipage}[t]{0.5\textwidth}
\begin{tabular}{@{}lp{4cm}l@{}}
\toprule
Symbol & Meaning & Ref. \\
\midrule
\grouprow{Operators and pencil}
\addlinespace[4pt]
$\mathrm A,\mathrm B,\mathrm C$ & Continuous $5\times5$ operators, wall-normal coordinate & \eqref{eqn:LHNS} \\
$\mathbf A,\mathbf B,\mathbf C$ & Banded wall-normal discretisations & \eqref{eqn:two-way2} \\
$\mathbf\Pi$ & Pencil $(\mathbf B,\mathbf A)$; $\mathbf C$ retained in march, not in projector & \S\ref{sec:pencil} \\
$\mathbf\Pi^\pm$ & Projected pencils $(\mathbf B\mathbf P^\pm,\mathbf A)$ & \S\ref{sec:projection} \\
$\phi$ & Discrete disturbance state at a station & \eqref{eqn:two-way2} \\
$\psi_k$ & Intermediate state, $k$th OWNS-S solve; depends on $\phi$ alone & \eqref{eqn:owns_s_solve} \\
$\phi_k$ & Intermediate state, $k$th OWNS-R stage; depends on $\phi_{k-1}$ & \eqref{eqn:owns_r_apply} \\
$\mathbf M_k$ & OWNS-R stage operator $(\mathbf B-i\beta_k^*\mathbf A)^{-1}(\mathbf B-i\beta_k^-\mathbf A)$ & \eqref{eqn:stage_op} \\
$\mathbf E_k$ & Finite-precision error of the $k$th OWNS-R stage & \eqref{eqn:stage_op} \\
\addlinespace[4pt]
\grouprow{Spectral objects}
\addlinespace[4pt]
$\alpha$ & Streamwise wavenumber; spectral variable & \eqref{eq:gevp} \\
$\alpha_k, v_k$ & Eigenvalue / right eigenvector of $\mathbf\Pi$ & \eqref{eq:gevp} \\
$\sigma(\mathbf\Pi)$ & Finite spectrum; infinite eigenvalues excluded & \S\ref{sec:pencil} \\
$\Gamma^+$ & Contour enclosing only downstream eigenvalues & \S\ref{sec:projection} \\
$\Omega^+,\Omega^-$ & Interior of $\Gamma^+$, and $\mathbb C\setminus\Omega^+$ & \S\ref{sec:projection} \\
$\alpha_\mathrm{mid}$ & Spectral reference point (subsonic/supersonic differ) & \eqref{eqn:mid_app} \\
$\xi(\alpha)$ & Cayley map, $\alpha_\mathrm{mid}\mapsto \xi_\mathrm{mid}=i$; diagnostics only & App.~\ref{app:Cayley} \\
\addlinespace[4pt]
\grouprow{Projection-quality metrics}
\addlinespace[4pt]
$\mathcal E_\mathrm{idem}$ & Departure from idempotency, $\lVert(\mathbf P^+_N)^2-\mathbf P^+_N\rVert_\mathrm F$ & \eqref{eqn:idem_metric} \\
$\mathcal E_\mathrm{retain}$ & Displacement of downstream eigenvalues & \eqref{eqn:retain_metric} \\
$\mathcal E_\mathrm{leak}$ & Surviving magnitude of upstream eigenvalues & \eqref{eqn:leak_metric} \\
$\lVert\cdot\rVert_\mathrm F$ & Normalised Frobenius norm & \S\ref{sec:synthetic} \\
\bottomrule
\end{tabular}
\end{minipage}%
\hfill
\begin{minipage}[t]{0.5\textwidth}
\begin{tabular}{@{}lp{4cm}l@{}}
\toprule
Symbol & Meaning & Ref. \\
\midrule
\addlinespace[4pt]
\grouprow{Filter and parameters}
\addlinespace[4pt]
$\mathcal P^+$ & Characteristic function: $1$ on $\Omega^+$, $0$ elsewhere & \eqref{eqn:char_func} \\
$\mathbf P^+$ & Exact spectral projector, downstream subspace & \eqref{eqn:contour} \\
$\mathcal P^+_N$ & Order-$N$ rational filter approximating $\mathcal P^+$ & \eqref{eqn:quadrature_scalar} \\
$\mathbf P^+_N$ & Matrix lift of $\mathcal P^+_N$ & \eqref{eqn:quadrature_matrix} \\
$N$ & Approximation order: pairs, poles, solves & \eqref{eqn:quadrature_scalar} \\
$\beta_k^+$ & Retention parameters, in $\Omega^+$ & \S\ref{sec:implicit} \\
$\beta_k^-$ & Removal parameters, $\mathrm T(\beta_k^+)$ & \eqref{eqn:involution} \\
$\beta_k^*$ & Auxiliary poles actually solved at & \eqref{eqn:poly_general} \\
$\mathrm T$ & Pairing involution: rotation (subsonic) / conjugation (supersonic) & \eqref{eqn:involution} \\
$c$ & Balance parameter, $c>0$ & \eqref{eqn:poly_general} \\
$R(\alpha)$ & $\prod_k(\alpha-\beta_k^+)/(\alpha-\beta_k^-)$; filter $=(1+cR)^{-1}$ & \eqref{eqn:R_def} \\
$w_0,w_k$ & Partial-fraction weights & \eqref{eqn:standard_weights} \\
$\mathbf D$ & Cauchy-like matrix, LS weight fit & \eqref{eqn:ls_matrix} \\
\addlinespace[4pt]
\grouprow{Parameter selection}
\addlinespace[4pt]
$\zeta_m$ & Test point for separation conditions & \eqref{eqn:interp_conditions} \\
$\mathcal Z,\mathcal Z^\pm$ & Test set / partition; also candidate pool & \S\ref{sec:greedy} \\
$N_\mathrm{c}$ & Candidate-pool size, $5$--$20\,N$ & \S\ref{sec:greedy_ours} \\
$\mathcal S^{(n)}$ & Pairs selected so far & \S\ref{sec:greedy} \\
$e(\zeta;\mathcal S^{(n)})$ & Departure from target value at $\zeta$ & \eqref{eqn:residual} \\
$\mathcal J_\infty,\mathcal J_2$ & Worst-case / aggregate objectives & \eqref{eqn:J_inf} \\
\addlinespace[4pt]
\grouprow{Flow and discretisation}
\addlinespace[4pt]
$\omega,\beta$ & Frequency, spanwise wavenumber & \eqref{eqn:dist_decomp} \\
$R,R_\delta(x)$ & Inlet / local Reynolds number & \S\ref{sec:numerics} \\
$f,b$ & Non-dim. $\omega/R$, $\beta/R$ & \S\ref{sec:numerics} \\
$M$ & Freestream Mach number & \S\ref{sec:numerics} \\
$M_\mathrm{2D}$ & Streamwise-edge Mach; governs acoustic continuum & \eqref{eq:mach2d3d} \\
$M_\mathrm{3D}$ & Total-edge Mach; governs base-flow supersonicity & \eqref{eq:mach2d3d} \\
$n_y,n_x$ & Wall-normal / streamwise grid counts & \S\ref{sec:numerics} \\
$n_\mathrm{iter}$ & PSE iterations per station & \S\ref{sec:solve_count} \\
$n_\mathrm{t}$ & Threads across OWNS-S solves & \S\ref{sec:solve_count} \\
\bottomrule
\end{tabular}
\end{minipage}
\caption{Notation table.}
\label{tab:notation}
\end{table}

\bibliographystyle{cas-model2-names} 
\bibliography{bib/owns}
\end{document}

%% file: figs2/tikz/spec1.tex
\begin{tikzpicture}
\begin{axis}[
    xlabel={$\mathrm{Re}(\alpha)$},
    ylabel={$\mathrm{Im}(\alpha)$},
    xmin=-0.14, xmax=0.1,
    ymin=-0.1, ymax=0.1,
    axis lines=center,
    axis line style={-,line width=1.5pt}, 
    width=6cm,
    height=6cm,
    scale only axis,
    tick align=outside,
    xtick=\empty,
    ytick=\empty,
    xticklabels={},
    yticklabels={},
    xlabel style={at={(axis cs:0.11,0)}, anchor=north east}, 
    ylabel style={at={(axis cs:-0.0025,0.09)}, anchor=south east}, 
]
\addplot [-stealth, domain=0:0.1, samples=2, blue, line width=4.0pt] coordinates {(-0.06,0) (-0.06,0.1)};
\addplot [-stealth, domain=0:0.1, samples=2, blue, line width=4.0pt] coordinates {(0.01,0) (-0.06,0.0)};

\addplot [-stealth, domain=-0.1:0, samples=2, red, line width=4.0pt] coordinates {(-0.06,0) (-0.06,-0.1)};
\addplot [-stealth, domain=0:0.1, samples=2, red, line width=4.0pt] coordinates {(-0.13,0) (-0.06,0.0)};
\addplot [-stealth, domain=0:0.1, samples=2, blue, line width=4.0pt] coordinates {(0.035,0) (0.035,0.1)};
\addplot [only marks, mark=o, blue, mark size=4pt, mark options={fill=white, line width=2.0pt}] coordinates {
    (0.05, 0.01)
};

\node[align=center] (mach) at (axis cs:0.05,-0.06) {$M=0.5$\\$\beta=0$};

\node[blue, right=2pt, font=\bfseries] at (axis cs:0.04,0.09) {$\alpha_{1,2,3}(\eta)$};
\node[blue, right=2pt, font=\bfseries] at (axis cs:-0.105,0.09) {$\alpha_{4}(\eta)$};
\node[red, right=2pt, font=\bfseries] at (axis cs:-0.105,-0.09) {$\alpha_{5}(\eta)$};

\node[align=center, text width=4cm] (discrete_label) at (axis cs:0.08,0.05) {T-S};
\draw[-{stealth}, line width=1.5pt, dashed, black!70!black] (discrete_label) -- (axis cs:0.048,0.0085);

\node[align=center, text width=4cm] (mid) at (axis cs:-0.10,0.04) {$\alpha_\text{mid}$};
\draw[-{stealth}, line width=1.5pt, dashed, black!70!black] (mid) -- (axis cs:-0.065,0.005);

\end{axis}
\end{tikzpicture}

%% file: figs2/tikz/spec3.tex
\begin{tikzpicture}
\begin{axis}[
    xlabel={$\mathrm{Re}(\alpha)$},
    ylabel={$\mathrm{Im}(\alpha)$},
    xmin=-0.072, xmax=0.16,
    ymin=-0.06, ymax=0.1,
    axis lines=center,
    axis line style={-,line width=1.5pt}, 
    width=6cm,
    height=6cm,
    scale only axis,
    tick align=outside,
    xtick=\empty,
    ytick=\empty,
    xticklabels={},
    yticklabels={},
    xlabel style={at={(axis cs:0.15,-0.002)}, anchor=north east}, 
    ylabel style={at={(axis cs:-0.0025,0.09)}, anchor=south east}, 
]

\addplot [-stealth, domain=0:0.1, samples=2, blue, line width=4.0pt] coordinates {(0.01,0) (-0.072,0.0)};

\addplot [-stealth, domain=0:0.1, samples=2, blue, line width=4.0pt] coordinates {(0.09,0) (0.16,0.0)};

\addplot [-stealth, domain=0:0.1, samples=2, blue, line width=4.0pt] coordinates {(0.05,0) (0.05,0.1)};

\node[blue, right=2pt, font=\bfseries] at (axis cs:0.055,0.09) {$\alpha_{1,2,3}(\eta)$};
\node[blue, right=2pt] at (axis cs:0.085,0.01) {Slow Branch};
\node[blue, right=2pt] at (axis cs:-0.08,0.01) {Fast Branch};


\addplot [only marks, mark=o, blue, mark size=4pt, mark options={fill=white, line width=2.0pt}] coordinates {
    (0.072, 0.008)
};
\node[align=center, text width=4cm] (discrete_label) at (axis cs:0.11,0.05) {Slow Mode};
\draw[-{stealth}, line width=1.5pt, dashed, black!70!black] (discrete_label) -- (axis cs:0.072,0.0085);

\addplot [only marks, mark=o, blue, mark size=4pt, mark options={fill=white, line width=2.0pt}] coordinates {
    (0.03, 0.008)
};
\node[align=center, text width=4cm] (discrete_label) at (axis cs:-0.04,0.05) {Fast Mode};
\draw[-{stealth}, line width=1.5pt, dashed, black!70!black] (discrete_label) -- (axis cs:0.03,0.0085);

\node[align=center, text width=4cm] (mid) at (axis cs:0.080,-0.025) {$\alpha_\text{mid}$};
\draw[-{stealth}, line width=1.5pt, dashed, black!70!black] (mid) -- (axis cs:0.05,-0.005);

\addplot [only marks, mark=o, red, mark size=4pt, mark options={fill=white, line width=2.0pt}] coordinates {
    (-0.035, -0.02)
    (-0.033, -0.035)  
    (-0.036, -0.05)  
};
\node[align=center, text width=3cm] (discrete_label) at (axis cs:0.07,-0.05) {Upstream Discrete \\ Modes};
\draw[-{stealth}, line width=1.5pt, dashed, black!70!black] (discrete_label) -- (axis cs:-0.035,-0.02);
\draw[-{stealth}, line width=1.5pt, dashed, black!70!black] (discrete_label) -- (axis cs:-0.036,-0.05);

\node[align=center] (mach) at (axis cs:-0.05, 0.075) {$M=4.5$\\$\beta=0$};

\end{axis}
\end{tikzpicture}

%% file: figs2/tikz/spec2.tex
\begin{tikzpicture}
\begin{axis}[
    xlabel={$\mathrm{Re}(\xi)$},
    ylabel={$\mathrm{Im}(\xi)$},
    xmin=-3.5, xmax=1.3,
    ymin=-2.1, ymax=2.1,
    axis lines=center,
    axis line style={-,line width=1.5pt}, 
    width=6cm,
    height=6cm,
    scale only axis,
    tick align=outside,
    xtick=\empty,
    ytick=\empty,
    xticklabels={},
    yticklabels={},
    xlabel style={at={(axis cs:-3.1,0)}, anchor=north east}, 
    ylabel style={at={(axis cs:0,2.)}, anchor=south east}, 
]

\node[align=center, text width=0.3cm] (mid) at (axis cs:-1.5,0.7) {$\xi_\text{mid}$};
\draw[-{stealth}, line width=1.5pt, dashed, black!70!black] (mid) -- (axis cs:-0.1,1.05);

\addplot [dashed, gray, line width=1.5pt, domain=0:360, samples=100] 
    ({cos(x)}, {sin(x)});

\addplot [-stealth, blue, line width=3.0pt, domain=45:315, samples=100] 
    ({-1 + sqrt(2)*cos(x)}, {sqrt(2)*sin(x)});
\addplot [-stealth, domain=-0.1:0, samples=2, blue, line width=3.0pt] coordinates {(0.3,1.3) (0,1)};

\addplot [-stealth, blue, line width=3.0pt, domain=45:315, samples=100] 
    ({-1 + 2.*cos(x)}, {2.*sin(x)});

\addplot [-stealth, red, line width=3.0pt, domain=45:-45, samples=50] 
    ({-1 + sqrt(2)*cos(x)}, {sqrt(2)*sin(x)});
\addplot [-stealth, domain=-0.1:0, samples=2, red, line width=3.0pt] coordinates {(-0.3,0.7) (0,1)};

\addplot [only marks, mark=o, blue, mark size=4pt, mark options={fill=white, line width=2.0pt}] coordinates {(0.5, 1.7)};
\node[align=center, text width=4cm] (discrete_label) at (axis cs:1.1,1) {T-S};
\draw[-{stealth}, line width=1.5pt, dashed, black!70!black] (discrete_label) -- (axis cs:0.5,1.7);

\node[blue, right=2pt, font=\bfseries] at (axis cs:0,-1.9) {$\xi_{1,2,3}(\eta)$};
\node[blue, right=2pt, font=\bfseries] at (axis cs:-1.2,-1.2) {$\xi_{4}(\eta)$};
\node[red, right=2pt, font=\bfseries] at (axis cs:0.3,-0.4) {$\xi_{5}(\eta)$};

\end{axis}
\end{tikzpicture}

%% file: figs2/tikz/spec4.tex
\begin{tikzpicture}
\begin{axis}[
    xlabel={$\mathrm{Re}(\xi)$},
    ylabel={$\mathrm{Im}(\xi)$},
    xmin=-1.8, xmax=1.8,
    ymin=-1.8, ymax=1.8,
    axis lines=center,
    axis line style={-,line width=1.5pt}, 
    width=6cm,
    height=6cm,
    scale only axis,
    tick align=outside,
    xtick=\empty,
    ytick=\empty,
    xticklabels={},
    yticklabels={},
    xlabel style={at={(axis cs:1.77,0.3)}, anchor=north east}, 
    ylabel style={at={(axis cs:0.7,1.4)}, anchor=south east}, 
]

\addplot [dashed, gray, line width=1.5pt, domain=0:360, samples=100] 
    ({cos(x)}, {sin(x)});

\addplot [-stealth, blue, line width=3.0pt, domain=120:265, samples=100] 
    ({ 1.1*cos(x)}, {1.1*sin(x)});

\addplot [-stealth, blue, line width=3.0pt, domain=120:265, samples=100] 
    ({- 1.1*cos(x)}, {1.1*sin(x)});    


\addplot [-stealth, domain=0:0.1, samples=2, blue, line width=4.0pt] coordinates {(0.0,1) (0.0,1.8)};
\addplot [-stealth, domain=0:0.1, samples=2, blue, line width=4.0pt] coordinates {(0.0,-1.8) (0.0,-1.1)};

\node[align=center, text width=0.3cm] (mid) at (axis cs:0.4,0.5) {$\xi_\text{mid}$};
\draw[-{stealth}, line width=1.5pt, dashed, black!70!black] (mid) -- (axis cs:0.,1.0);

\addplot [only marks, mark=o, red, mark size=4pt, mark options={fill=white, line width=2.0pt}] coordinates {
    (-0.3, 0.7)
    (-0.25, 0.3)  
    (-0.1, -0.1)  
    (-0.05, -0.5)  
    (-0.01, -0.8)  
};

\addplot [only marks, mark=o, blue, mark size=4pt, mark options={fill=white, line width=2.0pt}] coordinates {
    (0.3, 1.1)
    (-0.3, 1.1)  
};

\node[blue, right=2pt, font=\bfseries] at (axis cs:-0.85,1.55) {$\alpha_{1,2,3}(\eta)$};

\node[blue, right=2pt, font=\bfseries] at (axis cs:-0.85,-1.55) {$\alpha_{1,2,3}(\eta)$};

\node[blue, right=2pt] at (axis cs:0.63,-0.94) {Slow Branch};
\node[blue, right=2pt] at (axis cs:-1.85,-0.94) {Fast Branch};

\end{axis}
\end{tikzpicture}

%% file: latex/super_trans0.tex
\begin{tikzpicture}
\begin{axis}[
    xlabel={$\mathrm{Re}(\alpha)$},
    ylabel={$\mathrm{Im}(\alpha)$},
    xmin=-0.1, xmax=0.1,
    ymin=-0.1, ymax=0.1,
    axis lines=center,
    axis line style={-,line width=1.5pt}, 
    width=4.5cm,
    height=4.5cm,
    scale only axis,
    tick align=outside,
    xtick=\empty,
    ytick=\empty,
    xticklabels={},
    yticklabels={},
    xlabel style={at={(axis cs:0.11,0)}, anchor=north east}, 
    ylabel style={at={(axis cs:-0.00,0.08)}, anchor=south east}, 
]
\addplot [-stealth, domain=0:0.1, samples=2, blue, line width=4.0pt] coordinates {(0.02,0.05) (0.02,0.1)};

\addplot [-stealth, domain=0:0.1, samples=2, red, line width=4.0pt] coordinates {(0.02,-0.05) (0.02,-0.1)};

\addplot [-stealth, domain=0:0.1, samples=2, blue, line width=4.0pt] coordinates {(-0.07,0) (-0.07,0.1)};

\node[
 draw=black,
    fill=white,
    rounded corners=2pt,
    line width=0.5pt,
    align=center,
    font=\fontsize{10}{12}\selectfont,
    inner sep=3pt
] (mach) at (axis cs:-0.05,-0.06)
{$R_\delta\approx145$\\
$M_\mathrm{2D}\approx0.54$\\
$M_\mathrm{3D}\approx0.79$};

\addplot [only marks, mark=o, blue, mark size=4pt, mark options={fill=white, line width=2.0pt}] coordinates {
    (-0.055, 0.01)
};

\end{axis}
\end{tikzpicture}

%% file: latex/super_trans1.tex
\begin{tikzpicture}
\begin{axis}[
    xlabel={$\mathrm{Re}(\alpha)$},
    ylabel={$\mathrm{Im}(\alpha)$},
    xmin=-0.1, xmax=0.1,
    ymin=-0.1, ymax=0.1,
    axis lines=center,
    axis line style={-,line width=1.5pt}, 
    width=4.5cm,
    height=4.5cm,
    scale only axis,
    tick align=outside,
    xtick=\empty,
    ytick=\empty,
    xticklabels={},
    yticklabels={},
    xlabel style={at={(axis cs:0.11,0)}, anchor=north east}, 
    ylabel style={at={(axis cs:-0.00,0.08)}, anchor=south east}, 
]
\addplot [-stealth, domain=0:0.1, samples=2, blue, line width=4.0pt] coordinates {(0.04,0.015) (0.04,0.1)};

\addplot [-stealth, domain=0:0.1, samples=2, red, line width=4.0pt] coordinates {(0.04,-0.015) (0.04,-0.1)};

\addplot [-stealth, domain=0:0.1, samples=2, blue, line width=4.0pt] coordinates {(-0.05,0) (-0.05,0.1)};

\node[
 draw=black,
    fill=white,
    rounded corners=2pt,
    line width=0.5pt,
    align=center,
    font=\fontsize{10}{12}\selectfont,
    inner sep=3pt
] (mach) at (axis cs:-0.055,-0.055)
{$R_\delta\approx180$\\
$M_\mathrm{2D}\approx0.64$\\
$M_\mathrm{3D}\approx0.86$};

\addplot [only marks, mark=o, blue, mark size=4pt, mark options={fill=white, line width=2.0pt}] coordinates {
    (-0.035, 0.005)
};

\addplot [only marks, mark=*, red, mark size=3.5pt] coordinates {
    (0, -0.02)
};

\addplot [only marks, mark=*, blue, mark size=3.5pt] coordinates {
    (0, 0.02)
};

\end{axis}
\end{tikzpicture}

%% file: latex/super_trans2.tex
\begin{tikzpicture}
\begin{axis}[
    xlabel={$\mathrm{Re}(\alpha)$},
    ylabel={$\mathrm{Im}(\alpha)$},
    xmin=-0.06, xmax=0.14,
    ymin=-0.1, ymax=0.1,
    axis lines=center,
    axis line style={-,line width=1.5pt}, 
    width=4.5cm,
    height=4.5cm,
    scale only axis,
    tick align=outside,
    xtick=\empty,
    ytick=\empty,
    xticklabels={},
    yticklabels={},
    xlabel style={at={(axis cs:0.15,0)}, anchor=north east}, 
    ylabel style={at={(axis cs:-0.00,0.08)}, anchor=south east}, 
]
\addplot [-stealth, domain=0:0.1, samples=2, blue, line width=4.0pt] coordinates {(0.07,0.0) (0.07,0.1)};
\addplot [-stealth, domain=0:0.1, samples=2, blue, line width=4.0pt] coordinates {(0.03,0.0) (0.07,0.)};

\addplot [-stealth, domain=0:0.1, samples=2, red, line width=4.0pt] coordinates {(0.07,-0.0) (0.07,-0.1)};
\addplot [-stealth, domain=0:0.1, samples=2, red, line width=4.0pt] coordinates {(0.11,-0.0) (0.07,-0.0)};

\addplot [-stealth, domain=0:0.1, samples=2, blue, line width=4.0pt] coordinates {(-0.05,0) (-0.05,0.1)};

\node[
    draw=black,
    fill=white,
    rounded corners=2pt,
    line width=0.5pt,
    align=center,
    font=\fontsize{10}{12}\selectfont,
    inner sep=3pt
] (mach) at (axis cs:-0.016,-0.065)
{$R_\delta\approx320$\\
$M_\mathrm{2D}\approx0.88$\\
$M_\mathrm{3D}\approx1.06$};

\addplot [only marks, mark=o, blue, mark size=4pt, mark options={fill=white, line width=2.0pt}] coordinates {
    (-0.035, 0.005)
};

\addplot [only marks, mark=*, red, mark size=3.5pt] coordinates {
    (0, -0.02)
};

\addplot [only marks, mark=*, blue, mark size=3.5pt] coordinates {
    (0, 0.02)
};

\end{axis}
\end{tikzpicture}

%% file: latex/super_trans3.tex
\begin{tikzpicture}
\begin{axis}[
    xlabel={$\mathrm{Re}(\alpha)$},
    ylabel={$\mathrm{Im}(\alpha)$},
    xmin=-0.06, xmax=0.14,
    ymin=-0.1, ymax=0.1,
    axis lines=center,
    axis line style={-,line width=1.5pt}, 
    width=4.5cm,
    height=4.5cm,
    scale only axis,
    tick align=outside,
    xtick=\empty,
    ytick=\empty,
    xticklabels={},
    yticklabels={},
    xlabel style={at={(axis cs:0.15,0)}, anchor=north east}, 
    ylabel style={at={(axis cs:-0.00,0.08)}, anchor=south east}, 
]
\addplot [-stealth, domain=0:0.1, samples=2, blue, line width=4.0pt] coordinates {(0.09,0.0) (0.09,0.1)};
\addplot [-stealth, domain=0:0.1, samples=2, blue, line width=4.0pt] coordinates {(0.03,0.0) (0.09,0.)};

\addplot [-stealth, domain=0:0.1, samples=2, red, line width=4.0pt] coordinates {(0.09,-0.0) (0.09,-0.1)};
\addplot [-stealth, domain=0:0.1, samples=2, red, line width=4.0pt] coordinates {(0.14,-0.0) (0.09,-0.0)};

\addplot [-stealth, domain=0:0.1, samples=2, blue, line width=4.0pt] coordinates {(-0.05,0) (-0.05,0.1)};

\node[
    draw=black,
    fill=white,
    rounded corners=2pt,
    line width=0.5pt,
    align=center,
    font=\fontsize{10}{12}\selectfont,
    inner sep=3pt
] (mach) at (axis cs:-0.016,-0.065)
{$R_\delta\approx380$\\
$M_\mathrm{2D}\approx0.94$\\
$M_\mathrm{3D}\approx1.12$};

\addplot [only marks, mark=o, blue, mark size=4pt, mark options={fill=white, line width=2.0pt}] coordinates {
    (-0.035, 0.005)
};

\addplot [only marks, mark=*, red, mark size=3.5pt] coordinates {
    (0, -0.02)
};
\addplot [only marks, mark=*, blue, mark size=3.5pt] coordinates {
    (0, 0.02)
};

\addplot [only marks, mark=*, red, mark size=3.5pt] coordinates {
    (0.045, -0.04)
    (0.051, -0.06)
    (0.059, -0.08)
};
\addplot [only marks, mark=*, blue, mark size=3.5pt] coordinates {
    (0.045, 0.04)
    (0.051, 0.06)
    (0.059, 0.08)
};

\end{axis}
\end{tikzpicture}

%% file: latex/super_trans4.tex
\begin{tikzpicture}
\begin{axis}[
    xlabel={$\mathrm{Re}(\alpha)$},
    ylabel={$\mathrm{Im}(\alpha)$},
    xmin=-0.13, xmax=0.07,
    ymin=-0.1, ymax=0.1,
    axis lines=center,
    axis line style={-,line width=1.5pt}, 
    width=4.5cm,
    height=4.5cm,
    scale only axis,
    tick align=outside,
    xtick=\empty,
    ytick=\empty,
    xticklabels={},
    yticklabels={},
    xlabel style={at={(axis cs:0.08,-0.005)}, anchor=north east}, 
    ylabel style={at={(axis cs:-0.00,0.08)}, anchor=south east}, 
]
\addplot [-stealth, domain=0:0.1, samples=2, blue, line width=4.0pt] coordinates {(0.09,0.0) (0.09,0.1)};
\addplot [-stealth, domain=0:0.1, samples=2, blue, line width=4.0pt] coordinates {(0.03,0.0) (0.07,0.)};

\addplot [-stealth, domain=0:0.1, samples=2, blue, line width=4.0pt] coordinates {(-0.08,-0.0) (-0.13,-0.0)};

\addplot [-stealth, domain=0:0.1, samples=2, blue, line width=4.0pt] coordinates {(-0.05,0) (-0.05,0.1)};

\node[
    draw=black,
    fill=white,
    rounded corners=2pt,
    line width=0.5pt,
    align=center,
    font=\fontsize{10}{12}\selectfont,
    inner sep=3pt
] (mach) at (axis cs:-0.06,-0.065)
{$R_\delta>481$\\
$M_\mathrm{2D}>1.03$\\
$M_\mathrm{3D}>1.2$};

\addplot [only marks, mark=o, blue, mark size=4pt, mark options={fill=white, line width=2.0pt}] coordinates {
    (-0.035, 0.005)
};

\addplot [only marks, mark=*, red, mark size=3.5pt] coordinates {
    (0, -0.02)
};
\addplot [only marks, mark=*, blue, mark size=3.5pt] coordinates {
    (0, 0.02)
};

\addplot [only marks, mark=*, red, mark size=3.5pt] coordinates {
    (0.04, -0.035)
    (0.04, -0.055)
    (0.04, -0.075)
    (0.04, -0.095)

};
\addplot [only marks, mark=*, blue, mark size=3.5pt] coordinates {
    (0.04, 0.035)
    (0.04, 0.055)
    (0.04, 0.075)
    (0.04, 0.095)
};

\end{axis}
\end{tikzpicture}

%% file: latex/super_trans0_c.tex
\begin{tikzpicture}
\begin{axis}[
    xlabel={$\mathrm{Re}(\xi)$},
    ylabel={$\mathrm{Im}(\xi)$},
    xmin=-0.07, xmax=0.16,
    ymin=-0.08, ymax=0.15,
    axis lines=center,
    axis line style={-,line width=1.5pt}, 
    width=4.5cm,
    height=4.5cm,
    scale only axis,
    tick align=outside,
    xtick=\empty,
    ytick=\empty,
    xticklabels={},
    yticklabels={},
    xlabel style={at={(axis cs:0.17,0)}, anchor=north east}, 
    ylabel style={at={(axis cs:-0.00,0.13)}, anchor=south east}, 
]

\addplot [
    dashed,
    black,
    line width=2pt,
    domain=0:360,
    samples=180
]
({0.05*cos(x)}, {0.05*sin(x)});

\addplot [
    -stealth,
    blue,
    line width=3.5pt,
    domain=105:-135,
    samples=180
]
({0.05*(1 + 1.41421*cos(x))},
 {0.05*(1.41421*sin(x))});

\addplot [
    -stealth,
    red,
    line width=3.5pt,
    domain=165:225,
    samples=180
]
({0.05*(1 + 1.41421*cos(x))},
 {0.05*(1.41421*sin(x))});

\addplot [
    -stealth,
    blue,
    line width=3.5pt,
    domain=155:120,
    samples=180
]
({0.12*(1 + 1.41421*cos(x))},
 {0.12*(1.41421*sin(x))});

\addplot [only marks, mark=o, blue, mark size=4pt, mark options={fill=white, line width=1.5pt}] coordinates {
    (-0.02, 0.07)
};

\end{axis}
\end{tikzpicture}

%% file: latex/super_trans1_c.tex
\begin{tikzpicture}
\begin{axis}[
    xlabel={$\mathrm{Re}(\xi)$},
    ylabel={$\mathrm{Im}(\xi)$},
    xmin=-0.07, xmax=0.16,
    ymin=-0.08, ymax=0.15,
    axis lines=center,
    axis line style={-,line width=1.5pt}, 
    width=4.5cm,
    height=4.5cm,
    scale only axis,
    tick align=outside,
    xtick=\empty,
    ytick=\empty,
    xticklabels={},
    yticklabels={},
    xlabel style={at={(axis cs:0.17,0)}, anchor=north east}, 
    ylabel style={at={(axis cs:-0.00,0.13)}, anchor=south east}, 
]


\addplot [
    dashed,
    black,
    line width=2pt,
    domain=0:360,
    samples=180
]
({0.05*cos(x)}, {0.05*sin(x)});

\addplot [
    -stealth,
    blue,
    line width=3.5pt,
    domain=125:-135,
    samples=180
]
({0.05*(1 + 1.41421*cos(x))},
 {0.05*(1.41421*sin(x))});

\addplot [
    -stealth,
    red,
    line width=3.5pt,
    domain=145:225,
    samples=180
]
({0.05*(1 + 1.41421*cos(x))},
 {0.05*(1.41421*sin(x))});

\addplot [
    -stealth,
    blue,
    line width=3.5pt,
    domain=155:120,
    samples=180
]
({0.12*(1 + 1.41421*cos(x))},
 {0.12*(1.41421*sin(x))});

\addplot [only marks, mark=o, blue, mark size=4pt, mark options={fill=white, line width=1.5pt}] coordinates {
    (-0.02, 0.07)
};

\addplot [only marks, mark=*, red, mark size=3.1pt] coordinates {
    (-0.035, 0.04)
};

\addplot [only marks, mark=*, blue, mark size=3.1pt] coordinates {
    (0.0035, 0.075)
};

\end{axis}
\end{tikzpicture}

%% file: latex/super_trans2_c.tex
\begin{tikzpicture}
\begin{axis}[
    xlabel={$\mathrm{Re}(\xi)$},
    ylabel={$\mathrm{Im}(\xi)$},
    xmin=-0.07, xmax=0.16,
    ymin=-0.08, ymax=0.15,
    axis lines=center,
    axis line style={-,line width=1.5pt}, 
    width=4.5cm,
    height=4.5cm,
    scale only axis,
    tick align=outside,
    xtick=\empty,
    ytick=\empty,
    xticklabels={},
    yticklabels={},
    xlabel style={at={(axis cs:0.17,0)}, anchor=north east}, 
    ylabel style={at={(axis cs:-0.00,0.13)}, anchor=south east}, 
]


\addplot [
    dashed,
    black,
    line width=2pt,
    domain=0:360,
    samples=180
]
({0.05*cos(x)}, {0.05*sin(x)});

\addplot [
    -stealth,
    blue,
    line width=3.5pt,
    domain=135:-135,
    samples=180
]
({0.05*(1 + 1.41421*cos(x))},
 {0.05*(1.41421*sin(x))});

\addplot [
    -stealth,
    blue,
    line width=3.5pt,
    domain=70:45,
    samples=180
]
({0.05*(-1 + 1.41421*cos(x))},
 {0.05*(1.41421*sin(x))});

 \addplot [
    -stealth,
    red,
    line width=3.5pt,
    domain=15:45,
    samples=180
]
({0.05*(-1 + 1.41421*cos(x))},
 {0.05*(1.41421*sin(x))});

\addplot [
    -stealth,
    red,
    line width=3.5pt,
    domain=135:225,
    samples=180
]
({0.05*(1 + 1.41421*cos(x))},
 {0.05*(1.41421*sin(x))});

\addplot [
    -stealth,
    blue,
    line width=3.5pt,
    domain=155:120,
    samples=180
]
({0.14*(1 + 1.41421*cos(x))},
 {0.14*(1.41421*sin(x))});

\addplot [only marks, mark=o, blue, mark size=4pt, mark options={fill=white, line width=1.5pt}] coordinates {
    (-0.03, 0.08)
};

\addplot [only marks, mark=*, red, mark size=3.1pt] coordinates {
    (-0.04, 0.045)
};

\addplot [only marks, mark=*, blue, mark size=3.1pt] coordinates {
    (-0.0040, 0.09)
};

\end{axis}
\end{tikzpicture}

%% file: latex/super_trans3_c.tex
\begin{tikzpicture}
\begin{axis}[
    xlabel={$\mathrm{Re}(\xi)$},
    ylabel={$\mathrm{Im}(\xi)$},
    xmin=-0.07, xmax=0.16,
    ymin=-0.08, ymax=0.15,
    axis lines=center,
    axis line style={-,line width=1.5pt}, 
    width=4.5cm,
    height=4.5cm,
    scale only axis,
    tick align=outside,
    xtick=\empty,
    ytick=\empty,
    xticklabels={},
    yticklabels={},
    xlabel style={at={(axis cs:0.17,0)}, anchor=north east}, 
    ylabel style={at={(axis cs:-0.00,0.13)}, anchor=south east}, 
]


\addplot [
    dashed,
    black,
    line width=2pt,
    domain=0:360,
    samples=180
]
({0.05*cos(x)}, {0.05*sin(x)});

\addplot [
    -stealth,
    blue,
    line width=3.5pt,
    domain=135:-135,
    samples=180
]
({0.05*(1 + 1.41421*cos(x))},
 {0.05*(1.41421*sin(x))});

\addplot [
    -stealth,
    blue,
    line width=3.5pt,
    domain=75:45,
    samples=180
]
({0.05*(-1 + 1.41421*cos(x))},
 {0.05*(1.41421*sin(x))});

 \addplot [
    -stealth,
    red,
    line width=3.5pt,
    domain=10:45,
    samples=180
]
({0.05*(-1 + 1.41421*cos(x))},
 {0.05*(1.41421*sin(x))});

\addplot [
    -stealth,
    red,
    line width=3.5pt,
    domain=135:225,
    samples=180
]
({0.05*(1 + 1.41421*cos(x))},
 {0.05*(1.41421*sin(x))});

\addplot [
    -stealth,
    blue,
    line width=3.5pt,
    domain=158:120,
    samples=180
]
({0.17*(1 + 1.41421*cos(x))},
 {0.14*(1.41421*sin(x))});

\addplot [only marks, mark=o, blue, mark size=4pt, mark options={fill=white, line width=1.5pt}] coordinates {
    (-0.045, 0.07)
};

\addplot [only marks, mark=*, red, mark size=3.1pt] coordinates {
    (-0.05, 0.035)
    (-0.04, 0.01)
    (-0.035, -0.01)
    (-0.025, -0.03)

};

\addplot [only marks, mark=*, blue, mark size=3.1pt] coordinates {
    (-0.010, 0.09)
    ( 0.015, 0.10)
    ( 0.05, 0.11)
    ( 0.09, 0.115)
};

\end{axis}
\end{tikzpicture}

%% file: latex/super_trans4_c.tex
\begin{tikzpicture}
\begin{axis}[
    xlabel={$\mathrm{Re}(\xi)$},
    ylabel={$\mathrm{Im}(\xi)$},
    xmin=-0.07, xmax=0.13,
    ymin=-0.07, ymax=0.13,
    axis lines=center,
    axis line style={-,line width=1.5pt}, 
    width=4.5cm,
    height=4.5cm,
    scale only axis,
    tick align=outside,
    xtick=\empty,
    ytick=\empty,
    xticklabels={},
    yticklabels={},
    xlabel style={at={(axis cs:0.14,0)}, anchor=north east}, 
    ylabel style={at={(axis cs:-0.00,0.11)}, anchor=south east}, 
]


\addplot [
    dashed,
    black,
    line width=2pt,
    domain=0:360,
    samples=180
]
({0.05*cos(x)}, {0.05*sin(x)});

\addplot [
    -stealth,
    blue,
    line width=3.5pt,
    domain=190:270,
    samples=180
]
({0.055*(cos(x))},
 {0.055*(sin(x))});

 {0.055*(sin(x))});

 \addplot [
    -stealth,
    blue,
    line width=3.5pt,
    domain=50:-90,
    samples=180
]
({0.055*(cos(x))},
 {0.055*(sin(x))});

 {0.055*(sin(x))});

\addplot [only marks, mark=o, blue, mark size=4pt, mark options={fill=white, line width=1.5pt}] coordinates {
    (0.01, 0.05)
};

\addplot [-stealth, domain=0:0.1, samples=2, blue, line width=3.5pt] coordinates {(0,0.055) (0,0.13)};

\addplot [only marks, mark=*, red, mark size=3.1pt] coordinates {
    (0.012, 0.04)
    (+0.02, 0.03)
    (+0.014, 0.02)
    (+0.009, 0.01)
    (+0.003, -0.0)
    (+0.002, -0.01)
    (+0.001, -0.02)
    (0., -0.03)
    (0., -0.04)

};

\addplot [only marks, mark=*, blue, mark size=3.1pt] coordinates {
    (0.02, 0.06)
    ( 0.04, 0.055)
    ( 0.055, 0.07)
    ( 0.08, 0.085)
    ( 0.11, 0.1)    
};

\end{axis}
\end{tikzpicture}